\documentclass[prb,aps,showpacs,twocolumn,preprintnumbers,amsmath,amssymb,superscriptaddress,longbibliography,nofootinbib]{revtex4-2}
\usepackage[english]{babel}
\usepackage{amsmath,amssymb,bbm,mathrsfs,mathtools,multirow,diagbox,xcolor,graphicx,tabularx,comment,amsfonts,dsfont,lettrine,units,comment}

\newcolumntype{L}{>{\centering\arraybackslash}X}

\usepackage{graphicx}
\usepackage[colorlinks=True,linkcolor=blue,citecolor=blue,urlcolor=blue]{hyperref}
\usepackage{bookmark}
\usepackage{xcolor}
\usepackage{chngcntr}
\usepackage{braket}
\usepackage{nicefrac}
\usepackage{bm}
\usepackage{bbm}
\usepackage{braket}
\usepackage{lineno}
\usepackage{array}
\newcolumntype{C}{>{$}c<{$}}
\usepackage{newtxtext} 
\usepackage[normalem]{ulem}
\usepackage{titletoc}

\titlecontents{section}[1.5em]{}{\contentslabel{1.5em}}{\hspace*{-2.3em}}{}

\renewcommand{\dag}{^{\dagger}}

\usepackage{gensymb}

\begin{document}

\title{Twist-angle evolution of the intervalley-coherent antiferromagnet in twisted WSe$_2$}

\author{Daniel Mu\~noz-Segovia}
\email{dm3886@columbia.edu}
\affiliation{Department of Physics, Columbia University, New York, NY 10027, USA}
\author{Valentin Cr\'epel}
\affiliation{Center for Computational Quantum Physics, Flatiron Institute, New York, NY 10010, USA}
\author{Raquel Queiroz}
\affiliation{Department of Physics, Columbia University, New York, NY 10027, USA}
\author{Andrew J. Millis}
\affiliation{Department of Physics, Columbia University, New York, NY 10027, USA}
\affiliation{Center for Computational Quantum Physics, Flatiron Institute, New York, NY 10010, USA}

\date{\today}

\begin{abstract}
Recent experimental reports of correlated physics in twisted homobilayer WSe$_2$ have spurred interest in the interplay of electronic interactions and topology in this system. Here, we explore its phase diagram using the Hartree-Fock approximation within a three-orbital Wannier model of the bilayer. 
Our analysis reveals a dominant intervalley-coherent antiferromagnetic instability, whose stability in the space of twist angle, interaction strength, out-of-plane displacement field, and hole density is primarily set by nesting and commensurability. 
At large angles or low interaction-to-bandwidth ratios, the instability arises at hole densities above half filling near a van-Hove line where the strong Fermi surface nesting occurs due to the flatness of the band in a region enclosing the van-Hove and $\kappa$ points. 
Increasing interaction strength or decreasing the twist angle gradually shifts the ordered phase toward half filling, where the strongest antiferromagnetic order gets pinned due to commensurability effects that enable a full gap opening. 
The antiferromagnetic order parameter strongly couples to the layer polarization, which makes its transition to the normal state sharp in the strong-coupling limit and carries implications for collective modes.
Our Hartree-Fock phase diagram reproduces key aspects of recent experiments and the reconstructed Fermi surfaces and DOS in the antiferromagnetic phase account for subtle transport signatures observed in these studies. 
\end{abstract}

\maketitle

\section{Introduction}

Twisted transition metal dichalcogenides (TMDs) provide a tunable platform to realize a variety of correlated and topological phases, from correlated insulators and Wigner crystals~\cite{wang_correlated_2020,xu_correlated_2020,huang_correlated_2021,ghiotto_stoner_2024,knuppel_mak_correlated_2024} to integer and fractional quantum anomalous (spin) Hall states~\cite{tao2024valley,kang2024double,zeng2023thermodynamic,cai2023signatures,crepel2023anomalous,xu2023observation,park2023observation,foutty_mapping_2024,kang2024evidence,crepel2024spinon}. 
Tuning parameters that have been used to control the physics include twist angle, carrier concentration, and ``displacement field" (interlayer potential difference). Among the variety of twisted TMDs, rhombohedral-stacked twisted WSe$_2$ (tWSe$_2$) realizes a rich phase diagram upon hole doping, with coexisting correlated states and superconductivity~\cite{xia_superconductivity_2024,guo_superconductivity_2025}, whose evolution with experimental parameters remains to be understood. The  $\theta=5^\circ$ twist-angle device of Ref.~\cite{guo_superconductivity_2025} exhibits correlated metal (``Fermi surface reconstructed") and superconducting phases appearing as the density and displacement field are tuned to place the Fermi level near a van Hove singularity (VH) of the band, which points to a description based on weak-coupling physics. At a smaller twist $\theta=3.5^\circ$~\cite{xia_superconductivity_2024}, a correlated insulator is observed at half filling for a range of displacement fields, which is apparently not tied to a van Hove singularity and transitions into a superconductor for small displacement fields, perhaps suggesting intermediate or strong coupling physics. In an intermediate regime $\theta=4.2^\circ$~\cite{ghiotto_stoner_2024}, the correlated state has been observed to continuously evolve from an insulator at half filling to a metal with a Fermi surface reconstruction along the VH. Finally, a sample with even smaller twist $\theta=2.7^\circ$ shows a correlated insulator at half filling extending from vanishing displacement field, as well as signs of ferromagnetism below half filling~\cite{knuppel_mak_correlated_2024}.

Several theoretical scenarios have been proposed to explain the origin and properties of these phases~\cite{bi_Fu_excitonic_tWSe2_2021,schrade_Fu_SC_tWSe2_2024,wietek2022tunable,crepel_topological_2023,kim2024theory,myerson-jain_superconductor-insulator_2024,zhu2024theory,christos2024approximate,xie2024superconductivity,guerci2024topologicalSC,tuo2024theorytopo,qin2024kohn,chubukov2024quantum,fischer2024theory,peng_magnetism_2025}. A conclusion raised from comparing these more recent works with the previous literature on tWSe$_2$ based on the one-band moir\'e Hubbard model~\cite{pan_band_2020,zang_hartree-fock_2021,klebl2023competition} is the importance of a proper description of the noninteracting model, including the quantum geometry and Berry curvature. On the one hand, the evolution with displacement field of the topmost band dispersion, especially around the VH points, crucially determines the weak-coupling phase diagram. On the other hand, due to the link between the topology and the real-space structure of the wavefunctions~\cite{wu2019topological,zhang_direct_2024,crepel2024chiral}, interactions projected to the topmost band acquire nontrivial form factors which seem to be required to correctly reproduce the ground state in the intermediate-coupling regime~\cite{qiu_interaction-driven_2023,li_electrically_2024}.
Significant progress in capturing this physics was made in Ref.~\cite{fischer2024theory}, which presented consistent evidence from both single-particle susceptibility analysis and full-fledged functional renormalization group calculations, providing a comprehensive picture of the weak-coupling instabilities {including superconductivity} without explicitly accessing the ordered phases or incorporating self-energy corrections. 
Important aspects of the experiments that have still not been fully theoretically elucidated include the intricate fermiology near the simple and higher-order van Hove points as well as extended regions of enhanced density of states, the interplay between magnetic order and layer polarization and the evolution of the physics with twist angle.

In this paper, we study the zero-temperature Hartree-Fock phase diagram of tWSe2 as a function of twist angle, interaction strength, displacement field and hole density. This paper is organized as follows. Sec.~\ref{sec:model} describes the three-orbital model used in this work, which is obtained from the projected Wannierization of the continuum model~\cite{wu2019topological,devakul2021magic}, and correctly captures the topology and interaction form factors of the topmost two moir\'e bands~\cite{qiu_interaction-driven_2023,li_electrically_2024,crepel_bridging_2024,fischer2024theory}. Motivated by the $\theta=5^\circ$ experiment~\cite{guo_superconductivity_2025}, in Sec.~\ref{sec:fermiology}
we analyze the fermiology and bare susceptibility of the model, focusing on the physics close to the VH. The flatness of the bands in a region between the VH and $\kappa$ points of the moir\'e Brillouin zone causes a large intervalley susceptibility around the $\kappa$ point of the moir\'e Brillouin zone~\cite{bi_Fu_excitonic_tWSe2_2021,schrade_Fu_SC_tWSe2_2024}. This motivates the study of the Hartree-Fock instabilities in a $\sqrt{3}\times\sqrt{3}$ supercell, described in Sec.~\ref{sec:HF_phase_diag}. We find that the leading instability is an intervalley-coherent antiferromagnet (IVC-AFM), compatible with previous theory works~\cite{zang_hartree-fock_2021,tuo2024theorytopo,fischer2024theory}. The IVC-AFM appears along the VH at densities larger than half filling, and with decreasing twist angle it continuously extends to lower densities until reaching half filling, where the opening of a spectral gap strongly pins it, in agreement with the experiments of Refs.~\cite{ghiotto_stoner_2024,xia_superconductivity_2024,guo_superconductivity_2025,knuppel_mak_correlated_2024,knuppel_mak_correlated_2024}. In Sec.~\ref{sec:FSR}, we examine the reconstructed band structure, Fermi surface and density of states (DOS) within the IVC-AFM state. Notably, we have found a strong coupling between the IVC-AFM and the layer polarization. In Sec.~\ref{sec:coupling_AFM-LP}, we describe this strong magneto-electric coupling, which is enhanced by the interlayer interactions and results in sharp transitions from the normal state at small twist angles. Finally, in Sec.~\ref{sec:experiments} we summarize the experimental findings of Refs.~\cite{ghiotto_stoner_2024,xia_superconductivity_2024,guo_superconductivity_2025,knuppel_mak_correlated_2024,knuppel_mak_correlated_2024}, and show that the parameter dependence and transport fingerprints of the correlated state can be explained by our results.

\section{Model}
\label{sec:model}

Upon weak hole-doping, H-polytype transition metal dichalcogenides (TMDs) exhibit degenerate spin-valley locked pockets at the $K$ and $K'$ valleys~\cite{manzeli20172d,xiao2012coupled,dey2017gate}. When two such TMD monolayers are stacked with a small twist angle, a long-period hexagonal moir\'e pattern emerges, over which non-uniform interlayer hybridization and electrostatic potentials develop~\cite{wu2019topological}. The electrostatic potential is minimal at the vertices of the honeycomb lattice formed by the so-called MX and XM points of the moir\'e superlattice~\cite{carr2018relaxation,devakul2021magic}. Inter-layer hybridization, on the other hand, is maximal at the triangular lattice formed by the centers of the hexagons, the so-called MM points~\cite{tong2017topological}. The essential ingredients to describe the physics of twisted TMDs are, therefore, contained in the local properties of the system near the XM, MX, and MM points. 

\begin{figure}[!t]
    \centering
    \includegraphics[width=0.6\linewidth]{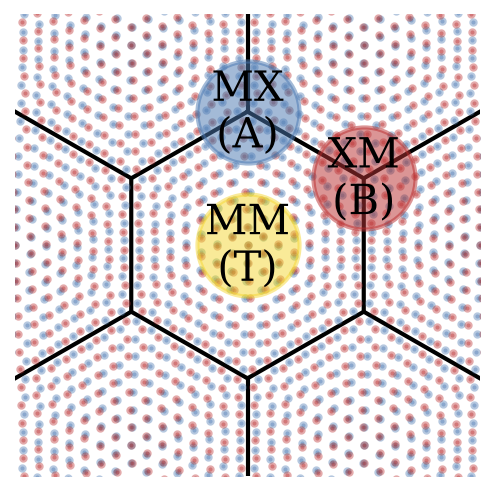}
    \caption{Sketch of the moir\'e structure of tWSe$_2$, highlighting the high-symmetry positions (MM, MX, XM) where the Wannier functions (T, A, B) of the moir\'e three-orbital model are centered. Small blue and red dots represent the individual metal atoms from the top and bottom layers, respectively.}
    \label{fig:model}
\end{figure}

This intuition was placed on firmer ground in Ref.~\cite{crepel_bridging_2024}, where a three orbital tight-binding model faithfully capturing the dispersion and topology of twisted TMDs was derived. 
It contains two orbitals $|\phi_{\mathrm{A}/\mathrm{B}}\rangle$ centered at the MX and XM honeycomb sites and nearly layer-polarized in the top and bottom layers, respectively, and another orbital $|\phi_{\mathrm{T}}\rangle$ located at the MM triangular sites, with comparable weight in both layers (see Fig.~\ref{fig:model}). 
In this work, we use this model to describe the topmost valence bands of twisted WSe$_2$ AA-homobilayers. 
For a single spin-valley $\sigma$, the model Hamiltonian reads:
\begin{equation}
    H_0^\sigma = \sum_{i,j=1}^N \sum_{\alpha,\beta=T,A,B} t_{i\alpha,j\beta}^{\sigma} \phi\dag_{i\alpha\sigma} \phi_{j\beta\sigma},
\label{eq:H0}
\end{equation}
where $i,j$ label the unit cell, and $\alpha,\beta={\rm T,A,B}$ label the orbitals. We extract all tight-binding parameters by single-shot Wannierization of the continuum model from Ref.~\cite{devakul2021magic} by the projection of Gaussian trial wavefunctions~\cite{Pizzi2020}, as detailed in Ref.~\cite{fischer2024theory} and App.~\ref{app:wannier}. The hopping amplitudes $t_{i\alpha,j\beta}^{\sigma}$ are set to zero if $|\boldsymbol{r}_i^{\alpha}-\boldsymbol{r}_j^{\beta}|>R_c$, with $R_c$ a cutoff equal to 9 times the moir\'e lattice constant. We extract the tight-binding parameters for a set of displacement fields, which are modeled as an interlayer potential difference $E_z$ in the continuum model~\cite{devakul2021magic}.

The tight-binding model inherits the symmetry properties of the continuum model. At vanishing displacement field, the system has time-reversal symmetry $\mathcal{T}$, $U_{\mathrm{V}}(1)$ valley symmetry and point group $D_{3d}$, generated by the threefold rotational symmetry $C_{3z}$ about the out-of-plane axis and the twofold rotational symmetry $C_{2x}$ about the $x$ axis. The model possesses an additional intravalley inversion symmetry $i$, resulting from retaining only the first harmonic of the moir\'e potential and tunneling in the continuum model~\cite{wang2021chiral,song2021twisted,crepel2023chiral,christos2024approximate}. While this symmetry is broken by higher harmonics and is therefore not exactly present in the physical system, it remains a good approximation and has been widely assumed in the literature~\cite{wang2021chiral,song2021twisted,crepel2023chiral,christos2024approximate}. The combination of this intravalley inversion together with $\mathcal{T}$ forces the degeneracy between opposite spins at each $\boldsymbol{k}$-point. A finite displacement field breaks $C_{2x}$ and $i$, which lifts the spin degeneracy, but preserves the mirror symmetry $m_x = i C_{2x}$, thus reducing the point group to $C_{3v}$ (see App.~\ref{app:symmetry_analysis} for the group-theory analysis).

\begin{figure*}[!t]
    \centering
    \includegraphics[width=1\linewidth]{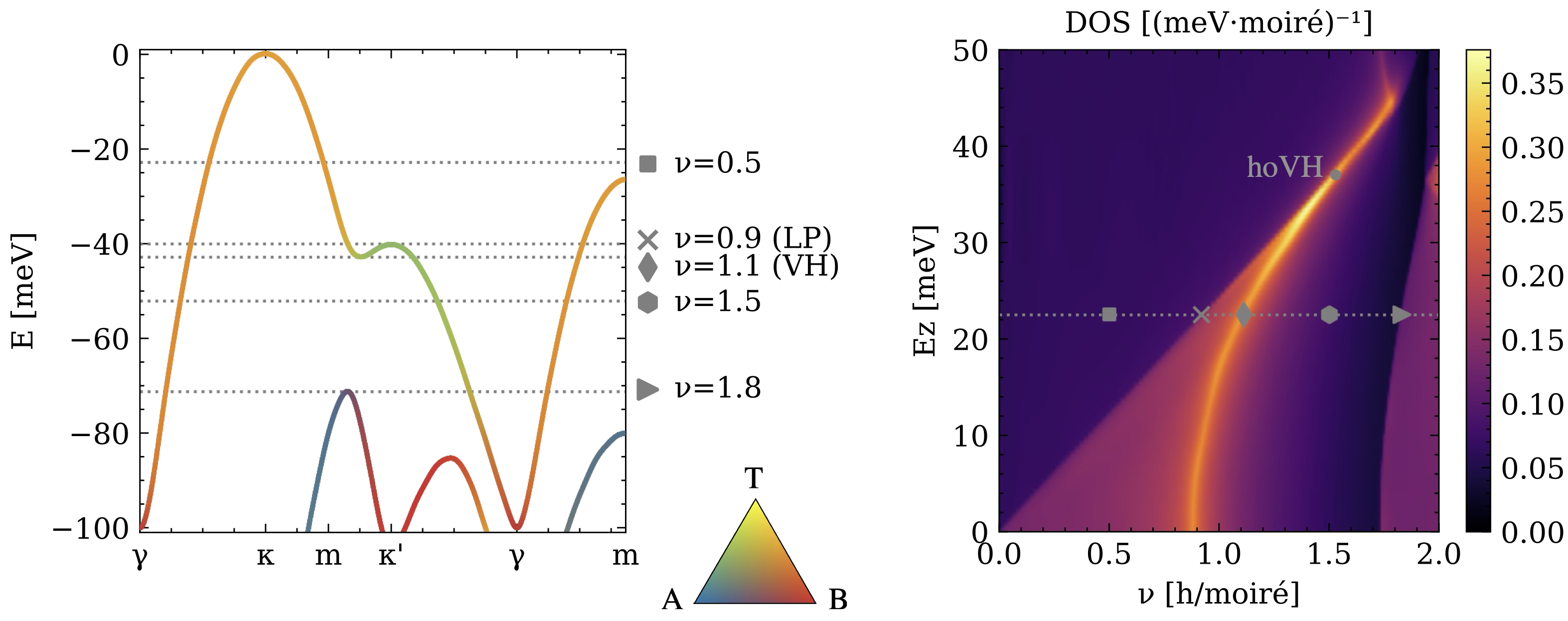}
    \caption{(a) Orbital-resolved band structure for the valley $K'$ of $\theta=5^\circ$ and $E_z=22.5\mathrm{meV}$. Due to the layer polarization properties of the Wannier orbitals, red/orange represents predominance of the bottom layer, while blue/green indicates top layer polarization. The horizontal dotted lines indicate the chemical potential for four representative fillings: $\nu=0.5$ (square), deep in the layer-polarized regime with one hole pocket around $\kappa$; $\nu=0.9$ (cross), at the layer-polarization point where the hole pocket around $\kappa'$ onsets; $\nu=1.1$ (diamond), at the VH; $\nu=1.5$ (hexagon), where the Fermi level crosses one electron pocket around $\gamma$; and $\nu=1.8$ (triangle), where the second bands starts to be filled. 
    (b) DOS as a function of filling $\nu$ and interlayer potential $E_z$ for $\theta=5^\circ$ and smearing $\eta=0.1\mathrm{meV}$. The horizontal dotted line signals $E_z=22.5\mathrm{meV}$, whose band structure is plotted in (a). The four representative chemical potentials highlighted in (a) are signaled by the corresponding markers. The gray circle marked hoVH denotes the point where the VH and the Lifshitz transition lines meet ($E_z=37\mathrm{meV}$, $\nu=1.5$).} 
    \label{fig:bands_nonint}
\end{figure*}

To study the possible instabilities, we project the long-range Coulomb interaction into the Wannier basis. The resulting leading interaction terms in the tight-binding model are onsite Hubbard $U_\alpha$ and nearest-neighbor density-density $V_{\alpha\beta}$ interactions, which are larger than other quartic terms connecting nearest neighbors by a factor $>10$ (see App.~\ref{app:interactions}). Moreover, the distances to the gates in Ref.~\cite{xia_superconductivity_2024} are relatively small compared to the moir\'e lattice constant, which will further suppress longer-range interaction terms. The interaction Hamiltonian thus reads:
\begin{equation}
H_{\mathrm{int}} = \frac{1}{\epsilon} \left[ \sum_{i\alpha} U_\alpha n_{i\alpha\uparrow} n_{i\alpha\downarrow} + \sum_{\langle i\alpha, j\beta \rangle} V_{\alpha\beta} n_{i\alpha} n_{j\beta} \right],
\label{eq:Hint}
\end{equation}
where $\epsilon$ is the dielectric constant of the surrounding environment, and $n_{i\alpha\sigma} = \phi\dag_{i\alpha\sigma} \phi_{i\alpha\sigma}$ and $n_{i\alpha} = \sum_\sigma n_{i\alpha\sigma}$ are the spin-resolved and total onsite densities, respectively. In practice, we consider equal onsite $U_\mathrm{A} = U_\mathrm{B}$ and nearest-neighbor $V_{\mathrm{TA}} = V_{\mathrm{TB}}$ interactions for the honeycomb lattice, which is exactly true in the absence of displacement field and a good approximation for finite $E_z$. In App.~\ref{app:interactions}, we show that, for fixed twist angle, $U_\alpha$ and $V_{\alpha\beta}$ do not significantly change with $E_z$; we fix them to their values at $E_z=0$ throughout our study. 

Our theory therefore has two control parameters, the dielectric constant $\epsilon$ and the twist angle, which control the ratio between the energy scales of the problem. A larger $\epsilon$ decreases the interaction coefficients $U,V$ compared to the bandwidth. Smaller twist angles generically result in more localized Wannier functions, thereby reducing the $V/U$ ratio. We refer to App.~\ref{app:interactions} for the values of the interactions used throughout our study.

\section{Fermiology analysis}
\label{sec:fermiology}

\subsection{Band structure and density of states}

\begin{figure*}[!t]
    \centering
    \includegraphics[width=\linewidth]{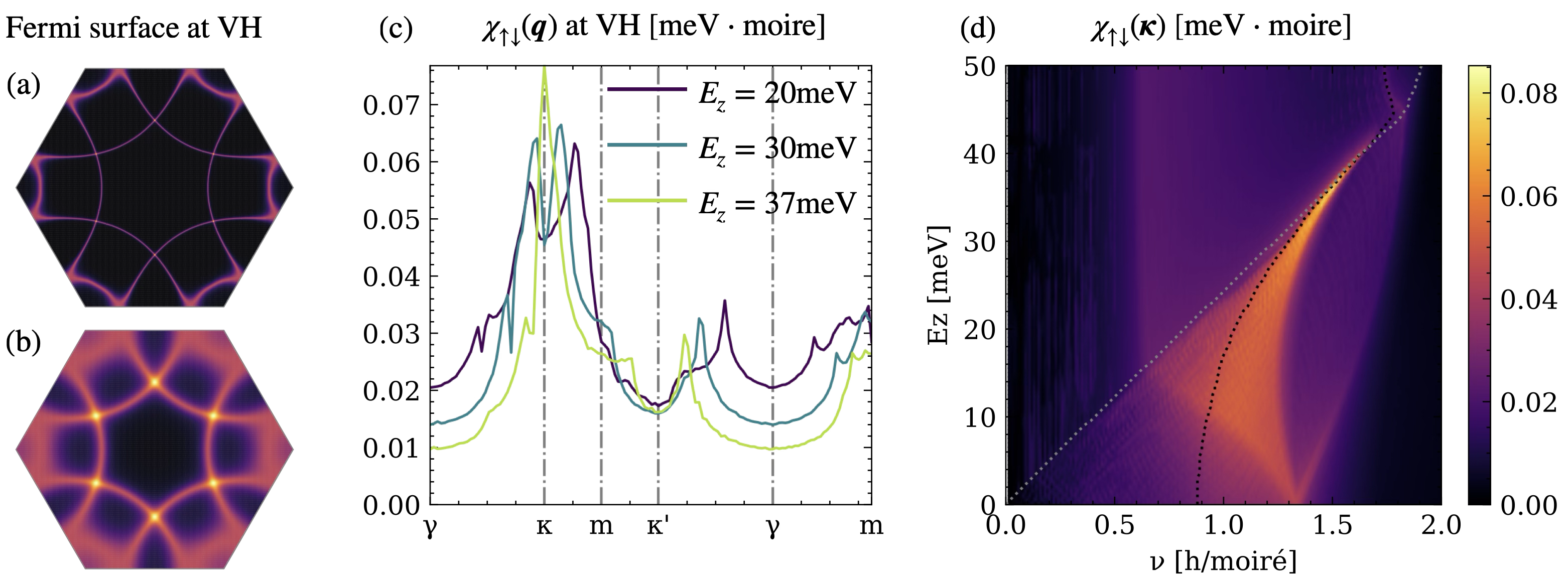}
    \caption{Fermi surface and bare particle-hole susceptibility of the topmost band for $\theta=5^\circ$. 
    (a,b) Spectral function at the Fermi level for the VH point $E_z=22.5\mathrm{meV}$, $\nu=1.1$, determined using a smearing $\eta=0.5\mathrm{meV}$ (a) and $\eta=5\mathrm{meV}$ (b). 
    (c) Intervalley susceptibility $\chi_{\uparrow\downarrow}(\boldsymbol{q})$ as a function of momentum $\boldsymbol{q}$ for several $E_z$ at the VH filling.
    (d) Intervalley susceptibility $\chi_{\uparrow\downarrow}(\kappa)$ at momentum $\kappa$ as a function of density and displacement field. 
    The smearing parameter used in (c) and (d) is $\eta=0.1\mathrm{meV}$.}
    \label{fig:chi_nonint}
\end{figure*}

We first focus on the topmost band only, which is separated by an indirect gap from the second band. 
In the presence of a small twist $\theta$, the top layer $K$ and $K'$ points are downfolded onto the $\kappa$ and $\kappa'$ corners of the mini-Brillouin zone, respectively, where the topmost valence band hence exhibits a strong top layer-polarization. In each valley $K/K'$, the intravalley inversion symmetry $i$ enforces a degeneracy and opposite layer polarization between $\kappa$ and $\kappa'$. A non-zero displacement field $E_z$ breaks $i$ and shifts the bottom layer (and therefore the B sublattice), upward in energy with respect to the top layer (A sublattice), thus controlling the energy difference between $\kappa$ and $\kappa'$. This is exemplified by Fig.~\ref{fig:bands_nonint}(a), which displays the resulting band structure in valley $K'$ for twist angle $\theta = 5^{\mathrm{o}}$ at a displacement field corresponding to an interlayer potential difference $E_z = 22.5 \mathrm{meV}$. 

As a function of the hole density $\nu$ at non-zero $E_z$, valley $K'$ starts with a single hole pocket at the $\kappa$ point of the moir\'e Brillouin zone, which is partially layer polarized in the bottom layer, with weight on the B and T sublattices (square in Fig.~\ref{fig:bands_nonint}). A smaller hole pocket at $\kappa'$, dominated by the A and T sublattices, onsets at larger $\nu$ (cross in Fig.~\ref{fig:bands_nonint}), which is reflected in Fig.~\ref{fig:bands_nonint}(b) as a step increase of the density of states (DOS) at the point marked by a gray cross. This Lifshitz transition has been loosely referred to as ``layer polarization'' in the literature due to the layer structure of the hole pockets at $\kappa$ and $\kappa'$. 
Despite this naming, we point out that both layers still have sizable occupancy in the ``layer polarized'' regime and the only signature of crossing the Lifshitz transition is a change in the slope of the layer polarization with $E_z$.

Further increasing the hole doping, the large and small hole pockets touch at three inequivalent van Hove singularities (VHs) per each valley, signaled by a peak in the DOS marked by a diamond in Fig.~\ref{fig:bands_nonint}(b). The three VH touching points continuously evolve from the $m$ points for $E_z=0$ (at $\nu\sim0.9$) towards the $\kappa'$ point as $E_z$ increases, until they meet at a higher-order VH point (hoVH) at $\kappa'$ for $E_z^{\mathrm{hoVH}}\sim 37\mathrm{meV}$ (at $\nu\sim1.5$)~\cite{bi_Fu_excitonic_tWSe2_2021,hsu_spin-valley_2021}. In the phase diagram $E_z$-$\nu$ of Fig.~\ref{fig:bands_nonint}(b), the hoVH corresponds to the crossing between the VH peak and the layer-polarization line, marked by a circle. At $E_z$ higher than $E_z^{\mathrm{hoVH}}$, the band at $\kappa'$ becomes a relative minimum, and the VHs move towards the $\gamma$ point along the $\gamma\kappa'$ lines. For a fixed $E_z<E_z^{\mathrm{hoVH}}$, increasing the hole doping beyond the VH filling gives rise to a single electron pocket around $\gamma$ per each valley (hexagon in Fig.~\ref{fig:bands_nonint}). Finally, we point out that the step increase in the DOS of Fig.~\ref{fig:bands_nonint}(b) at $E_z<40\mathrm{meV}$ and $\nu$ close to 2 (triangle in Fig.~\ref{fig:bands_nonint}) corresponds to the onset of the second band, only separated from the topmost one by an indirect gap.

We note that smaller twist angles show qualitatively similar behavior, with the energy scales scaled by $\simeq \theta^2$ due to their smaller bandwidth (see App.~\ref{app:DOS_angles}). Besides this, the whole VH line is slightly pushed to lower density, including the hoVH point. Additionally, the indirect gap to the second band increases with decreasing twist angle, until it becomes a full gap at $E_z=0$ for $\theta \sim 3.5^\mathrm{o}$.

\subsection{Weak-coupling analysis: Susceptibility} \label{sec:susceptibility}

Motivated by the larger twist angle experiment with $\theta=5^\circ$ \cite{guo_superconductivity_2025}, where the correlated metal appears along the VH, we now analyze the weak-coupling instabilities of the system by computing the noninteracting static particle-hole susceptibility of the topmost band:
\begin{equation}
    \chi_{\sigma\tau}(\boldsymbol{q}) = \frac{1}{N} \sum_{\boldsymbol{k}} \frac{n_{1\sigma}(\boldsymbol{k}-\tfrac{\boldsymbol{q}}{2})-n_{1\tau}(\boldsymbol{k}+\tfrac{\boldsymbol{q}}{2})}{\varepsilon_{1\sigma}(\boldsymbol{k}-\tfrac{\boldsymbol{q}}{2})-\varepsilon_{1\tau}(\boldsymbol{k}+\tfrac{\boldsymbol{q}}{2})+i\eta},
\label{eq:susceptibility}
\end{equation}
where $\varepsilon_{n\sigma}(\boldsymbol{k})$ is the energy of band $n=1,2,3$ at valley $\sigma$, $n_{n\sigma}(\boldsymbol{k}) = \{1+\exp[(\varepsilon_{n\sigma}(\boldsymbol{k})-\mu)/(k_B T)]\}^{-1}$ is the Fermi occupation, and $\eta$ is the smearing parameter, which normalizes the possible divergences and controls the contribution from states away from the Fermi level. The intravalley $\chi_{\uparrow\uparrow}(\boldsymbol{q})$ and intervalley $\chi_{\uparrow\downarrow}(\boldsymbol{q})$ susceptibilities quantify the tendency towards valley-polarized and intervalley-coherent orders, respectively. Due to the divergent DOS at the VHs, $\chi_{\uparrow\uparrow}(\boldsymbol{q})$ and $\chi_{\uparrow\downarrow}(\boldsymbol{q})$ are expected to be divergent for the VH filling at wavevectors connecting VHs of the same and opposite valleys, respectively. We have found that $\chi_{\uparrow\downarrow}(\boldsymbol{q})$ is generically stronger than $\chi_{\uparrow\uparrow}(\boldsymbol{q})$ for tWSe$_2$. 

At zero displacement field, the VHs are located at the $m$ points, so that the nesting occurs at $\boldsymbol{q}=m$ for both the intravalley and intervalley channels. With increasing $E_z$, the maximum of the intravalley $\chi_{\uparrow\uparrow}(\boldsymbol{q})$ shifts to $\boldsymbol{q}=0$, indicating tendency towards out-of-plane ferromagnetism (see Fig.~\ref{Sfig:Intra_and_Inter_Susceptibility_vs_n_eta=01}(a) of App.~\ref{app:susceptibility}). On the other hand, the intervalley susceptibility $\chi_{\uparrow\downarrow}(\boldsymbol{q})$ peaks at several momenta corresponding to the different nesting wavevectors between the VHs of opposite valleys, see Fig.~\ref{fig:chi_nonint}(a). Its maximum evolves from $\boldsymbol{q}=m$ at $E_z=0$ to $\boldsymbol{q}=\kappa$ point at $E_z^{\mathrm{hoVH}}$ as shown in Fig.~\ref{fig:chi_nonint}(c). Notably, the intervalley susceptibility is more divergent than the intravalley one for $E_z<E_z^{\mathrm{hoVH}}$ (see App.~\ref{app:susceptibility}), signaling a weak coupling instability towards an in-plane spin density wave with in general incommensurate $\boldsymbol{q}$. The intervalley susceptibility has a sharp decrease after the hoVH, where the $\kappa$ ($\kappa'$) points at valley $K$ ($K'$) become unoccupied by holes, so that the $\boldsymbol{q}=0$ intravalley susceptibility becomes the leading one at $E_z>E_z^{\mathrm{hoVH}}$.

\begin{figure*}[!t]
    \centering
    \includegraphics[width=1\linewidth]{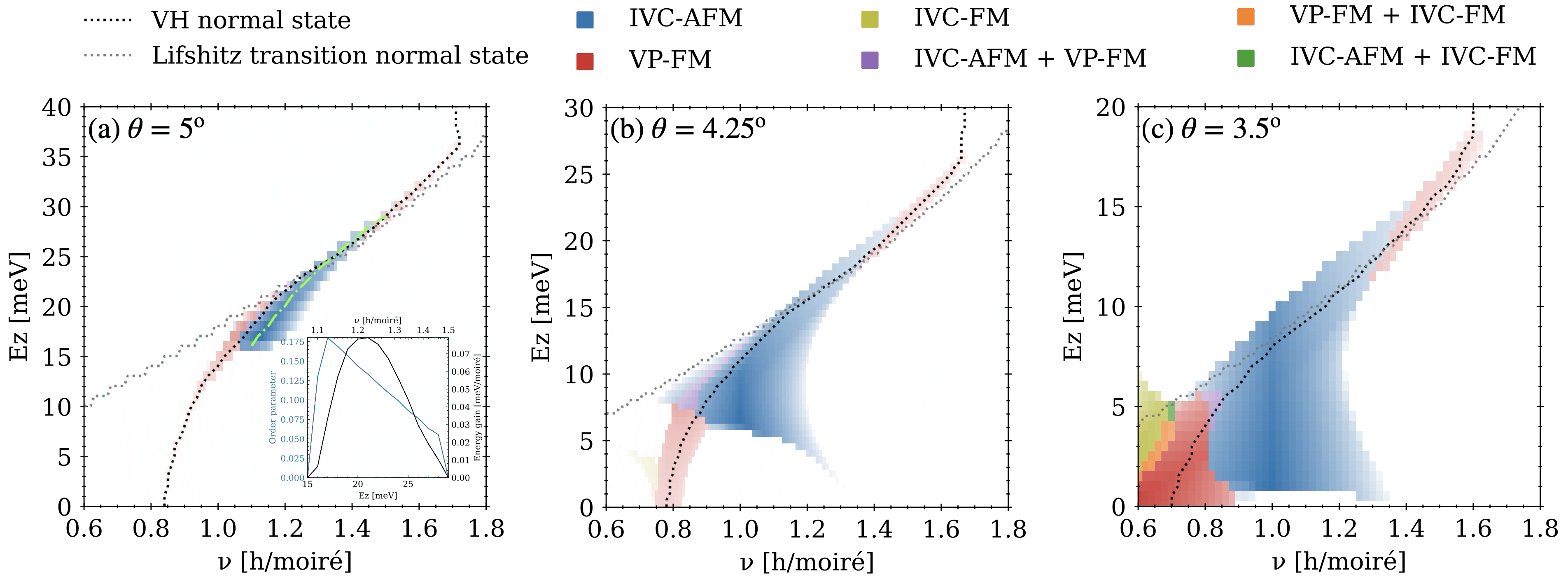}
    \caption{Hartree-Fock phase diagram of tWSe2 in a $\sqrt{3}\times\sqrt{3}$ supercell for dielectric constant $\epsilon=48$ and decreasing twist angle $\theta$: $\theta=5^\circ$ (a), $\theta=4.25^\circ$ (b), and $\theta=3.5^\circ$ (c). Different phases are indicated by different colors: IVC-AFM (blue), VP-FM (red), IVC-FM (yellow), coexisting IVC-AFM and VP-FM (purple), coexisting IVC-AFM and IVC-FM (green), and coexisting VP-FM and IVC-FM (orange). The intensity of each color is proportional to the sum of the order parameters (see App.~\ref{app:symmetry_analysis} for their definition), normalized for each $\theta$ (see Fig.~\ref{Sfig:HF_phase_diagram_angle_energy} in App.~\ref{app:HF_phase_diagram_angle_energy} for phase diagrams representing the magnitude of the order parameters). The normal state is defined as the self-consistent Hartree-Fock state which does not spontaneously break any symmetry. Black dotted lines indicate the VH line in the normal state, determined as the maximum of the DOS for $\eta=0.1\mathrm{meV}$. Gray dotted lines signal the Lifshitz transition or ``layer-polarization line'' in the normal state. The inset of (a) shows the IVC-AFM order parameter (blue) and its energy gain with respect to the normal state (black) for $\theta=5^\circ$ and $\epsilon=48$ along the green dash-dot line in (a), defined as the $E_z$-$\nu$ line where the IVC-AFM energy gain with respect to the normal state is largest.}
    \label{fig:HF_phase_diagram_angle}
\end{figure*}

While the commensurate $\boldsymbol{q}=\kappa$ wavevector corresponds to exact VH nesting only at the hoVH, there are three facts that highlight the importance of intervalley instabilities with $\boldsymbol{q}=\kappa$ even for $E_z\le E_z^{\mathrm{hoVH}}$. First, while the maximum of $\chi_{\uparrow\downarrow}(\boldsymbol{q})$ occurs at an incommensurate $\boldsymbol{q}$, it is relatively close to $\kappa$ for a wide range of $E_z\le E_z^{\mathrm{hoVH}}$. Second, for a reasonable smearing parameter $\eta$ ($\eta=0.1\mathrm{meV}\equiv 1 \mathrm{K}$ in Fig.~\ref{fig:chi_nonint}(c)), $\chi_{\uparrow\downarrow}(\boldsymbol{q})$ has a comparably large value along a closed curve surrounding the $\kappa$ point (see also Fig.~\ref{Sfig:Intra_and_Inter_Susceptibility_vs_n_eta=01} of App.~\ref{app:susceptibility}). This implies a competition between different incommensurate orders with $\boldsymbol{q} = \kappa + \boldsymbol{\delta q}$ for a continuously varying direction of $\boldsymbol{\delta q}$. The fluctuations derived from this competition might favor commensurate $\boldsymbol{q}=\kappa$ instabilities. Finally, the portion of the band between the VHs and the $\kappa$ ($\kappa'$) points at the $K$ ($K'$) valley remains remarkably flat for an extensive range of displacement fields: the energy difference between the VH and $\kappa$ ($\kappa'$) points remains smaller than $3\mathrm{meV}$ for $20\mathrm{meV} \lesssim E_z \lesssim E_z^{\mathrm{hoVH}}$ (corresponding to $1 \lesssim \nu \lesssim 1.5$). 
This effect is best captured by plotting the spectral functions at the Fermi level for a different smearing $\eta$. The near-zero temperature Fermi surfaces displayed in Fig.~\ref{fig:chi_nonint}(a) drastically broaden into patches that encompass the $\kappa$ points and show an approximate $\boldsymbol{\kappa}$-nesting for the smearing parameter $\eta = 5\mathrm{meV}$ used in Fig.~\ref{fig:chi_nonint}(b). Consequently, as shown in Fig.~\ref{fig:chi_nonint}(c), even if the intervalley nesting wavevectors at the VH filling are incommensurate, the susceptibility at $\boldsymbol{q}=\boldsymbol{\kappa}$ is comparable for a small smearing of the Fermi surface. 

Commensurate $\boldsymbol{q}=\boldsymbol{\kappa}$ instabilities will be further favored as the interaction strength is increased. On the one hand, increasing the interaction enhances the umklapp terms, which might drive an incommensurate to commensurate transition \cite{mcmillan_landau_1975,pokrovsky_ground_1979,saito_commensurate-incommensurate_1980}. On the other hand, the states away from the Fermi level become more relevant for the energetics of the interacting ground state. The contribution from the aforementioned relatively flat portions of the band around the $\kappa$ points will therefore enhance commensurate ordering at the wavevector $\boldsymbol{q}=\boldsymbol{\kappa}$. This effect can be quantified by the dependence of the susceptibility on the smearing, which controls the contribution from states away from the Fermi level. As expected from the the flatness of the band between the VH and ($\kappa'$) points at the $K$ ($K'$) valley, increasing the smearing decreases the ratio between the maximum $\chi_{\uparrow\downarrow}(\boldsymbol{q})$ and $\chi_{\uparrow\downarrow}(\kappa)$, indicating that the latter is favored when the coupling is increased (see Fig.~\ref{Sfig:Susceptibility_eta_dependence} of App.~\ref{app:susceptibility}).

In order to have further insight into the evolution of the phase diagram from weak- to intermediate-coupling, we show in Fig.~\ref{fig:chi_nonint}(d) the intervalley susceptibility at $\boldsymbol{q}=\boldsymbol{\kappa}$ in the $\nu$-$E_z$ plane. As previously discussed, the maximum of $\chi_{\uparrow\downarrow}(\kappa)$ occurs at the hoVH, but it remains large along the VH line for $E_z\gtrsim10\mathrm{meV}$ due to the flatness of the band around the VH and $\kappa'$ points. In particular, for a smearing parameter $\eta=0.1\mathrm{meV}$, the maximum of $\chi_{\uparrow\downarrow}(\kappa)$ at $\nu=1$ is just $1.6$ times smaller than its value at the hoVH. Notably, $\chi_{\uparrow\downarrow}(\kappa)$ also retains a sizable magnitude away from the VH in a considerable range of the phase diagram, especially for $\nu>\nu_{\mathrm{VH}}$, where the Fermi surface consists of an electron pocket around $\gamma$ for each valley. In this region, $\chi_{\uparrow\downarrow}(\kappa)$ is set by the approximate nesting of the hexagonal-shaped electron pockets, which requires the pockets having a size $\sim\kappa$ and their edges being approximately straight. For a given $E_z<E_z^{\mathrm{hoVH}}$ in this electron pocket regime, the size of the pockets decreases and their edges become less rounded with increasing $\nu$ (see App.~\ref{app:susceptibility}), explaining both the increase of $\chi_{\uparrow\downarrow}(\kappa)$ with $\nu$ and its sudden drop at a certain $\nu>1.2$. Analogously, for a fixed $1<\nu<1.2$, they explain the first increase of $\chi_{\uparrow\downarrow}(\kappa)$ with $E_z$, which then remains approximately constant until reaching the VH, after which it drops. This dependence of $\chi_{\uparrow\downarrow}(\kappa)$ on $E_z$ and $\nu$ will determine the evolution with twist angle of the Hartree-Fock phase diagram, and in particular how far down can the strong-coupling insulator extend in the $E_z$ plane.

\section{Hartree-Fock phase diagram: IVC-AFM}
\label{sec:HF_phase_diag}

\subsection{Twist-angle evolution of the IVC-AFM}
\label{subsec:HF_phase_diag}

Motivated by the large intervalley susceptibility at $\boldsymbol{q}=\kappa$, we study the zero-temperature Hartree-Fock phase diagram of the Hamiltonian of Eqs. \eqref{eq:H0} and \eqref{eq:Hint} in a $\sqrt{3}\times\sqrt{3}$ supercell as a function of displacement field, hole density, twist angle, and interaction strength. In line with the weak-coupling intuition developed in Sec.~\ref{sec:susceptibility}, we find that the leading instability in tWSe$_2$ is a $\boldsymbol{q}=\boldsymbol{\kappa}$ intervalley-coherent in-plane antiferromagnet (IVC-AFM). We remark that capturing the physics that drives this commensurate IVC-AFM requires a description of the energy dispersion of the bands beyond patches around the VHs, perhaps explaining differences with previous 'patch' renormalization group studies~\cite{hsu_spin-valley_2021,shtyk_chamon_hoVH_2017,isobe_liang_fu_supermetal_hoVH_2019,you_viswanath_kohn-luttinger_hoVH_2022,lee_chubukov_hoVH_2024}. Within the range of twist angles considered, the phase diagram in the displacement field - hole density space is controlled by the ratio of interaction strength to bandwidth, so that the effect of increasing the interaction strength is qualitatively similar to that of decreasing the twist angle (see App.~\ref{app:HF_interactions}). In the main text, we fix the dielectric constant $\epsilon=48$, whose value has been chosen to reproduce the experiments~\cite{ghiotto_stoner_2024,xia_superconductivity_2024,guo_superconductivity_2025}, and study the dependence of the phase diagram on the twist angle. In App.~\ref{app:HF_interactions}, we show that an analogous evolution of the phase diagram is obtained when decreasing the dielectric constant at fixed twist angle.

Fig.~\ref{fig:HF_phase_diagram_angle} displays the Hartree-Fock phase diagrams of tWSe$_2$ for decreasing twist angle, \textit{i.e.} from weak to strong coupling. The blue phase represents the IVC-AFM. For a finite interaction strength and $\nu>1$, the IVC-AFM tracks the VH line but appears at slightly smaller $E_z$, in agreement with the maximum of the bare susceptibility $\chi_{\uparrow\downarrow}(\kappa)$, compare with Fig.~\ref{fig:chi_nonint}. The IVC-AFM is anisotropic with respect to the hoVH, extending only to lower densities when decreasing the twist angle or increasing the interaction strength due to the sharp decrease of $\chi_{\uparrow\downarrow}(\kappa)$ at the hoVH (see Fig.~\ref{fig:chi_nonint}(d) and Fig.~\ref{Sfig:Intra_and_Inter_Susceptibility_vs_n_eta=01} of App.~\ref{app:susceptibility}). Notably, the magnitude of the IVC-AFM order parameter is maximized just before dying off at the smallest $\nu$ where it survives, which does not coincide with the point of maximum energy stability of the phase. Indeed, the inset of Fig.~\ref{fig:HF_phase_diagram_angle}(a) displays the order parameter along the line in the $E_z$ versus $\nu$ plane where the IVC-AFM is more stable for $\theta=5^\circ$, displayed with a dashed green line in Fig.~\ref{fig:HF_phase_diagram_angle}(a). Although the transitions to the normal state are continuous in this weak-coupling case, the IVC-AFM order parameter increases more abruptly in the transition at $\nu\sim1.1$ and $E_z\sim15\mathrm{meV}$, where it reaches its maximum, and then shows a long tail until disappearing slightly beyond the hoVH, at $\nu\sim1.45$ and $E_z\sim30\mathrm{meV}$. In contrast, the energy stability of the IVC-AFM, defined as its energy gain with respect to the normal state, is maximized around $\nu\sim1.2$, deep inside the phase.

The IVC-AFM continuously extends to lower densities approximately tracking the VH with decreasing twist angle until it eventually reaches $\nu=1$, where it is enhanced by commensurability effects. Due to the small $\chi_{\uparrow\downarrow}(\kappa)$ at $E_z=0$ and $\nu=1$ (see Fig.~\ref{fig:chi_nonint}), the IVC-AFM phase only develops above a certain critical $E_z^{\mathrm{crit}}$ which decreases with twist angle, until covering the full layer-hybridized region for strong coupling (see App.~\ref{app:HF_angle_3} and Refs.~\cite{qiu_interaction-driven_2023,li_electrically_2024}). Similarly, a gap opens in the IVC-AFM at $\nu=1$ only for sufficiently large $\chi_{\uparrow\downarrow}(\kappa)$ (see Fig.~\ref{fig:1D_cut}(b)), further stabilizing the IVC-AFM at half-filling. As a function of $E_z$, this generically gives rise to the sequence of phases at $\nu=1$: layer-hybridized normal metal $-$ IVC-AFM metal $-$ IVC-AFM insulator $-$ IVC-AFM metal $-$ layer-polarized normal metal. The insulating region increases with decreasing angle, until the full IVC-AFM region becomes gapped at strong coupling (see Fig.~\ref{Sfig:HF_theta=3} of App.~\ref{app:HF_angle_3}). 

The magnitude of the IVC-AFM order parameter for twist angles $\theta\leq4.25^\circ$ at fixed $\nu\lesssim1.3$ is approximately constant for a wide range of $E_z$ (see Figs.~\ref{fig:HF_phase_diagram_angle}(b,c)), which directly reflects the value of the intervalley susceptibility arising from the approximate nesting in this region (see Fig.~\ref{fig:chi_nonint}(d) and App.~\ref{app:susceptibility}). Nevertheless, the stability of the IVC-AFM, measured by the energy saved with respect to the normal state, is stronger both at $\nu=1$ and close to the VH at $\nu\geq1$ than in the rest of the phase diagram (see Fig.~\ref{Sfig:HF_phase_diagram_angle_energy} of App.~\ref{app:HF_phase_diagram_angle_energy}). This suggests that the IVC-AFM in this region is more robust against both thermal and quantum fluctuations. Analogously, while their order parameters are similar, Fig.~\ref{Sfig:HF_phase_diagram_angle_energy} also shows that the IVC-AFM is more stable than the other valley-polarized (VP-FM, red) and intervalley-coherent (IVC-FM, yellow) ferromagnetic orders obtained with Hartree Fock (see Sec.~\ref{sec:discussion}). As we will discuss in Sec.~\ref{sec:experiments}, the stronger stability of the IVC-AFM, especially at $\nu=1$ and close to the VH at $\nu>1$, is consistent with the experiments of Refs.~\cite{ghiotto_stoner_2024,xia_superconductivity_2024,guo_superconductivity_2025,knuppel_mak_correlated_2024}, where a VP-FM phase has only been observed for small angle ($\theta=2.7^\circ$).

\begin{figure*}[!t]
    \centering
    \includegraphics[width=1\linewidth]{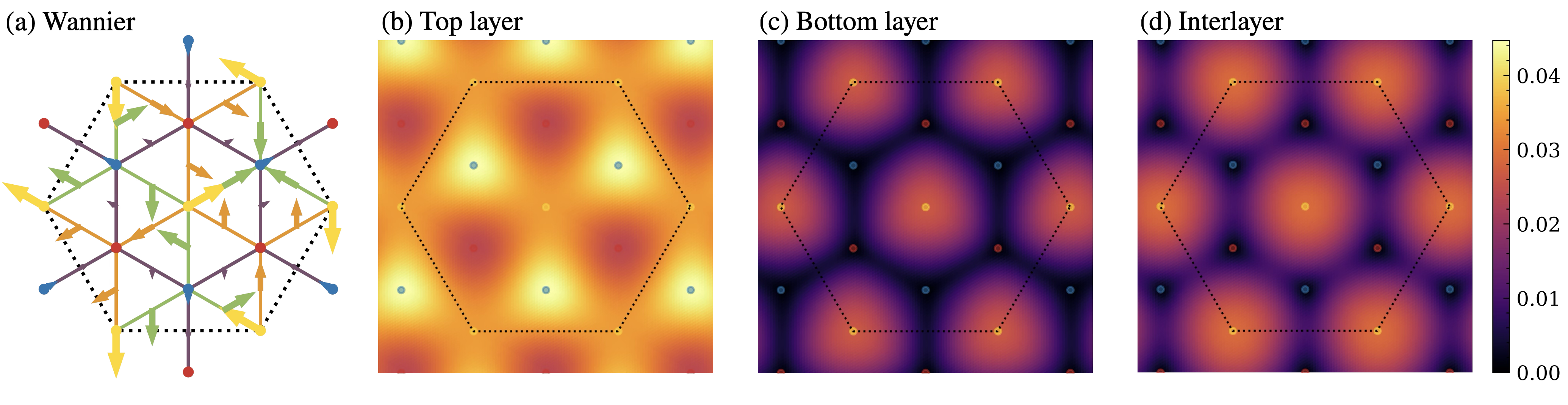}
    \caption{Real-space spin structure of the IVC-AFM ($\theta=4.25^\mathrm{o}$, $\epsilon=48$, $E_z=10\mathrm{meV}$, $\nu=1$). 
    (a) Onsite and bond spin expectation values in the Wannier orbital basis, $\langle \phi\dag_{i\alpha} \sigma_\mu \phi\dag_{j\beta}\rangle$, for $\mu=x,y$, with their direction and magnitude indicated by the direction and size of the arrows. Yellow, blue and red represent the position and onsite expectation values in the T, A and B sublattices, respectively, while green, orange and purple indicate the positions and spin expectation values in the TA, TB and BA bonds. The black dotted line corresponds to the $\sqrt{3}\times\sqrt{3}$ moir\'e supercell. 
    (b,c,d) Total spin density in real space in the top layer (b), bottom layer (c), and interlayer (d), defined as the expectation value $\sqrt{\sum_{\mu=x,y} |\langle c\dag_l(\boldsymbol{r}) \sigma_\mu c_{l'}(\boldsymbol{r})\rangle|^2}$, where $c_{l\sigma}(\boldsymbol{r})$ is the quantum field operator annihilating a particle at layer $l$, valley $\sigma$ and position $\boldsymbol{r}$.} 
    \label{fig:AFM_real_space}
\end{figure*}

\subsection{Symmetry and structure of the IVC-AFM}
\label{subsec:symmetry_IVC-AFM}

For a non-zero displacement field, the IVC-AFM order parameter transforms according to a two-dimensional irreducible representation (irrep) of the space group which breaks translational, time-reversal and $U_{\mathrm{V}}(1)$ valley symmetries (see App.~\ref{app:symmetry_analysis} for the symmetry analysis and definition of the order parameters). For $E_z>0$, which favors holes occupying the bottom layer (and thus the XM sites and the B orbital), it is invariant under the $C_{3z}$ centered at the honeycomb MX (A) sites. A displacement field in the opposite direction would invert the roles of the MX and XM sites (and the A and B orbitals), resulting in $C_{3z}$ invariance around the XM (B) site, and opposite chirality of the spins. At $E_z=0$, the two irreps characterizing the IVC-AFM at positive and negative $E_z$ become degenerate, forming a four-dimensional irrep. The IVC-AFM phase found in this work can therefore be identified as the ``O-$120^\circ$-AFM'' that Refs.~\cite{qiu_interaction-driven_2023,li_electrically_2024} found for tMoTe$_2$.

The IVC-AFM state has an intricate structure in real space, shown in Fig.~\ref{fig:AFM_real_space}, characterized by a $120^\mathrm{o}$ pattern of the in-plane spins. Similar $120^\mathrm{o}$-ordered spin density waves have also been proposed for twisted bilayer graphene at the VH filling \cite{isobe_TBG_VH_SDW_2018,chichinadze_TBG_VH_SDW_2020,lu_TBG_VH_SDW_2022} but the absence of spin-momentum locking in twisted bilayer graphene leads to some differences in the mechanism for ordering relative to what is found here. In terms of the Wannier orbitals (Fig.~\ref{fig:AFM_real_space}(a)), the IVC-AFM has contributions from both onsite spins (induced by the onsite Hubbard $U_\alpha$ interactions) and spins in the bonds (induced by the nearest-neighbor density-density interactions $V_{\alpha\beta}$). For $E_z>0$, the onsite spin has maximum weight on the T sublattice, smaller weight on the A sublattice and vanishes at the B sublattice, as indicated by the size of the arrows in Fig.~\ref{fig:AFM_real_space}(a). The spin order also has strong contributions from the TA inter-orbital bond expectation values, and smaller weight on the TB and BA bonds. Due to the different transformation properties of the orbitals under $C_{3z}$ (see App.~\ref{app:symmetry_analysis}), the spin expectation values involving different orbitals transform differently under $C_{3z}$, in a way that the pattern shown in Fig.~\ref{fig:AFM_real_space}(a) is invariant under $C_{3z}$ around the A site. The stronger magnetization in the T and A sites and bonds compared to the B ones can be understood from the dominant T and A orbital weight on the flat portion of the band between the VH and $\kappa$ ($\kappa'$) points in the $K$ ($K'$) valley (see Fig.~\ref{fig:bands_nonint}(a)), which, as discussed in Sec. \ref{sec:susceptibility}, is the region of the band that mainly determines the energetics (see also Sec. \ref{sec:FSR}).

Using the Wannier functions, we obtain the spin density in real space, whose magnitude is shown in Figs.~\ref{fig:AFM_real_space}(b,c,d) for the top layer, bottom layer, and interlayer, respectively. The magnitude of the magnetization is translation and threefold symmetric. In the top layer, it is nonzero over the full unit cell, and it is maximized at the MX (A) sites. The bottom layer and interlayer magnetizations, on the other hand, vanish at the MX and XM sites and are maximized at the MM (T) sites. The magnetization is larger on the top layer due to the larger spin expectation values on the A sites and TA bonds than on the B sites and TB bonds. We remark that $E_z>0$ favors the bottom-layer polarization, where the hole density is larger (see Fig.~\ref{Sfig:charge_density} of App.~\ref{app:charge_density}), while the magnetization in the IVC-AFM is stronger on the top layer.

\begin{figure*}[!t]
    \centering
    \includegraphics[width=1\linewidth]{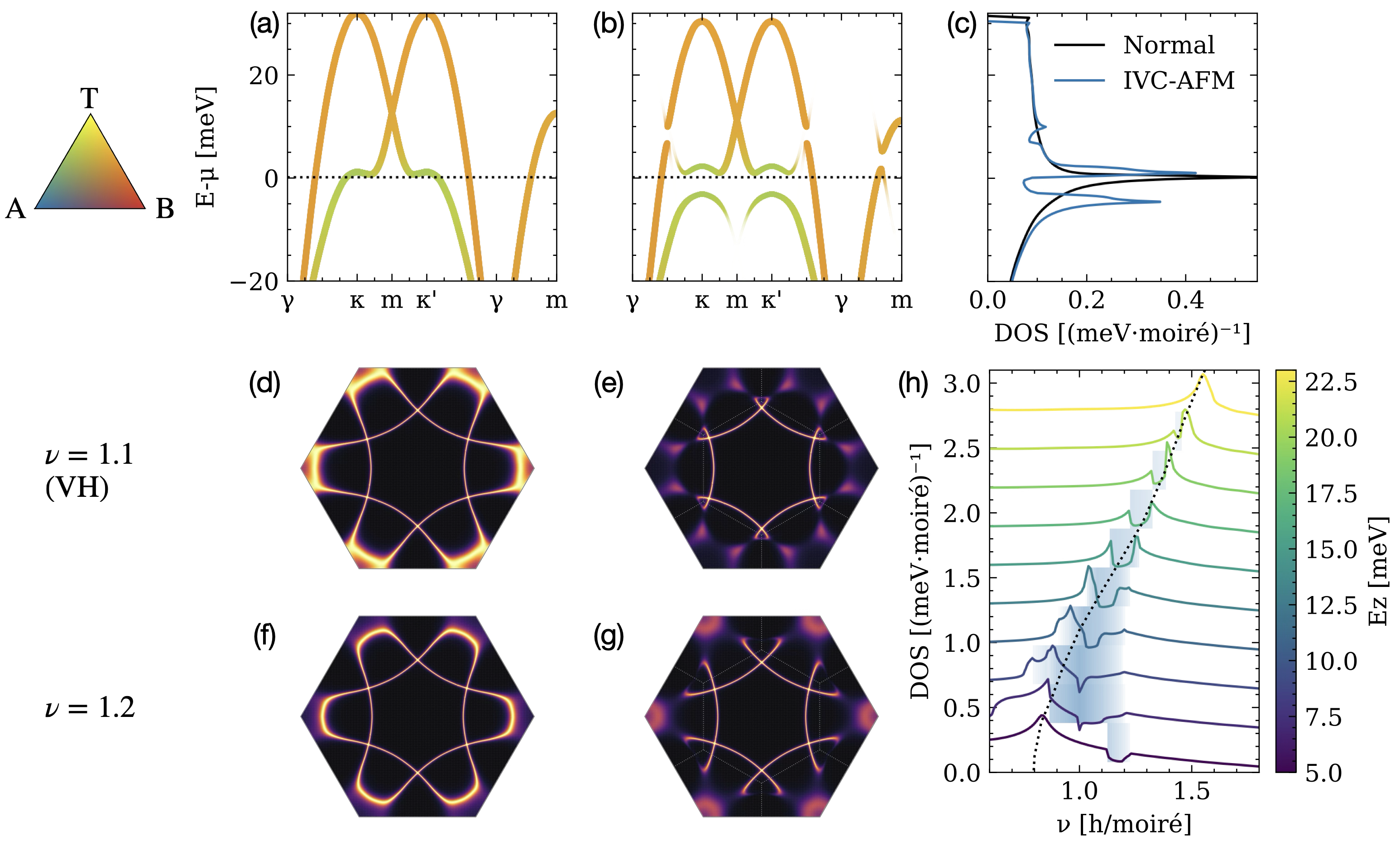}
    \caption{Reconstructed band structure, Fermi surface and DOS in the IVC-AFM phase for $\theta=4.25^\mathrm{o}$ and $\epsilon=48$. 
    (a,b) Normal state (a) and IVC-AFM (b) band structure at the VH point $E_z=13.5\mathrm{meV}$, $\nu=1.1$. 
    (c) DOS as a function of energy for the normal state (black) and IVC-AFM phase (blue), corresponding to the bands plotted in (a) and (b), respectively. 
    (d-g) Spectral function at the Fermi level in the normal state (d,f) and in the IVC-AFM backfolded to the original Brillouin zone (e,g) for $E_z=13.5\mathrm{meV}$, $\nu=1.1$ (d,e) and $\nu=1.2$ (f,g). A smearing parameter $\eta=0.5\mathrm{meV}$ has been used for the spectral functions in (d-g).
    (h) DOS as a function of $\nu$ for several $E_z$ computed using the self-consistent IVC-AFM order parameter at each point. Different $E_z$, indicated by different colors, are shifted vertically for visualization purposes. The black dotted line corresponds to the VH line in the normal state. The intensity of the blue shadowed area is proportional to the IVC-AFM order parameter.}
    \label{fig:HF_bands_FS_DOS}
\end{figure*}

\section{Fermi surface reconstruction in the IVC-AFM} \label{sec:FSR}

\begin{figure*}[!t]
    \centering
    \includegraphics[width=1\linewidth]{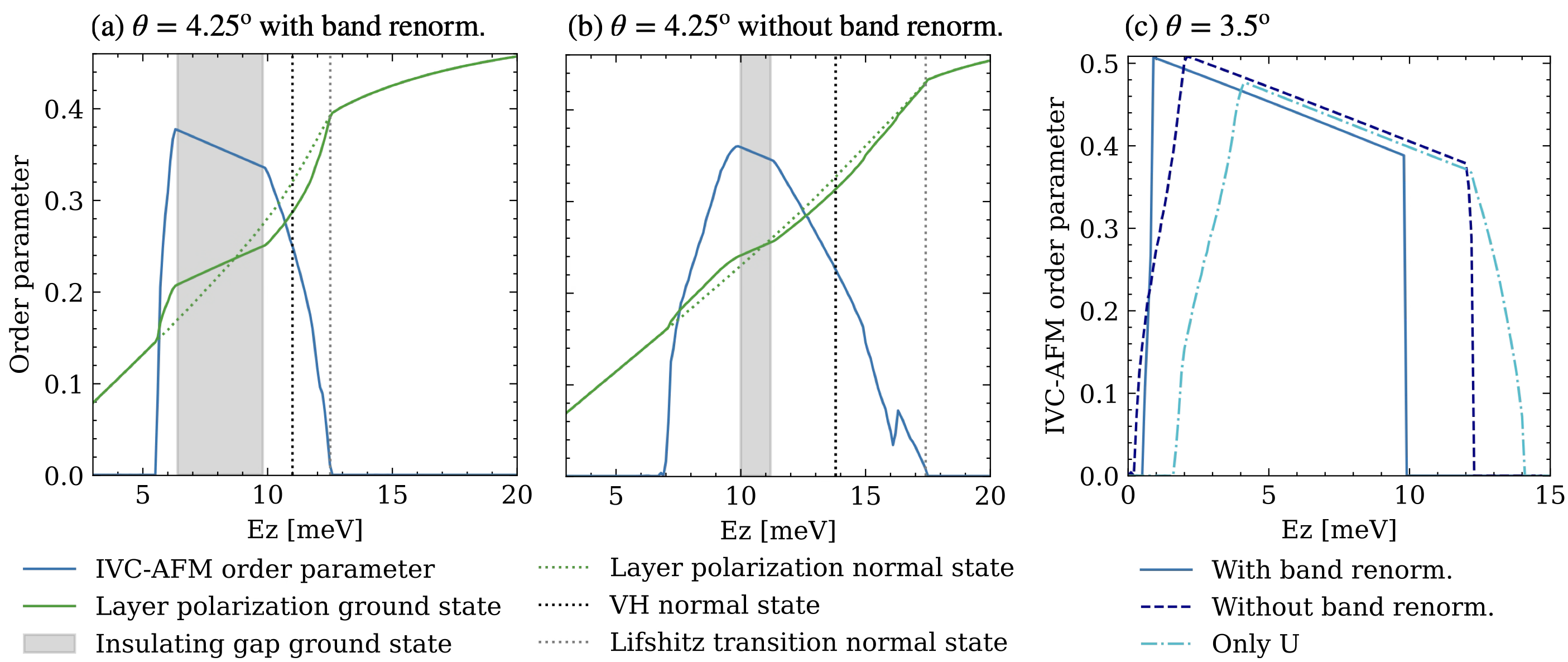}
    \caption {Order parameters at half filling.
    (a) IVC-AFM order parameter (blue), and layer polarization in the ground (solid green) and normal states (dotted green) as a function of $E_z$ at $\nu=1$ for $\theta=4.25^\mathrm{o}$, $\epsilon=48$. Based on the layer structure of the Wannier functions in our model, we define the layer polarization as the supercell average of $(n_B-n_A)/(n_B+n_A+n_T)$, where $n_\alpha$ is the onsite density at orbital $\alpha$. The gray shadowed area indicates the $E_z$ where a full gap opens. The black and gray dotted vertical lines indicate the VH and Lifshitz transition in the normal state, respectively. 
    (b) Same as (a), but excluding the interaction-induced renormalization of the bands that does not break symmetries, i.e., including only the renormalization due to the IVC-AFM. This excludes the renormalization of the bands by the layer polarization due to the nearest-neighbor interactions and, therefore, the direct coupling between the IVC-AFM and the layer polarization is highly suppressed. The nonmonotonic behavior of the IVC-AFM order parameter in (b) around $E_z=16\mathrm{meV}$ is due to small coexisting VP-FM correlations in this region (see Sec.~\ref{sec:discussion}).
    (c) IVC-AFM order parameter as a function of $E_z$ at $\nu=1$ for $\theta=3.5^\mathrm{o}$ (i) including the nearest-neighbor density-density interactions and the symmetric band renormalizations (solid blue), (ii) including the nearest-neighbor density-density interactions but neglecting the symmetric band renormalizations (dashed dark blue), and (iii) including only the onsite Hubbard interactions (dash-dot cyan). When including the nearest-neighbor density-density interactions, we have used $\epsilon=48$, while in the case with only onsite Hubbard interactions we have used a smaller $\epsilon=29$ so that the magnitude of the IVC-AFM order parameter is similar in both cases.}
    \label{fig:1D_cut}
\end{figure*}

Let us now analyze the fermiology inside the IVC-AFM phase. Figs. \ref{fig:HF_bands_FS_DOS}(a, b) show the band structures of the (a) normal and (b) IVC-AFM states for $\theta=4.25^{\mathrm{o}}$ at the VH point $E_z=13.5\mathrm{meV}$, $\nu=1.1$. For better comparison, both bands are displayed in the original Brillouin zone, with the transparency in the IVC-AFM bands proportional to the backfolded spectral weight. The main effect of the IVC-AFM is to open a partial gap on the lower hole pockets around the $\kappa$ points, asymmetrically splitting the VH singularities, as shown in the DOS of Fig.~\ref{fig:HF_bands_FS_DOS}(c). The energy gain in the IVC-AFM mainly comes from this splitting, with the chemical potential lying within the split VHs. Figs. \ref{fig:HF_bands_FS_DOS}(d, e) show the corresponding normal state (d) and IVC-AFM (e) reconstructed Fermi surfaces in the original Brillouin zone, showing the partial gap around the VH points. The Fermi surface in the normal state consists of hole pockets around the $\kappa$ points touching at the VH points. In the IVC-AFM phase, the bands around the VH points are gapped, and the Fermi surface consists of hole-like lines, which arise from the backfolding of small hole pockets around the $m'$ points of the Brillouin zone of the $\sqrt{3}\times\sqrt{3}$ supercell. This demonstrates a large decrease in the carrier density of the bands crossing the Fermi level. Increasing the chemical potential, the size of the hole pockets around the $m'$ points increases, and new hole pockets around the $\kappa$ points appear (see \ref{fig:HF_bands_FS_DOS}(f, g)). Further increasing the doping, these two sets of hole pockets touch at the second set of VH, finally giving rise to two big electron pockets around the $\gamma$ point, as in the normal state.

Besides the Fermi surface reconstruction, the other remarkable consequence of the IVC-AFM is the reconstruction of the DOS. Following the VH line for a fixed IVC-AFM order parameter, at small $E_z$ the part of the bands around the VH is only weakly renormalized, resulting in only one peak in the DOS. Increasing $E_z$, the IVC-AFM reconstructs the VH, shifting it to slightly lower $\nu$. Moreover, a second VH peak arises at larger $\nu$ (see Fig.~\ref{fig:HF_bands_FS_DOS}(c)), whose magnitude increases and becomes stronger than that of the previous VH peak. Fig.~\ref{fig:HF_bands_FS_DOS}(h) shows that this behavior survives in the DOS computed for the self-consistent IVC-AFM order parameter at each point of the $E_z$-$\nu$ phase diagram. Due to the decreasing order parameter with $E_z$, the splitting of the VHs decreases, until they merge in the normal state. Therefore, for this range of displacement fields the transitions from the IVC-AFM phase to the normal state are determined by the energy of the peaks in the DOS with respect to the Fermi energy.

\section{Coupling of the IVC-AFM to the layer polarization}
\label{sec:coupling_AFM-LP}

As described above, the physics of the IVC-AFM is determined by the part of the band between the VH and $\kappa$ ($\kappa'$) points at valley $K$ ($K'$), which are approximately nested by $\kappa$ and contribute with a large DOS. Indeed, the energy gain in the IVC-AFM mainly comes from the splitting of this part of the band when the resulting chemical potential lies within these split bands. Therefore, the stability of the IVC-AFM requires this part of the band to be occupied by holes in the normal state, i.e., it requires to be in the layer-hybridized side of the Lifshitz transition. This is also reflected in the sharp decrease of the susceptibility $\chi_{\uparrow\downarrow}(\kappa)$ at the Lifshitz transition in Fig. \ref{fig:chi_nonint}(d). Accordingly, as shown in Figs.~\ref{fig:HF_phase_diagram_angle} and \ref{fig:1D_cut}, the IVC-AFM dies soon after crossing the Lifshitz transition of the normal state (gray dotted line). Due to the link between the Lifshitz transition and the layer polarization, this implies that the IVC-AFM and the layer polarization are coupled. Moreover, when the IVC-AFM survives after the Lifshitz transition of the normal state, the transition from the IVC-AFM to the `layer-polarized'' normal state is sharp, compatible with a first-order transition at strong coupling, as signaled by the jumps in the IVC-AFM order parameter (see blue solid and dark blue dashed lines in Fig.~\ref{fig:1D_cut}(c)).

The coupling between the IVC-AFM and the layer polarization is further enhanced by the nearest-neighbor density-density interactions $V$, whose Hartree shift favors layer polarization. One consequence from the Hartree shift of $V$ is to renormalize the normal state band structure, bringing the layer polarization line and the VH to lower displacement fields (see App.~\ref{app:DOS_angles}). More importantly, within the IVC-AFM, the layer polarization is strongly renormalized with respect to the normal state. This can be deduced by comparing the solid and dotted green lines in Fig.~\ref{fig:1D_cut}(a), which show the layer polarization in the ground and normal states, respectively, for $\theta=4.25^\circ$ at half filling. The layer polarization in the ground state changes its slope with $E_z$ in the transitions from layer-hybridized metal to IVC-AFM and from IVC-AFM to ``layer-polarized'' metal. For $E_z$ inside the IVC-AFM close to these transitions, where the IVC-AFM order parameter (blue in Fig.~\ref{fig:1D_cut}(a)) behaves approximately as $\sim\sqrt{E_z}$, the layer polarization in the IVC-AFM increases relatively abruptly compared to the one in the normal state. When a full gap opens in the IVC-AFM (gray region in Fig.~\ref{fig:1D_cut}(a)), the layer polarization changes its slope again, and grows only slightly with $E_z$ (the linearly decreasing IVC-AFM order parameter in the gapped region stems from the fact that the ratio of magnetization to charge remains constant in this region, and the charge in the T sites decreases with $E_z$). This behavior indicates that there is an optimal layer polarization for the stability of the IVC-AFM and the opening of a gap, which is both non-zero and non-maximal, and stems from the layer structure of the bands at the $\kappa$ ($\kappa'$) points in the $K$ ($K'$) valleys. 
This underscores the importance of interlayer hybridization for the emergence of the IVC-AFM order.

To show that this behavior of the layer polarization is mainly induced by the Hartree shift of the nearest-neighbor density-density interactions $V$, we have determined the Hartree-Fock ground state neglecting the renormalization of the bands by order parameters that respect the point group $C_{3v}$, $U_{\mathrm{V}}(1)$ and TRS, and therefore including only the renormalization of the bands due to the IVC-AFM. This excludes the renormalization of the bands by the layer polarization due to $V$ and, therefore, the direct coupling between the IVC-AFM and the layer polarization is highly suppressed. Fig.~\ref{fig:1D_cut}(b) displays the resulting phase diagram for $\theta=4.25^\circ$ at $\nu=1$. The overall phase diagram is shifted to higher $E_z$ compared to Fig.~\ref{fig:1D_cut}(a) due to neglecting the Hartree shift of $V$. More importantly, while the slope of the layer polarization in the IVC-AFM in Fig.~\ref{fig:1D_cut}(b) still changes at the transitions to the normal state and at the gap opening points, the renormalization is much weaker compared to Fig.~\ref{fig:1D_cut}(a), demonstrating the main role of the Hartree shift of $V$ in the coupling of the IVC-AFM to the layer polarization. Comparing Figs.~\ref{fig:1D_cut}(a) and (b), we conclude that this coupling strongly favors the opening of a gap, which further stabilizes the IVC-AFM. Notably, for a given twist angle, the maximum stability of the IVC-AFM, which is reached in the middle of the insulating region, occurs at approximately the same value of the layer polarization both including (Fig.~\ref{fig:1D_cut}(a)) and neglecting (Fig.~\ref{fig:1D_cut}(b)) the Hartree shift (see also App.~\ref{app:IVC-AFM-LP}). This again demonstrates the importance of the layer structure for the stability of the IVC-AFM.

Another effect derived from the strong coupling between the layer polarization and the IVC-AFM order parameter induced by $V$ is the sharpening of the transition at small $E_z$ from the layer-hybridized normal state to the IVC-AFM at fixed $\nu$, which becomes compatible with a first-order transition at strong coupling (see the blue solid line of Fig.~\ref{fig:1D_cut}(c) for $\theta=3.5^\circ$ at $\nu=1$). To show that this is indeed induced by the Hartree shift of $V$, in Fig.~\ref{fig:1D_cut}(c) we compare the IVC-AFM order parameters for $\theta=3.5^\circ$ at $\nu=1$ including (blue solid line) and neglecting (dark blue dashed line) the symmetry-allowed renormalization of the bands, as well as using only onsite Hubbard interactions (cyan dash-dot line). The transition from the layer-hybridized normal state to the IVC-AFM is discontinuous for the former but continuous in the two latter cases. This suggests that the sharpness of the transition from the layer-hybridized normal state to the IVC-AFM can be experimentally controlled by the distance to the gates, which controls the screening of the interaction. 

Finally, we mention that at smaller twist angles where the IVC-AFM extends to $E_z=0$, it induces spontaneous ferroelectricity (see App.~\ref{app:HF_angle_3}), with the spin chirality controlled by the sign of $E_z$, which identifies the IVC-AFM as a type-II multiferroic~\cite{cheong_multiferroics_2007,khomskii_classifying_2009,song_evidence_2022,qiu_interaction-driven_2023,li_electrically_2024}. This once again shows the magneto-electric coupling between the antiferromagnetic order and layer-polarization ferroelectricity of the bilayer~\cite{crepel_bridging_2024}.

\section{Comparison of the IVC-AFM to experiments}
\label{sec:experiments}

The twist-angle evolution of the IVC-AFM within Hartree-Fock with a fixed dielectric constant $\epsilon=48$ is compatible with the transport experiments of Refs.~\cite{ghiotto_stoner_2024,xia_superconductivity_2024,guo_superconductivity_2025}. These experiments have studied the longitudinal and Hall transport as a function of displacement field and hole density of the topmost moir\'e band for the twist angles $\theta=5^\circ$~\cite{guo_superconductivity_2025}, $\theta=4.2^\circ$~\cite{ghiotto_stoner_2024}, and $\theta=3.5^\circ$~\cite{xia_superconductivity_2024}. We first point out that these experiments can identify both the VH and Lifshitz transition lines, since the longitudinal resistivity $\rho_L$ peaks at the VH and presents an increase when entering the layer-hybridized region. This can be explained within the semiclassical transport theory in the weak-scattering limit, where the impurity-scattering rate between two states $\boldsymbol{k}$ and $\boldsymbol{k}'$ is given by the Fermi golden rule, $\tau_{\boldsymbol{k},\boldsymbol{k}'}^{-1} \sim |\langle \boldsymbol{k}|V_{\mathrm{imp}}|\boldsymbol{k}'\rangle|^2 \delta(\varepsilon_{\boldsymbol{k}}-\varepsilon_{\boldsymbol{k}'})$, so that the average scattering rate is approximately proportional to the DOS, and therefore $\rho_L \propto \tau^{-1} \propto \mathrm{DOS}$~\cite{herman_deviation_2019}.

The experiments of Refs.~\cite{ghiotto_stoner_2024,xia_superconductivity_2024,guo_superconductivity_2025} also observe signatures of correlated states in different regions of the phase diagram. The $\theta=5^\circ$ device of Ref.~\cite{guo_superconductivity_2025} shows a metal with enhanced longitudinal and Hall resistivities close to the VH line at $1.1 \lesssim \nu \lesssim 1.25$, and a transition to a superconductor at $\nu\sim1.1$. The enhanced resistivities were interpreted in Ref.~\cite{guo_superconductivity_2025} as a reconstruction of the Fermi surface induced by some antiferromagnetic order, which would dramatically reduce the hole density of the bands crossing the Fermi level. The phase diagram of the IVC-AFM found in this work for $\theta=5^\circ$ (see Fig.~\ref{fig:HF_phase_diagram_angle}(a)), with its decrease of the transport hole density discussed in Fig.~\ref{fig:HF_bands_FS_DOS}, is therefore compatible with this experiment. Within the reconstructed phase in Ref.~\cite{guo_superconductivity_2025}, the resistivities increase sharply at $\nu \sim 1.1$, and then gradually decrease with increasing density approximately following the VH. This is also compatible with the behavior of the IVC-AFM order parameter shown in the inset of Fig.~\ref{fig:HF_phase_diagram_angle}(a). One of the findings of Ref.~\cite{guo_superconductivity_2025} that had remained unexplained until now was the structure of the longitudinal resistivity within the reconstructed phase, which showed an evolution from a single peak at the lowest densities to two peaks at larger densities. We claim that this can be explained by the DOS dependence of the scattering rate, with the reconstructed DOS within the IVC-AFM showing a similar evolution from a one-peak to a two-peak structure, as shown in Fig.~\ref{fig:HF_bands_FS_DOS}(h). We note that the Hall resistivity in the weak-field limit has a much weaker dependence on the DOS, explaining the presence of only one peak for this quantity.

The $\theta=4.2^{\mathrm{o}}$ device of Ref.~\cite{ghiotto_stoner_2024} displays analogous features along the VH for $\nu>1$: increased longitudinal and Hall resistivities with a double- and single-peak structure, respectively. In this case, the instability survives until $\nu=1$, where it becomes insulating and extends to lower displacement fields. Due to the device architecture, Ref.~\cite{ghiotto_stoner_2024} could not reach zero displacement field and only analyzed displacement fields above a certain value, where the correlated phase remained insulating at $\nu=1$. The crossover from Fermi reconstructed phase along the VH to the insulating state at $\nu=1$ is compatible with the IVC-AFM at $\theta=4.25^\circ$ (see Figs.~\ref{fig:HF_phase_diagram_angle}(b), \ref{fig:HF_bands_FS_DOS}, and \ref{fig:1D_cut}(a,b)).

The $\theta=3.5^{\mathrm{o}}$ device of Ref.~\cite{xia_superconductivity_2024} displays a strong insulating region at $\nu=1$ between the layer polarization and a small non-zero displacement field, below which the systems transitions to a superconducting state. The insulating state is compatible with the IVC-AFM for $\theta=3.5^{\mathrm{o}}$ (see Figs.~\ref{fig:HF_phase_diagram_angle}(c) and \ref{fig:1D_cut}(c)), and the disappearance of the IVC-AFM at small $E_z$ can be explained by the suppressed intervalley susceptibility (Fig.~\ref{fig:chi_nonint}(d)). Experimentally, the enhanced fluctuations due to the higher symmetry at $E_z=0$, where the IVC-AFM irrep becomes four-dimensional, might further suppress the order.

The study of the transition from the IVC-AFM to the superconducting state at its low displacement field boundary in both the $\theta=5^{\mathrm{o}}$~\cite{guo_superconductivity_2025} and $\theta=3.5^{\mathrm{o}}$~\cite{xia_superconductivity_2024} devices lies beyond the scope of this work. However, we mention that the suppression of the IVC-AFM at this transition due to lower bare susceptibility, together with the fact that the maximum IVC-AFM order parameter lies exactly at this transition, suggests strong fluctuations of the order around this point, which could be compatible with the fluctuation-driven superconductivity scenario~\cite{chubukov2024quantum,fischer2024theory}. These fluctuations might be further enhanced at strong coupling close to $E_z=0$, where the symmetry of the IVC-AFM is enhanced. The higher superconducting critical temperature in the $\theta=5^{\mathrm{o}}$ device~\cite{guo_superconductivity_2025} might be explained by the higher DOS in this case.

Finally, we can compare our results to the optical experiment of Ref.~\cite{knuppel_mak_correlated_2024} performed in a $\theta=2.7^\circ$ device. The reflection contrast of an exciton sensor reveals an insulating correlated state for all $E_z$ in the layer-hybridized region at $\nu=1$, which is consistent with our fully gapped IVC-AFM for $\theta=3^\circ$ shown in App.~\ref{app:HF_angle_3}. Furthermore, Ref.~\cite{knuppel_mak_correlated_2024} observes a non-zero magnetic circular dichroism signal for small $E_z$ around the VH, which are compatible with our valley-polarized ferromagnet (see Sec.~\ref{sec:discussion} and App.~\ref{app:HF_angle_3}). We also mention that our work is compatible with recent theory works studying the IVC-AFM in tWSe$_2$ \cite{bi_Fu_excitonic_tWSe2_2021,tuo2024theorytopo,fischer2024theory,peng_magnetism_2025} and tMoTe$_2$ \cite{qiu_interaction-driven_2023,li_electrically_2024} (see App.~\ref{app:comparison_theory}).

\section{Discussion}
\label{sec:discussion}

In this work, we have performed zero-temperature Hartree-Fock calculations of tWSe$_2$ in a $\sqrt{3}\times\sqrt{3}$ supercell. We have analyzed the displacement field, hole density, twist-angle and interaction-strength dependence of the IVC-AFM state. In the weak-coupling limit, the IVC-AFM state appears at a high density closely following the van Hove singularity, and it continuously moves towards half filling ($\nu=1$) with decreasing twist angle, where the gap opening strongly stabilizes it, in agreement with the experiments \cite{ghiotto_stoner_2024,xia_superconductivity_2024,guo_superconductivity_2025,knuppel_mak_correlated_2024}. Notably, at strong coupling, the maximum IVC-AFM order parameter at $\nu=1$ occurs at the $E_z$ where the order onsets, signaling strong fluctuations, which might therefore drive the superconductivity also in the strong-coupling limit \cite{chubukov2024quantum,fischer2024theory}.  {The weak-coupling limit of our results is broadly consistent with previous work ~\cite{bi_Fu_excitonic_tWSe2_2021,schrade_Fu_SC_tWSe2_2024,wietek2022tunable,crepel_topological_2023,kim2024theory,myerson-jain_superconductor-insulator_2024,zhu2024theory,christos2024approximate,xie2024superconductivity,guerci2024topologicalSC,tuo2024theorytopo,qin2024kohn,chubukov2024quantum,fischer2024theory,peng_magnetism_2025}, but going beyond this work we provide an extension to strong coupling, demonstrating how the region of magnetic order detaches from the van Hove singularity as the twist angle decreases and highlighting the role of interlayer interactions in driving the transitions first order and enhancing the dependence of results on the displacement field.} {We further associate the subtle transport features observed experimentally \cite{ghiotto_stoner_2024,guo_superconductivity_2025} to the properties of the reconstructed band structure within the ordered state.}

We have found that due to the strong interlayer character of the nearest-neighbor density-density interactions the IVC-AFM order strongly couples to the layer polarization. This has important consequences including a very sharp transition from the normal state. The sharpness of the transition and the resulting step in the layer polarization could be controlled by the twist angle and the distance to the gates, and detected experimentally in low-temperature transport measurements~\cite{zheng2020unconventional,yasuda2021stacking,niu2022giant}, scanning-probe microscopy~\cite{vizner2021interfacial}, near-field optics~\cite{zhang2023visualizing}, and nanoscale electrometry with color center~\cite{dolde_electric-field_2011,block_optically_2021,bian_nanoscale_2021}. We also anticipate that the coupling between the IVC-AFM and the layer polarization will strongly affect the collective mode spectrum by hybridizing the interlayer charge collective modes with the spin waves, which can facilitate the optical detection of the latter. The strong coupling between the IVC-AFM and the layer polarization further enriches the tunability of twisted TMDs, where other multiferroic phases have been proposed \cite{haavisto_topological_2022,abouelkomsan_multiferroicity_2024,qiu_interaction-driven_2023,li_electrically_2024}.

The Hartree-Fock calculation in the $\sqrt{3}\times\sqrt{3}$ supercell also predicts a weaker $\boldsymbol{q}=0$ intravalley out-of-plane ferromagnet (VP-FM) in some regions of the phase diagram (red in Fig.~\ref{fig:HF_phase_diagram_angle}). On the one hand, a weak VP-FM phase appears along the VH line after the hoVH, in agreement with the divergent $\chi_{\uparrow\uparrow}(\boldsymbol{q}=0)$. On the other hand, a VP-FM also appears at small displacement fields and densities close to the VH. There, the intervalley nesting wavevector is closer to $\boldsymbol{q}=m$, explaining the leading intravalley instability in the $\sqrt{3}\times\sqrt{3}$ supercell. 
At the transition between the two ordered phases, we observe a small region of coexisting IVC-AFM and VP-FM (purple in Fig.~\ref{fig:HF_phase_diagram_angle}).
Remarkably, this VP-FM phase close to $E_z=0$ is compatible with the magnetic circular dichroism observed in the $\theta=2.7^\mathrm{o}$ device of Ref.~\cite{knuppel_mak_correlated_2024}. Nevertheless, a calculation in a supercell allowing for orders with $\boldsymbol{q}=m$ should be carried out to reliably determine the ground state in this region. Finally, we mention that, at strong coupling, an intervalley-coherent in-plane ferromagnet (IVC-FM, yellow in Fig.~\ref{fig:HF_phase_diagram_angle}(c)) appears at the border of the low-density VP-FM region, which is associated to the large $\chi_{\uparrow\uparrow}(\boldsymbol{q}=0)$ close to the VH at small $E_z$. However, the validity of Hartree-Fock for metals at strong coupling away from commensurate filling or VH singularities remains to be checked.

Our work highlights the interplay between topology, layer polarization and correlations in TMD bilayers, which has enabled the experimental stabilization of both superconducting~\cite{guo_superconductivity_2025,xia_superconductivity_2024} and topologically ordered phases~\cite{zeng2023thermodynamic,cai2023signatures} in these materials. This makes them a promising platform to study their interplay, realize novel phases of matter such as paired non-Abelian phases~\cite{haldane1988spin,moore1991nonabelions,moller2008paired,crepel2019matrixhr}, and manipulate anyons~\cite{vaezi2014superconducting,mong2014universal,repellin2018numerical,crepel2024attractive}. Examining the physics of smaller twist angle systems with beyond Hartree-Fock methods is an important open question.

\section{Acknowledgements}

We are grateful to A. N. Pasupathy, Y. Guo, A. Ghiotto, A. B. Tazi, D. Ostrom, D. Guerci, and C. De Beule for interesting discussions on our work. D.M-S. acknowledges financial support from the National Science Foundation (NSF) Materials Research Science and Engineering Centers (MRSEC) program through Columbia University under the Precision-Assembled Quantum Materials (PAQM) Grant No. DMR-2011738. The Flatiron Institute is a division of the Simons Foundation.

\begin{figure*}
    \centering
    \includegraphics[width=\linewidth]{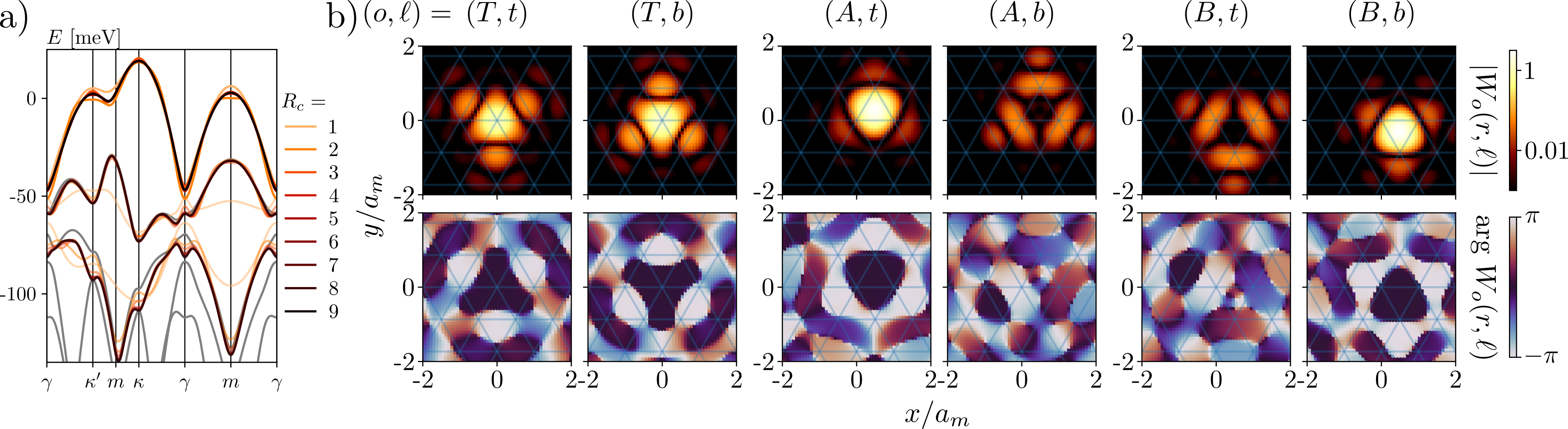}
    \caption{a) Band structure of the continuum model (gray) compared to that of the wannierized model where we only keep tight-binding parameters between orbitals distant by at most $R_c$ in units of the moir\'e lattice constant $a_m$. The dispersion of the two topmost bands is perfectly reproduced, and the details of the wannierized do not change for $R_c > 4$. b) Amplitude (top row, log scale) and phase (bottom row, linear scale) of the Wannier orbitals in real space. }
    \label{fig_appwannierfunction}
\end{figure*}

\newpage
\section*{Appendix}
\appendix

\makeatletter
\def\l@subsection#1#2{}
\makeatother
\noindent \textbf{Table of contents:} \vspace{3pt}
\startcontents[sections]
\printcontents[sections]{l}{1}{}

\section{Wannierization procedure} \label{app:wannier}

Our microscopic modeling of tWSe$_2$ starts from the continuum model of Ref.~\cite{devakul2021magic}, which depends on three parameters: $v$ the intra-layer moir\'e potential strength, $w$ the inter-layer moir\'e tunneling amplitude, and an angle $\psi$ measuring the difference between the intra-layer potential minima in the two layers. 
Once diagonalized, it gives rise to the band dispersion $\varepsilon_{k,n}^{\rm cont}$ shown in gray in Fig.~\ref{fig_appwannierfunction}a and provides us with the corresponding Bloch eigenvectors $\psi_{k,n}^{\rm cont}(r; \ell)$, where $r$ and $\ell \in \{t,b\}$ respectively denote position and layer.

We wannierize the model by single-shot projection~\cite{Pizzi2020} of three trial Gaussian wavefunctions with centers and layer distributions determined from the generic symmetry properties of the bilayer, see Ref.~\cite{crepel_bridging_2024}. 
In practice, we choose 
\begin{align}
\psi_{\mathrm{A}}^{\rm gauss}(r; \ell) &= \frac{\delta_{\ell,t} }{2\pi \sigma_h} \exp\left[ - \frac{(r-r_{\rm MX)})^2}{2\sigma_h^2} \right] , \\
\psi_{\mathrm{B}}^{\rm gauss}(r; \ell) &= \frac{\delta_{\ell,b} }{2\pi \sigma_h} \exp\left[ - \frac{(r-r_{\rm XM)})^2}{2\sigma_h^2} \right] , \\
\psi_{\mathrm{T}}^{\rm gauss}(r; \ell) &= \frac{f_{\ell}(E_z) }{2\pi \sigma_t} \exp\left[ - \frac{(r-r_{\rm MM)})^2}{2\sigma_t^2} \right] , 
\end{align}
where $r_{\rm MX,XM,MM}$ denote the position of the MX, XM, and MM sites in the unit cell (see Fig.~\ref{fig:model}), while the phenomenological parameters $\sigma_{h,t}$ and $f_{\ell}$ capture the expected width and layer-polarization of the Wannier orbitals at the honeycomb and triangular sites. 
Guesses for the width are obtained by expanding the continuum model to quadratic order around the $r_{\rm MX,XM,MM}$ and extracting the harmonic oscillator length of the resulting restoring force
\begin{equation}
\sigma_h^{-4} = 3\alpha^2 v \cos (\psi -2\pi/3) \frac{m^* g^2}{\hbar^2} , \quad \sigma_t^{-4} = w \frac{m^* g^2}{2\hbar^2} , 
\end{equation}
with $g$ the norm of the lowest moir\'e reciprocal lattice vector, $m^*$ the effective mass of monolayer WSe$_2$, and $\alpha = 3/4$ an arbitrary scaling factor chosen for numerical stability. 
A similar analysis provides the effective form for the layer polarization at the MM point $f_t^2(E_z) = 1 -\frac{1}{2} \tanh(E_z/\beta w)$ and $f_b^2 = 1-f_t^2$, where the arbitrary factor is fixed to $\beta = 9$. The function $f_t$ interpolates between the equal polarization expected at $E_z=0$ and the fully layer-polarized orbitals expected at very large $E_z$ while capturing the correct slope of the layer polarization at small $E_z$. 
The symmetry, shape, and typical extends of the obtained Wannier orbitals are largely independent of these specific values, and they should only be seen as educated guesses for the specific form of the continuum model. 
For instance, similar results were obtained by fixing $\sigma_{t,h}$ to $30\%$ of the moir\'e period and keeping $f_t = f_b = 1/\sqrt{2}$ in Ref.~\cite{fischer2024theory}. 

The momentum-eigenstates describing the final states are obtained at each $k$ through the transformation 
\begin{equation}
\psi_{k,o}^{\rm wan} = \sum_{n} (\mathcal{U}_k)_{o,n} \psi_{k,n}^{\rm cont} , 
\end{equation}
where the transformation matrix $\mathcal{U}_k = U_k V_k$ is obtained by singular value decomposition of the overlap matrix 
\begin{align}
(\mathcal{O}_k)_{o,n} & = \gamma_n \sum_\ell \int {\rm d}^2 r [ \psi_{k,n}^{\rm gauss} (r; \ell) ]^*  \psi_{k,n}^{\rm cont} (r; \ell) \\ 
& = U_k S_k V_k ,
\end{align}
where the sequence of $\gamma_n$ tunes the relative contribution of the various bands and is, in our calculations, set to $(\gamma_n)_{n} = (1,1,3/4,1/2,1/4,0,0,\cdots)$ in order to perfectly describe the two topmost hole bands while overcoming potential topological obstruction due to non-vanishing Chern numbers. 

The real-space representation of the Wannier orbitals is finally obtained as 
\begin{equation}
W_{o, R} (r, \ell) = \sum_{k} e^{i k \cdot R} \psi_{k,o}^{\rm wan} (r, \ell) , 
\end{equation}
which is used to compute the interaction matrix elements. We plot them in Fig.~\ref{fig_appwannierfunction}b for $E_z = 10$meV and $\theta = 4.25$ to provide an example. We note that we use the gauge of the continuum model that is explicitly threefold symmetric (see App.~\ref{app:IVC_real_space}).
The Bloch Hamiltonian in the resulting three-orbital tight-binding model is obtained at each momentum $k$ as $\mathcal{H}_k = \mathcal{U}_k^* {\rm diag} (\varepsilon_{k,n}) \mathcal{U}_k^T$, which is Fourier transformed to provide the tunneling amplitudes between orbitals in real-space as written in Eq.~\ref{eq:H0}.
We compare its spectrum to that of the continuum in Fig.~\ref{fig_appwannierfunction}a.

\section{$E_z$ and $\theta$ dependence of the projected interactions} \label{app:interactions}

\begin{figure}[!t]
    \centering
    \includegraphics[width=\linewidth]{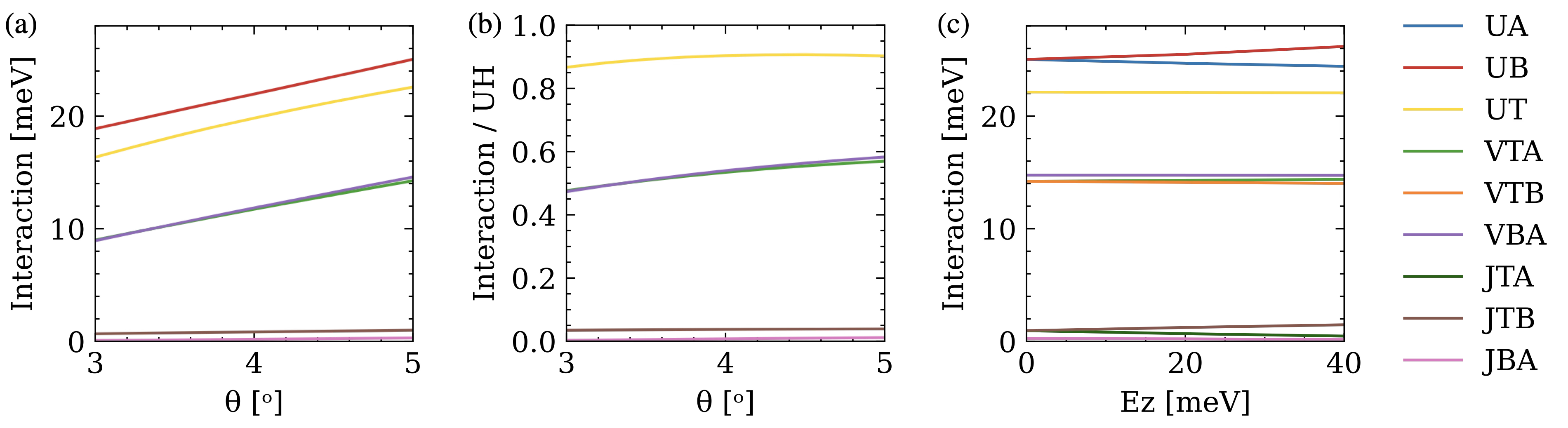}
    \caption{Projected onsite and nearest-neighbor interactions in the Wannier three-orbital model for a dielectric constant $\epsilon=48$.  
    (a) Interactions as a function of twist angle for fixed displacement field $E_z=0$. 
    (b) Ratio of the interactions to $U_H \coloneq U_\mathrm{A} = U_\mathrm{B}$ at each twist angle for fixed displacement field $E_z=0$.
    (c) Interactions as a function of displacement field for fixed twist angle $\theta=5^\circ$.}
    \label{Sfig:interactions_Ez_theta}
\end{figure}

Fig.~\ref{Sfig:interactions_Ez_theta} shows the $E_z$ and $\theta$ dependence of the onsite Hubbard $U_\alpha$, nearest-neighbor density-density $V_{\alpha\beta}$, and nearest-neighbor pair-hopping $J_{\alpha\beta}$ interactions in the Wannier three-orbital model~\cite{crepel_bridging_2024}, obtained from the projection of the long-range Coulomb interaction with dielectric constant $\epsilon=48$. We can characterize the interactions by the overall magnitude of $U_\mathrm{B}$, which is controlled by the dielectric constant, and the ratios of the rest of the interactions with respect to $U_\mathrm{B}$. We first note that the pair-hopping interactions remain smaller than $5\%$ $U_\mathrm{B}$, and we neglect them in this work. 

\begin{table*}[!t]
\caption{Onsite Hubbard interaction $U_{\mathrm{A}}=U_{\mathrm{B}}$ used in this work for different twist angles and dielectric constants.}
\begin{tabularx}{1\linewidth}{LLLLLLLLLLL}
\hline \hline
$\theta$ & $5^\circ$ & $4.25^\circ$ & $3.5^\circ$ & $3^\circ$ & $5^\circ$ & $5^\circ$ & $5^\circ$ & $5^\circ$ & $4.25^\circ$ & $3.5^\circ$\\
$\epsilon$ & 48 & 48 & 48 & 48 & 60 & 40 & 30 & 24 & 29 & 29 \\
$U_A \hspace{5pt} [\mathrm{meV}]$ & 25 & 22.7 & 20.4 & 18.8 & 20 & 30 & 40 & 50 & 45 & 33.3 \\ \hline \hline
\end{tabularx}
\label{table:UH}
\end{table*}

\begin{table}
\caption{Ratios of the interactions $U_\mathrm{T}$, $V_{\mathrm{TA}}=V_{\mathrm{TB}}$, and $V_{\mathrm{BA}}$ to $U_{\mathrm{A}}=U_{\mathrm{B}}$ used in this work for different twist angles.}
\begin{tabularx}{1\linewidth}{LLLL}
\hline \hline
                    & $U_\mathrm{T}/U_\mathrm{A}$ & $V_{\mathrm{TA}}/U_\mathrm{A}$ & $V_{\mathrm{BA}}/U_\mathrm{A}$ \\ \hline
$\theta=5^\circ$    & 0.9       & 0.57         & 0.58         \\
$\theta=4.25^\circ$ & 0.9       & 0.55         & 0.55         \\
$\theta=3.5^\circ$  & 0.89      & 0.51         & 0.51         \\ 
$\theta=3^\circ$    & 0.87      & 0.48         & 0.47         \\ \hline \hline
\end{tabularx}
\label{table:Ratios_interaction}
\end{table}

Fig.~\ref{Sfig:interactions_Ez_theta}(a) displays the dependence of the interactions with twist angle at $E_z=0$, where the interlayer $C_2$ symmetry relates the A and B sublattices, and therefore $U_\mathrm{A}=U_\mathrm{B}$, $V_{\mathrm{T}\mathrm{A}}=V_{\mathrm{T}\mathrm{B}}$, and $J_{\mathrm{T}\mathrm{A}}=J_{\mathrm{T}\mathrm{B}}$. As expected from the proportionality between the inverse moir\'e length scale and the twist angle, the interactions increase approximately linearly with $\theta$. Fig.~\ref{Sfig:interactions_Ez_theta}(b) shows that the ratios of the interactions with respect to $U_\mathrm{B}$ remain approximately constant, increasing only slightly with twist angle.

A finite $E_z>0$ breaks the degeneracy between the A and B sublattices, and the wavefunctions increase their weight on the bottom layer. Therefore, the B Wannier function becomes more localized, with the corresponding increase of $U_\mathrm{B}$ and decrease of $V_{\mathrm{T}\mathrm{B}}$, whereas $U_\mathrm{A}$ decreases and $V_{\mathrm{T}B}$ increases. However, the relative differences $2(f_\mathrm{B}-f_\mathrm{A})/(f_\mathrm{B}+f_\mathrm{A})$, for $f_\alpha = U_\alpha, V_{\mathrm{T}\alpha}$, are smaller than $5\%$, as shown in Fig.~\ref{Sfig:interactions_Ez_theta}(c). Moreover, the overall values of the interactions change by a factor smaller than $5\%$ within the range of relevant $E_z$. Therefore, for a given twist angle, in this work we have used the $E_z=0$ values of the interactions, changing only the hopping values when varying $E_z$. For completeness, we provide the values of the interactions used in this work in Tables~\ref{table:UH} and~\ref{table:Ratios_interaction}.

\section{Fermiology for different twist angles and effect of $V$ in the normal state}
\label{app:DOS_angles}

\begin{figure}[!t]
    \centering
    \includegraphics[width=1\linewidth]{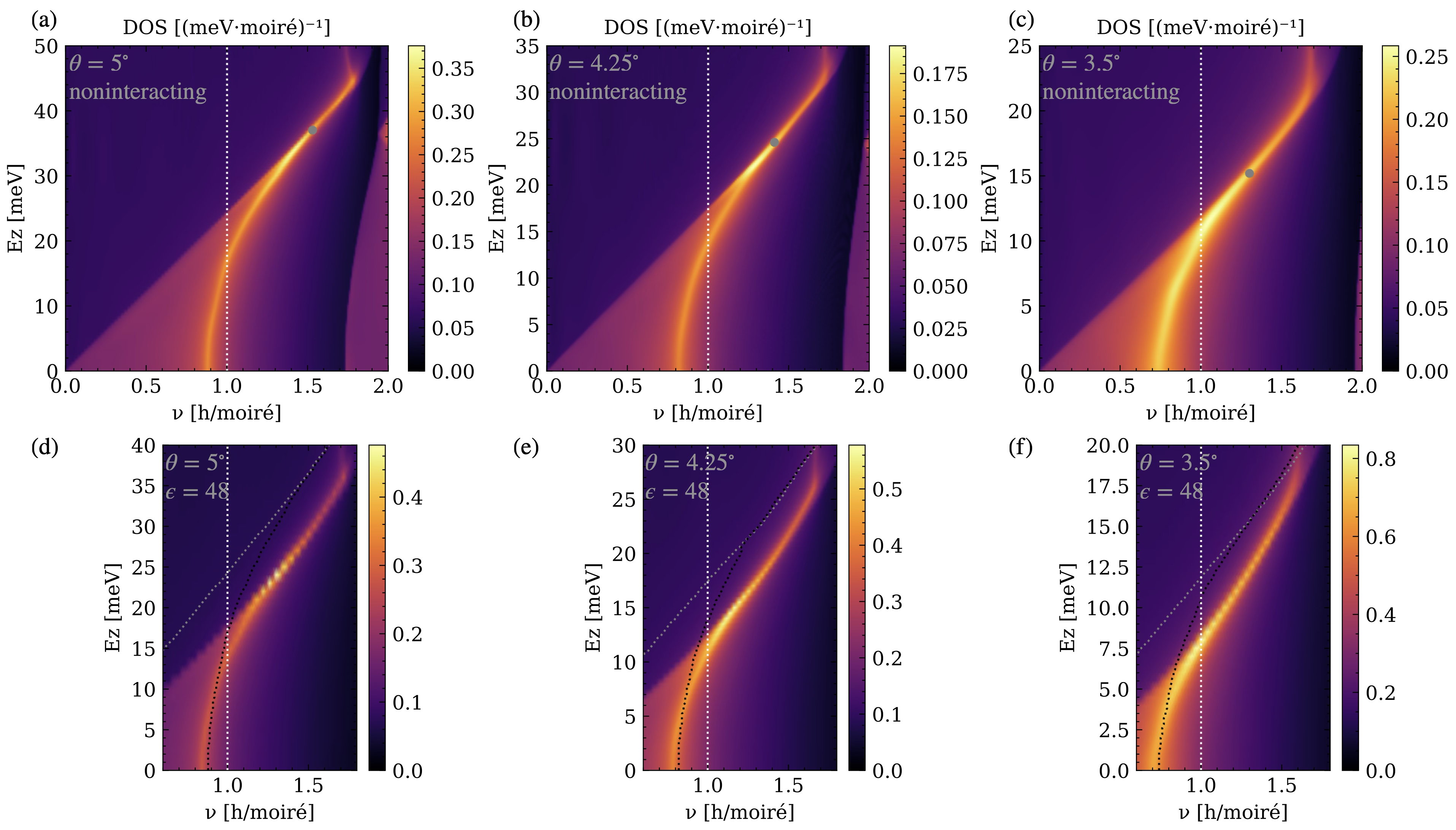}
    \caption{DOS as a function of hole filling $\nu$ and displacement field $E_z$ for twist angle $\theta=5^\circ$ (a,d), $\theta=4.25^\circ$ (b,e), and $\theta=3.5^\circ$ (c,f). 
    The first row (a,b,c) corresponds to the DOS of the noninteracting model. The gray dots mark the hoVH points, and the white vertical dotted lines indicate half filling. 
    The second row (d,e,f) shows the DOS of the Hartree-Fock normal state using a dielectric constant $\epsilon=48$. The black and gray lines represent the VH and Lifshitz transition lines in the noninteracting case.}
    \label{Sfig:DOS_angles}
\end{figure}

The first row of Fig.~\ref{Sfig:DOS_angles} shows the noninteracting DOS as a function of hole density and displacement field for three different twist angles. All show the same qualitative behavior, with the energy scales scaled by $\simeq \theta^2$ due to their smaller bandwidth. Besides this, the whole VH line is slightly pushed to lower density, including the hoVH point. Additionally, the indirect gap to the second band increases with decreasing twist angle, until it becomes a full gap at $E_z=0$ for $\theta \sim 3.5^\mathrm{o}$.

The second row of Fig.~\ref{Sfig:DOS_angles} displays the corresponding DOS in the Hartree-Fock normal state for $\epsilon=48$, defined as the self-consistent state that does not spontaneously break any symmetry. The VH and Lifshitz transition lines of the noninteracting state are also plotted for reference. Although the overall structure of the DOS is maintained, due to the Hartree shift of the nearest-neighbor interactions $V$, which favor layer polarization, the full spectrum is shifted to lower $E_z$, including the VH and Lifshitz transition lines. This can be approximately accounted for by an effective interlayer potential difference which is renormalized by a term proportional to $V$ and the self-consistently determined layer polarization, $E_z^{\mathrm{eff}} - E_z \propto V \langle n_B - n_A \rangle$. This shift affects more to the region $\nu>1$. The other effect of including the interactions in the normal state is that the VH line gets closer to the Lifshitz transition line at smaller densities, indicating that the flat portion of the band between the VH and $\kappa$ points becomes flatter, thus favoring the IVC-AFM instability. 

\begin{figure*}[!t]
    \centering
    \includegraphics[width=1\linewidth]{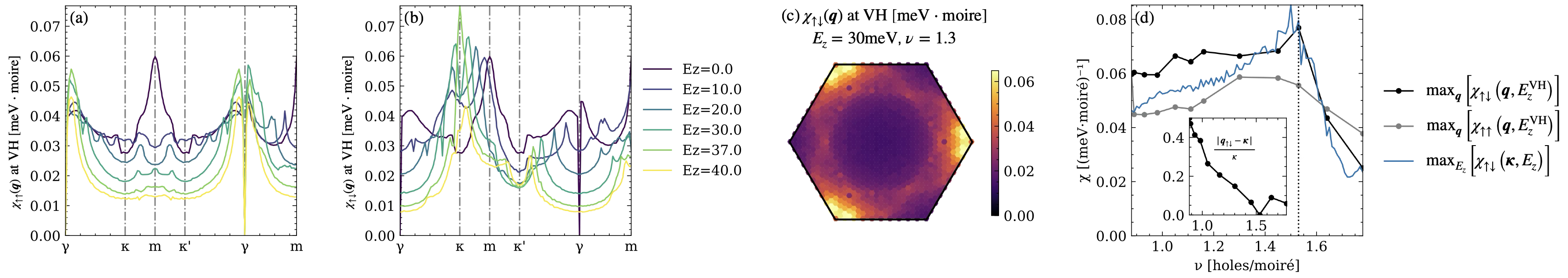}
    \caption{Momentum and density dependence of the susceptibility for $\theta=5^\circ$ and smearing $\eta=0.1$.
    (a,b) Intravalley (a) and intervalley (b) susceptibilities as a function of momentum $\boldsymbol{q}$ for several $E_z$ along the VH. The sudden drop of the intravalley susceptibility at $\boldsymbol{q}\simeq0$ stems from the finite momentum grid ($243\times243$ k-points) used to compute it. 
    (c) Intervalley susceptibility as a function of momentum $\boldsymbol{q}$ for $E_z=30\mathrm{meV}$ at the corresponding VH filling $\nu=1.3$. 
    (d) Susceptibility as a function of density. The black curve is the maximum intervalley susceptibility along the VH $\mathrm{max}_{\boldsymbol{q}}\left[\chi_{\uparrow\downarrow}\left(\boldsymbol{q},E_z^{\mathrm{VH}}\right)\right]$ (evaluated at the intervalley nesting wavevector). 
    The gray curve is the maximum intravalley susceptibility along the VH $\mathrm{max}_{\boldsymbol{q}}\left[\chi_{\uparrow\uparrow}\left(\boldsymbol{q},E_z^{\mathrm{VH}}\right)\right]$ (evaluated at the intravalley nesting wavevector).  
    The blue curve is the maximum intervalley susceptibility at $\boldsymbol{q}=\boldsymbol{\kappa}$, $\mathrm{max}_{E_z}\left[\chi_{\uparrow\downarrow}\left(\boldsymbol{\kappa},E_z\right)\right]$ (evaluated at the $E_z$ where $\chi_{\uparrow\downarrow}(\boldsymbol{\kappa})$ is maximum for each density). 
    The inset shows the density dependence of the distance between the intervalley nesting wavevector $\boldsymbol{q}_{\uparrow\downarrow}$ and the $\kappa$, in units of $\kappa$.}
    \label{Sfig:Intra_and_Inter_Susceptibility_vs_n_eta=01}
\end{figure*} 

\section{Intravalley versus intervalley susceptibility and smearing dependence}
\label{app:susceptibility}

Figs.~\ref{Sfig:Intra_and_Inter_Susceptibility_vs_n_eta=01}(a,b) shows the intravalley $\chi_{\uparrow\uparrow}(\boldsymbol{q})$ (a) and intervalley $\chi_{\uparrow\downarrow}(\boldsymbol{q})$ (b) susceptibilities as a function of the momentum $\boldsymbol{q}$ for several $E_z$ along the VH for $\theta=5^\circ$. Due to the inversion symmetry $i$ at $E_z=0$, the intravalley and intervalley susceptibilities are equal, with a relative maximum at $\boldsymbol{q=0}$ and the absolute maximum at $\boldsymbol{q}=\boldsymbol{m}$. At finite $E_z<E_z^{\mathrm{hoVH}}$, the divergence of the intervalley susceptibility at an incommensurate $\boldsymbol{q}$ is stronger, with its strength increasing approaching the hoVH from below (black curve in Fig.~\ref{Sfig:Intra_and_Inter_Susceptibility_vs_n_eta=01}(d)). This explains that, in the weak-coupling limit, the IVC-AFM develops close to the hoVH (see Figs.~\ref{fig:HF_phase_diagram_angle} and~\ref{Sfig:HF_phase_diagram}). 

As shown by the black curve in Fig.~\ref{Sfig:Intra_and_Inter_Susceptibility_vs_n_eta=01}(d), the intervalley susceptibility has a sharp decrease after the hoVH, where the $\kappa$ ($\kappa'$) points at valley $K$ ($K'$) become unoccupied by holes. The gray curve in Fig.~\ref{Sfig:Intra_and_Inter_Susceptibility_vs_n_eta=01}(d) indicates that the $\boldsymbol{q}=0$ intravalley susceptibility becomes the leading one at $E_z>E_z^{\mathrm{hoVH}}$, although the strength of its divergence is weaker than that of the intervalley one for $E_z<E_z^{\mathrm{hoVH}}$. This explains the fact that the VP-FM phase along the VH for $E_z>E_z^{\mathrm{hoVH}}$ is weaker than the IVC-AFM for $E_z<E_z^{\mathrm{hoVH}}$ (see Figs.~\ref{fig:HF_phase_diagram_angle} and~\ref{Sfig:HF_phase_diagram}). 

\begin{figure}[!t]
    \centering
    \includegraphics[width=0.5\linewidth]{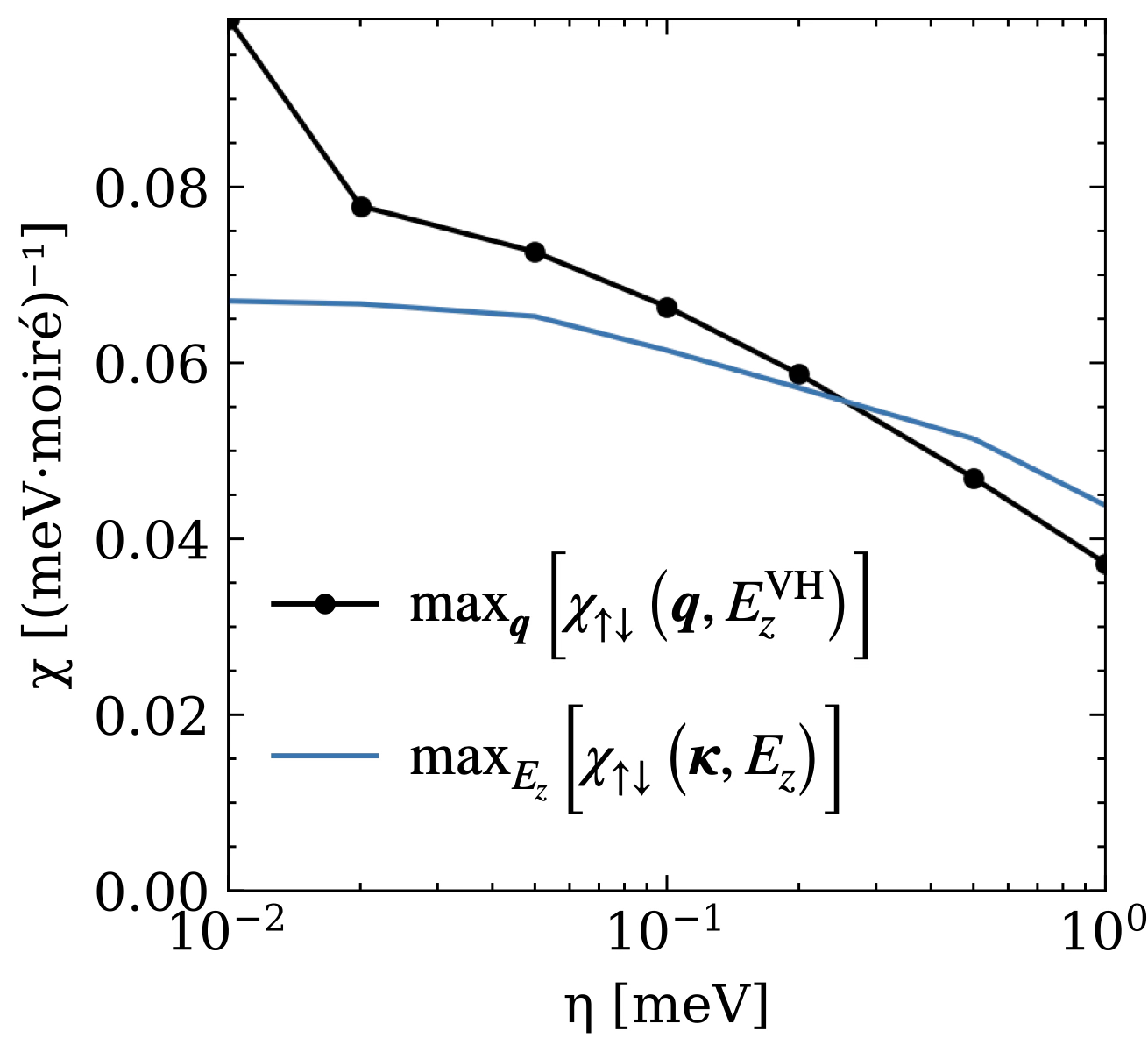}
    \caption{Dependence of the intervalley susceptibility on the smearing $\eta$ for $\theta=5^\circ$ at $\nu=1.3$. 
    The black line is the maximum intervalley susceptibility along the VH, evaluated at the intervalley nesting wavevector. 
    The blue line is the maximum intervalley susceptibility at $\boldsymbol{q}=\boldsymbol{\kappa}$, evaluated at the $E_z$ where $\chi_{\uparrow\downarrow}(\boldsymbol{\kappa})$ is maximum for each density.}
    \label{Sfig:Susceptibility_eta_dependence}
\end{figure}

In the region $E_z<E_z^{\mathrm{hoVH}}$ where the leading instability is intervalley, commensurate $\boldsymbol{q}=\boldsymbol{\kappa}$ orders will be favored for several reasons. First, the maximum value of $\chi_{\uparrow\downarrow}(\boldsymbol{\kappa})$, indicated by the blue line in Fig.~\ref{Sfig:Intra_and_Inter_Susceptibility_vs_n_eta=01}(d), is larger than $85\%$ times that at the VH nesting in a wide range of densities for $\eta=0.1\mathrm{meV} \equiv 1\mathrm{K}\equiv 10^{-3}\cdot$bandwidth. As explained in the main text, the reason behind this is the flatness of the band between the VH and $\kappa$ points. Second, the intervalley VH nesting vector is relatively close to $\kappa$, as displayed in the inset of Fig.~\ref{Sfig:Intra_and_Inter_Susceptibility_vs_n_eta=01}(d), so that commensurate $\boldsymbol{q}=\boldsymbol{\kappa}$ instabilities can be favored when increasing the interaction strength. Finally, as shown in Fig.~\ref{Sfig:Intra_and_Inter_Susceptibility_vs_n_eta=01}(c), the intervalley susceptibility for $E_z<E_z^{\mathrm{hoVH}}$ is maximum along a line encircling the $\kappa$ points. This implies a competition between different incommensurate orders with $\boldsymbol{q} = \kappa + \boldsymbol{\delta q}$ for a continuously varying direction of $\boldsymbol{\delta q}$. The fluctuations derived from this competition might further favor the commensurate $\boldsymbol{q}=\kappa$ instability.

In Fig.~\ref{Sfig:Susceptibility_eta_dependence} we show the dependence of the intervalley susceptibility with the smearing $\eta$. Increasing the smearing involves coupling to more states away from the Fermi level, and therefore cuts off the divergence of the susceptibility at the VH nesting, while at the same time allows coupling the flat portions of the band around the $\kappa$ points. Consequently, the ratio of the intervalley susceptibility at $\kappa$ to that at the VH nesting increases with $\eta$, being $>0.8$ for $\eta>0.05$ ($5\times10^{-4}$ times the bandwidth). This suggests that increasing the interaction strength will favor the commensurate instability.

\begin{figure}
    \centering
    \includegraphics[width=\linewidth]{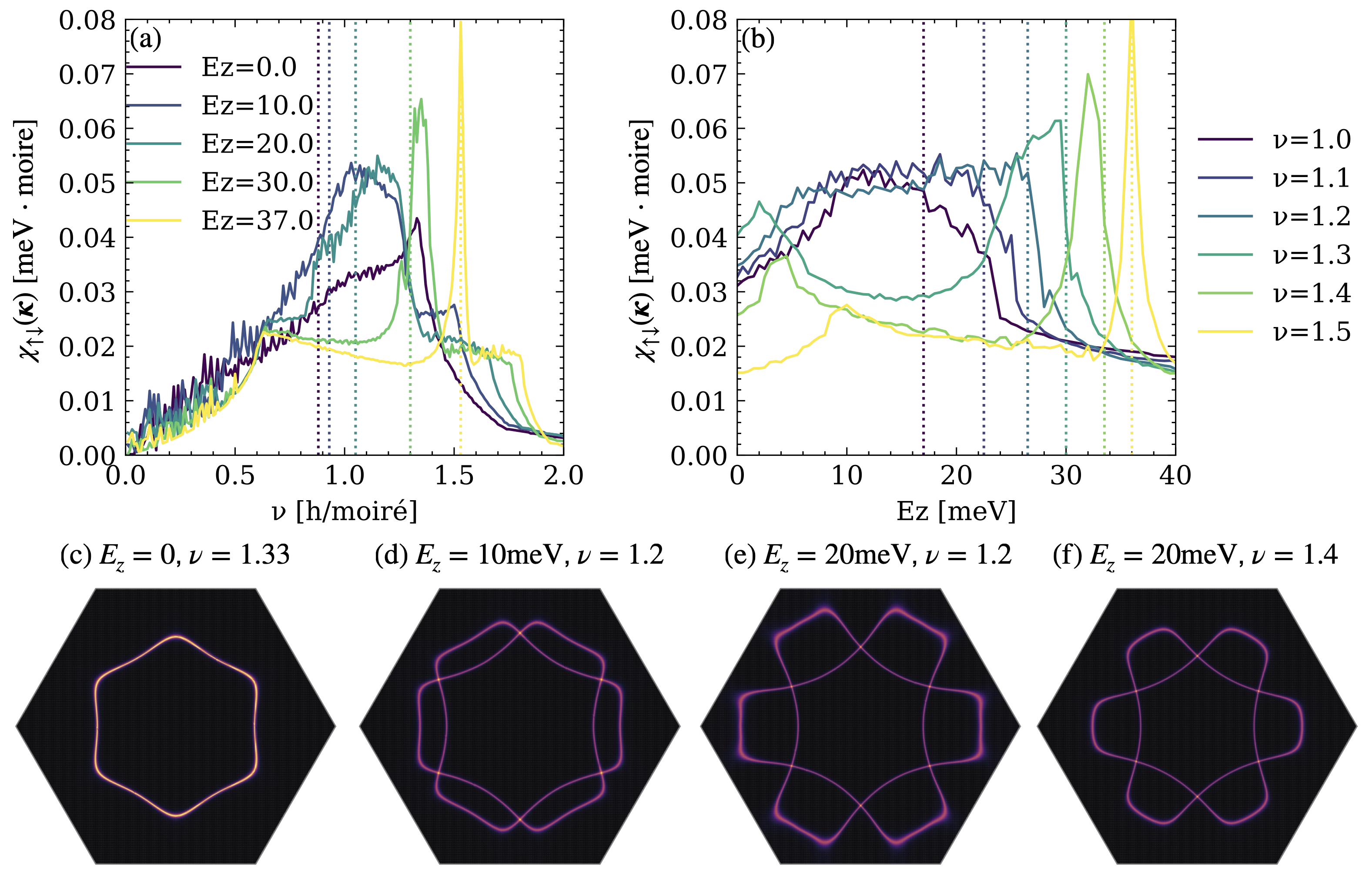}
    \caption{(a,b) Intervalley susceptibility at $\boldsymbol{q}=\boldsymbol{\kappa}$ for $\theta=5^\circ$ and smearing $\eta=0.1\mathrm{meV}$. (a) As a function of density for different displacement fields. (b) As a function of displacement field for different densities. The vertical dotted lines indicate the position of the VH for the displacement field (a) or density (b) corresponding to its color. 
    (c-f) Spectral function at the Fermi level for different $E_z$ and $\nu$ computed using $\eta=0.5\mathrm{meV}$.}
    \label{Sfig:Intervalley_susceptibility_at_kappa_line_cuts_Ez_n_FS}
\end{figure}

Figs.~\ref{Sfig:Intervalley_susceptibility_at_kappa_line_cuts_Ez_n_FS}(a,b) show line-cuts of the intervalley susceptibility at $\boldsymbol{q}=\boldsymbol{\kappa}$ of Fig.~\ref{fig:chi_nonint}(d) along different displacement fields or densities. The maximum $\chi_{\uparrow\downarrow}(\boldsymbol{\kappa})$ is reached at $E_z=E_z^{\mathrm{hoVH}}=37\mathrm{meV}$ and $\nu=\nu^{\mathrm{hoVH}}=1.5$, where the hoVHs are nested by $\kappa$. For a given $E_z<E_z^{\mathrm{hoVH}}$ in Fig.~\ref{Sfig:Intervalley_susceptibility_at_kappa_line_cuts_Ez_n_FS}(a), the peak of $\chi_{\uparrow\downarrow}(\boldsymbol{\kappa})$ becomes weaker and broader, and occurs at a density larger than the VH filling. The difference between this density and the VH filling increases with decreasing $E_z$, explaining the continuous detach of the IVC-AFM instability from the VH as the twist angle is decreased in Fig.~\ref{fig:HF_phase_diagram_angle}. The same conclusion can be reached from analyzing the $E_z$ dependence of $\chi_{\uparrow\downarrow}(\boldsymbol{\kappa})$ in Fig.~\ref{Sfig:Intervalley_susceptibility_at_kappa_line_cuts_Ez_n_FS}(b), where the peak of $\chi_{\uparrow\downarrow}(\boldsymbol{\kappa})$ shifts to increasingly smaller $E_z$ compared to the VH with decreasing density. The relatively large value of $\chi_{\uparrow\downarrow}(\boldsymbol{\kappa})$ for a range of displacement fields and densities away from the hoVH can be understood from the approximate intervalley Fermi surface nesting exemplified in Figs.~\ref{Sfig:Intervalley_susceptibility_at_kappa_line_cuts_Ez_n_FS}(c-e), which show different Fermi surfaces with the appropriate size and sharp edges nested by $\kappa$. This approximate nesting is lost at higher density due to the smaller size and round edges of the Fermi surface, as shown in Fig.~\ref{Sfig:Intervalley_susceptibility_at_kappa_line_cuts_Ez_n_FS}(f).

\section{Symmetry analysis and order parameters}
\label{app:symmetry_analysis}

\subsection{Transformation properties of Wannier functions and particle-hole bilinears}

In the normal state at finite displacement field, the point group is $C_{3v}$, generated by an intravalley $C_{3z}$ symmetry and an artificial $m_x$ mirror symmetry that mixes valleys. Taking into account the contributions from the spin $\bar{\sigma}$ in valley $\sigma$, from the $|d_\sigma\rangle = |d_{x^2-y^2}\rangle -\sigma i|d_{xy}\rangle$ orbital character at the valley $\sigma$ coming from the monolayer wavefunction at the $\sigma K$ valley, and from the envelope wavefunctions described by the TMD moir\'e continuum model~\cite{devakul2021magic}, the Wannier functions transform under $C_{3z}$ as~\cite{crepel_bridging_2024}: 
\begin{align}
    C_{3z} \phi\dag_{\mathrm{T}\sigma} &= e^{-i\sigma\pi/3} \phi\dag_{\mathrm{T}\sigma}, \\
    C_{3z} \phi\dag_{\mathrm{H}\sigma} &= - \phi\dag_{\mathrm{H}\sigma},
\end{align}
where $H=A,B$ labels the honeycomb orbitals. The mirror and time-reversal (TRS) symmetries for the three orbitals can be represented as $m_x = i \sigma_x$ and $\mathcal{T}=i\sigma_y K$, respectively, where $K$ denotes complex conjugation. 

The particle-hole bilinears $\phi\dag_{\alpha}\sigma_{\mu}\phi_{\beta}=\sum_{\sigma\sigma'}\phi\dag_{\alpha\sigma}\sigma_{\mu}^{\sigma\sigma'}\phi_{\beta\sigma'}$ therefore transform as:
\begin{align}
    & C_{3z} \phi\dag_{\mathrm{T}} \sigma_0 \phi_{\mathrm{T}} =  \phi\dag_{\mathrm{T}} \sigma_0 \phi_{\mathrm{T}}, \\
    & C_{3z} \phi\dag_{\mathrm{T}} \sigma_z \phi_{\mathrm{T}} =  \phi\dag_{\mathrm{T}} \sigma_z \phi_{\mathrm{T}}, \\
    & C_{3z} \begin{pmatrix}
        \phi\dag_{\mathrm{T}} \sigma_x \phi_{\mathrm{T}} \\
        \phi\dag_{\mathrm{T}} \sigma_y \phi_{\mathrm{T}}
    \end{pmatrix}  = \begin{pmatrix}
        -\tfrac{1}{2} & +\tfrac{\sqrt{3}}{2} \\
        -\tfrac{\sqrt{3}}{2} & -\tfrac{1}{2}
    \end{pmatrix} \begin{pmatrix}
        \phi\dag_{\mathrm{T}} \sigma_x \phi_{\mathrm{T}} \\
        \phi\dag_{\mathrm{T}} \sigma_y \phi_{\mathrm{T}}
    \end{pmatrix}, \\
    & C_{3z} \phi\dag_{\mathrm{H}} \sigma_\mu \phi_{\mathrm{H'}} =  \phi\dag_{\mathrm{H}} \sigma_\mu \phi_{\mathrm{H'}}, \\
    & C_{3z} \begin{pmatrix}
        \phi\dag_{\mathrm{T}} \sigma_0 \phi_{\mathrm{H}} \\
        -i\phi\dag_{\mathrm{T}} \sigma_z \phi_{\mathrm{H}}
    \end{pmatrix}  = \begin{pmatrix}
        -\tfrac{1}{2} & -\tfrac{\sqrt{3}}{2} \\
        +\tfrac{\sqrt{3}}{2} & -\tfrac{1}{2}
    \end{pmatrix} \begin{pmatrix}
        \phi\dag_{\mathrm{T}} \sigma_0 \phi_{\mathrm{H}} \\
        -i\phi\dag_{\mathrm{T}} \sigma_z \phi_{\mathrm{H}}
    \end{pmatrix}, \\
    & C_{3z} \begin{pmatrix}
        \phi\dag_{\mathrm{T}} \sigma_x \phi_{\mathrm{H}} \\
        \phi\dag_{\mathrm{T}} \sigma_y \phi_{\mathrm{H}}
    \end{pmatrix}  = \begin{pmatrix}
        -\tfrac{1}{2} & -\tfrac{\sqrt{3}}{2} \\
        +\tfrac{\sqrt{3}}{2} & -\tfrac{1}{2}
    \end{pmatrix} \begin{pmatrix}
        \phi\dag_{\mathrm{T}} \sigma_x \phi_{\mathrm{H}} \\
        \phi\dag_{\mathrm{T}} \sigma_y \phi_{\mathrm{H}}
    \end{pmatrix},
\end{align}
i.e., the in-plane TT spin rotates clockwise under a counterclockwise rotation, the HH' spins remain invariant under $C_{3z}$, and both the out-of-plane and in-plane TH spins rotate counterclockwise under a counterclockwise rotation. Taking into account these transformation properties, it is clear that the real space pattern of Fig.~\ref{fig:AFM_real_space}(a) is threefold symmetric around the A sites.

\subsection{Character table of the extended point group}

First, for concreteness, we define the original lattice vectors of the model in units of the lattice constant $a$ as $\boldsymbol{a}_1=(1,0)$, $\boldsymbol{a}_2=(-\tfrac{1}{2},\tfrac{\sqrt{3}}{2})$, and $\boldsymbol{a}_3=-\boldsymbol{a}_1-\boldsymbol{a}_2=(-\tfrac{1}{2},-\tfrac{\sqrt{3}}{2})$. The reciprocal lattice are therefore $\boldsymbol{g}_1=\tfrac{4\pi}{\sqrt{3}}(\tfrac{1}{2},\tfrac{\sqrt{3}}{2})$ and $\boldsymbol{g}_2=\tfrac{4\pi}{\sqrt{3}}(0,1)$, so that the $\kappa$ point momenta are $\boldsymbol{\kappa}=-\boldsymbol{\kappa}'=\tfrac{2\boldsymbol{g}_1-\boldsymbol{g}_2}{3}=(\tfrac{4\pi}{3},0)$. We also define the vectors $\boldsymbol{u}_1=\tfrac{1}{\sqrt{3}}(0,1)$, $\boldsymbol{u}_2=\tfrac{1}{\sqrt{3}}(-\tfrac{\sqrt{3}}{2},-\tfrac{1}{2})$, and $\boldsymbol{u}_3=\tfrac{1}{\sqrt{3}}(\tfrac{\sqrt{3}}{2},-\tfrac{1}{2})$, so that the positions of the orbitals within the unit cell are $\boldsymbol{d}_\mathrm{T}=0$, $\boldsymbol{d}_\mathrm{A}=-\boldsymbol{u}_2$, and $\boldsymbol{d}_\mathrm{B}=\boldsymbol{u}_1$. For the $\sqrt{3}\times\sqrt{3}$ supercell calculations, we choose the three sublattices centered at positions $\boldsymbol{h}_1=0$, $\boldsymbol{h}_2=-\boldsymbol{a}_2$ and $\boldsymbol{h}_3=\boldsymbol{a}_3$.

\begin{table}
	\begin{center}
    \caption{Character table of the extended point group $C_{3v}^{(K)}$. $C_{3\alpha}$ indicates the class of the $C_3$ symmetries around the orbital site $\alpha=\mathrm{T}, \mathrm{A}, \mathrm{B}$: $C_{3\mathrm{T}}=\{C_{3z},C_{3z}^{-1}\}$, $C_{3\mathrm{A}}=\{t_{\boldsymbol{a}_3}C_{3z}, t_{-\boldsymbol{a}_2}C_{3z}^{-1}\}$, and $C_{3\mathrm{B}}=\{t_{-\boldsymbol{a}_2}C_{3z}, t_{\boldsymbol{a}_3}C_{3z}^{-1}\}$, where $C_{3z}$ is centered at the T site. Within the $\sqrt{3}\times\sqrt{3}$ supercell, $t_{-\boldsymbol{a}_2}$ is the inverse operation of $t_{\boldsymbol{a}_3}$, so they behave as a $C_3$-like symmetry.}
	\begin{tabularx}{1\linewidth}{LLLLLLL}
		\hline \hline
		& $1E$ 	& $2t_{\boldsymbol{a}}$ 	& $2C_{3\mathrm{T}}$ 	& $2 C_{3\mathrm{A}}$ & $2 C_{3\mathrm{B}}$ & $9 m_x$	\\ \hline
		$\Gamma_1$	& 1 & 1	 & 1  & 1  & 1  & 1 \\ 
		$\Gamma_2$	& 1	& 1	 & 1  & 1  & 1  & -1 \\ 
		$\Gamma_3$	& 2	& 2	 & -1 & -1 & -1 & 0 \\ 
		$K_1$		            & 2	& -1 & 2  & -1 & -1 & 0 \\ 
		$K_2$	                & 2	& -1 & -1 & 2  & -1 & 0 \\ 
        $K_3$	                & 2	& -1 & -1 & -1 & 2  & 0 \\ \hline \hline
	\end{tabularx} 
	\label{table:character_table_extended_group}

    \caption{Representation matrices of the generators of the extended point group $C_{3v}^{(K)}$.}	
	\begin{tabularx}{1\linewidth}{LLLL}
		\hline \hline
		            & $t_{\boldsymbol{a}_3}$ & $C_{3z}$ & $m_{x}$	\\ \hline
		$\Gamma_1$	& 1		& 1		& 1		 			\\ 
		$\Gamma_2$	& 1		& 1		& -1	     		\\ 
		$\Gamma_3$	& $\begin{pmatrix} 1 & 0 \\ 0 & 1 \end{pmatrix}$
                    & $\begin{pmatrix} -\tfrac{1}{2} & -\tfrac{\sqrt{3}}{2} \\ \tfrac{\sqrt{3}}{2} & -\tfrac{1}{2} \end{pmatrix}$  
                    & $\begin{pmatrix} 1 & 0 \\ 0 & -1 \end{pmatrix}$ \\ 
		$K_1$       & $\begin{pmatrix} -\tfrac{1}{2} & -\tfrac{\sqrt{3}}{2} \\ \tfrac{\sqrt{3}}{2} & -\tfrac{1}{2} \end{pmatrix}$
                    & $\begin{pmatrix} 1 & 0 \\ 0 & 1 \end{pmatrix}$
	                & $\begin{pmatrix} 1 & 0 \\ 0 & -1 \end{pmatrix}$ \\ 
		$K_2$	    & $\begin{pmatrix} -\tfrac{1}{2} & \tfrac{\sqrt{3}}{2} \\ -\tfrac{\sqrt{3}}{2} & -\tfrac{1}{2} \end{pmatrix}$
	                & $\begin{pmatrix} -\tfrac{1}{2} & -\tfrac{\sqrt{3}}{2} \\ \tfrac{\sqrt{3}}{2} & -\tfrac{1}{2} \end{pmatrix}$
	                & $\begin{pmatrix} 1 & 0 \\ 0 & -1 \end{pmatrix}$ \\ 
        $K_3$	    & $\begin{pmatrix} -\tfrac{1}{2} & -\tfrac{\sqrt{3}}{2} \\ \tfrac{\sqrt{3}}{2} & -\tfrac{1}{2} \end{pmatrix}$
	                & $\begin{pmatrix} -\tfrac{1}{2} & -\tfrac{\sqrt{3}}{2} \\ \tfrac{\sqrt{3}}{2} & -\tfrac{1}{2} \end{pmatrix}$
	                & $\begin{pmatrix} 1 & 0 \\ 0 & -1 \end{pmatrix}$ \\ \hline \hline
	\end{tabularx}
	\label{table:representation_matrices_extended_group}
	\end{center}
\end{table}

To classify the symmetry of the order parameters, we use the irreducible representations (irreps) of the space group at the $\gamma$ and $\kappa$ points. These correspond to the irreps of the extended point group of a $\sqrt{3}\times\sqrt{3}$ supercell~\cite{basko_theory_2008,venderbos_symmetry_2016,venderbos_multi-q_2016,munoz-segovia_nematic_2023}, denoted $C_{3v}^{(K)}$. The extended point group $C_{3v}^{(K)}$ includes the point group operations of $C_{3v}$ as well as the translations by the original lattice vectors $\boldsymbol{a_3}$ and $-\boldsymbol{a}_2$, which relate the sublattices within the $\sqrt{3}\times\sqrt{3}$ supercell. The character table of the extended point group $C_{3v}^{(K)}$ is shown in Table~\ref{table:character_table_extended_group}. The intuition behind this character table is the following. The space group irreps at the $\gamma$ point are the 1D $\Gamma_1$ and $\Gamma_2$ and the 2D $\Gamma_3$, which correspond to the $A_1$, $A_2$ and $E$ irreps of the point group $C_{3v}$, and classify the $\boldsymbol{q}=0$ orders. The little group at $\kappa$ is $C_3$, which has three 1D irreps that get a phase $\exp[i2\pi n/3]$ under $C_{3z}$, with $n=0,1,-1$. The irreps of the little group at $\kappa$ acquire a phase $\exp[i\boldsymbol{\kappa}\cdot\boldsymbol{a}]$ under the translations by a original lattice vector $\boldsymbol{a}$. Time-reversal symmetry or $m_x$ impose the degeneracy between the irreps with $+n$ at $\kappa$ and $-n$ at $\kappa'$. Therefore, there are three 2D space group irreps at the $\kappa$ points, formed by the little group irreps $+n$ at $\kappa$ and $-n$ at $\kappa'$, which we denote $K_{\mathrm{mod}(n,3)+1}$, and classify the $\boldsymbol{q}=\boldsymbol{\kappa}$ orders. Physically, the three $K_n$ irreps are threefold symmetric, but around a different site: T for $K_1$, A for $K_2$, and B for $K_3$. The representation matrices for the generators of the extended point group $C_{3v}^{(K)}$ are shown in Table~\ref{table:representation_matrices_extended_group}. Particle-hole bilinears can be further classified by their parity under TRS, as well as by the $U_{\mathrm{V}}(1)$ valley symmetry, so that intravalley (intervalley) orders respect (break) $U_{\mathrm{V}}(1)$. 

Finally, we mention that at $E_z=0$, the point group becomes $D_{3d}$ due to the additional $C_{2x}$ and $i$ symmetries. In the resulting extended point group $D_{3d}^{(K)}$ for $E_z=0$, the $\Gamma_n$ and $K_1$ irreps carry an additional label indicating the $i$ parity, whereas the $K_2$ and $K_3$ irreps merge into a 4D irrep \cite{qiu_interaction-driven_2023,li_electrically_2024}. To consult the character table in this case, we refer to Ref.~\cite{basko_theory_2008}, where the extended point group $C_{6v}^{(K)}$, which is isomorphic to $D_{3d}^{(K)}$, was derived.

\subsection{Definition of the order parameters}

The IVC-AFM order parameter found in this work for $E_z>0$ transforms according to the intervalley TRS-odd $K_2$ irrep. Within the IVC-AFM phase, the system is invariant under the combination of the point group $C_3$, generated by the $C_{3z}$ symmetry centered at the A site, and the time-reversal-like combination of a $U_{\mathrm{V}}(1)$ valley rotation by $\pi$ and the physical TRS, which squares to $+1$. 
The case $E_z<0$ is related by a $C_{2x}$ transformation which reverses the roles of the $A$ and $B$ orbitals, and therefore the IVC-AFM transforms as the intervalley TRS-odd $K_3$ irrep in this case. The VP-FM phase transforms as the intravalley TRS-odd $\Gamma_2$ irrep, and the IVC-FM state transforms as the intervalley TRS-odd $\Gamma_3$ irrep. 

To define the order parameters we introduce the following notation. In the $\sqrt{3}\times\sqrt{3}$ supercell, the Wannier function $|\phi_{i,a,\alpha,\sigma\rangle}$, labeled by the supercell $i$, the sublattice $a$ within the $\sqrt{3}\times\sqrt{3}$ supercell, the orbital $\alpha$, and the spin/valley $\sigma$, is located at position $\boldsymbol{R}_i+\boldsymbol{h}_a+\boldsymbol{d}_\alpha$. To simplify the notation when defining the order parameters, we will use the notation $|\phi_{\alpha\sigma\boldsymbol{r}_\alpha}\rangle$ for the Wannier function centered at the position $\boldsymbol{r}_\alpha$. 

Up to a global $U_{\mathrm{V}}(1)$ phase, the operators transforming according to the TRS-odd $K_2$ IVC-AFM for each orbital pair read:
\begin{align}
    & \hat{\Delta}_{K_2^{\mathrm{IVC}-}}^{\mathrm{TT}} = \sum_{\boldsymbol{r}_\mathrm{T}} e^{i\boldsymbol{\kappa}\cdot\boldsymbol{r}_\mathrm{T}} 
    \phi\dag_{\mathrm{T}\uparrow\boldsymbol{r}_\mathrm{T}} \phi_{\mathrm{T}\downarrow\boldsymbol{r}_\mathrm{T}}+\mathrm{h.c.},\\  
    & \hat{\Delta}_{K_2^{\mathrm{IVC}-}}^{\mathrm{AA}} = \sum_{\boldsymbol{r}_\mathrm{T}} e^{i\boldsymbol{\kappa}\cdot\boldsymbol{r}_\mathrm{T}} \phi\dag_{\mathrm{A}\uparrow\boldsymbol{r}_\mathrm{T}+\boldsymbol{d}_\mathrm{A}} \phi_{\mathrm{A}\downarrow\boldsymbol{r}_\mathrm{T}+\boldsymbol{d}_\mathrm{A}}+\mathrm{h.c.},\\ 
    & \hat{\Delta}_{K_2^{\mathrm{IVC}-}}^{\mathrm{BB}} = 0,\\
    & \hat{\Delta}_{K_2^{\mathrm{IVC}-}}^{\mathrm{TA}} = \sum_{\boldsymbol{r}_\mathrm{T},j} e^{i\left[\boldsymbol{\kappa}\cdot\boldsymbol{r}_\mathrm{T}-\tfrac{2\pi}{3}(j+1)\right]} \phi\dag_{\mathrm{T}\uparrow\boldsymbol{r}_\mathrm{T}} \phi_{\mathrm{A}\downarrow\boldsymbol{r}_\mathrm{T}-\boldsymbol{u}_j}+\mathrm{h.c.},\\  
    & \hat{\Delta}_{K_2^{\mathrm{IVC}-}}^{\mathrm{TB}} = - \sum_{\boldsymbol{r}_\mathrm{T},j} e^{i\left[\boldsymbol{\kappa}\cdot\boldsymbol{r}_\mathrm{T}-\tfrac{2\pi}{3}(j+1)\right]} \phi\dag_{\mathrm{T}\uparrow\boldsymbol{r}_\mathrm{T}} \phi_{\mathrm{B}\downarrow\boldsymbol{r}_\mathrm{T}+\boldsymbol{u}_j}+\mathrm{h.c.}, \\
    & \hat{\Delta}_{K_2^{\mathrm{IVC}-}}^{\mathrm{BA}} = \sum_{\boldsymbol{r}_\mathrm{T},j} e^{i\left[\boldsymbol{\kappa}\cdot\boldsymbol{r}_\mathrm{T}-\tfrac{2\pi}{3}j\right]}  \phi\dag_{\mathrm{B}\uparrow\boldsymbol{r}_\mathrm{T}+\boldsymbol{d}_\mathrm{B}} \phi_{\mathrm{A}\downarrow\boldsymbol{r}_\mathrm{T}+\boldsymbol{d}_\mathrm{B}+\boldsymbol{u}_j}+\mathrm{h.c.},
\end{align}
where $j=1,2,3$ runs over the three nearest-neighbor bonds of the corresponding site. In the IVC-AFM phase found in this work, a linear combination of these operators for each orbital pair with real positive coefficients and an overall arbitrary phase condenses. 

The operators transforming according to the TRS-odd $\Gamma_2$ VP-FM are:
\begin{align}
    \hat{\Delta}_{\Gamma_2^{\mathrm{VP}-}}^{\mathrm{\alpha\alpha}} &= \sum_{\boldsymbol{r}_\alpha}  
    \phi\dag_{\alpha\uparrow\boldsymbol{r}_\alpha} \phi_{\alpha\uparrow\boldsymbol{r}_\alpha} - \phi\dag_{\alpha\downarrow\boldsymbol{r}_\alpha} \phi_{\alpha\downarrow\boldsymbol{r}_\alpha},\\  
    \hat{\Delta}_{\Gamma_2^{\mathrm{VP}-}}^{\mathrm{\alpha\beta}} &= \sum_{\langle\boldsymbol{r}_\alpha,\boldsymbol{r}_\beta\rangle} \phi\dag_{\alpha\uparrow\boldsymbol{r}_\alpha} \phi_{\beta\uparrow\boldsymbol{r}_\beta} - \phi\dag_{\alpha\downarrow\boldsymbol{r}_\alpha} \phi_{\beta\downarrow\boldsymbol{r}_\beta},
\end{align}
where, in the second line, $\beta\neq\alpha$ and $\langle\boldsymbol{r}_\alpha,\boldsymbol{r}_\beta\rangle$ indicates nearest neighbors.

\begin{figure*}[!t]
    \centering
    \includegraphics[width=1\linewidth]{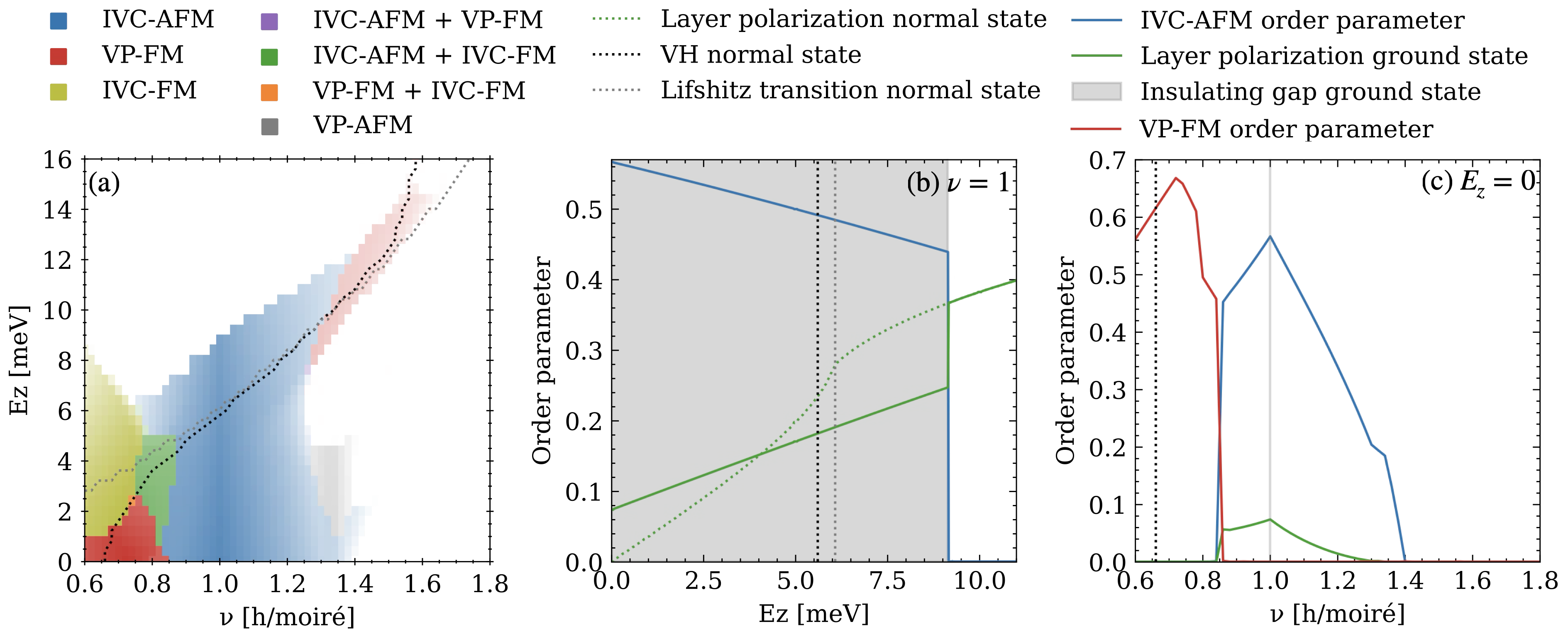}
    \caption{Hartree-Fock phase diagram of tWSe2 in a $\sqrt{3}\times\sqrt{3}$ supercell for $\theta=3^\circ$ and dielectric constant $\epsilon=48$. 
    (a) Phase diagram as a function of $\nu$ and $E_z$. Different phases are indicated by different colors: IVC-AFM (blue), VP-FM (red), IVC-FM (yellow), coexisting IVC-AFM and VP-FM (purple), coexisting IVC-AFM and IVC-FM (green), coexisting VP-FM and IVC-FM (orange), and VP-AFM (gray). The intensity of each color is proportional to the sum of the order parameters. Black dotted lines indicate the VH line in the normal state, determined as the maximum of the DOS for $\eta=0.1\mathrm{meV}$. Gray dotted lines signal the Lifshitz transition in the normal state. 
    (b) IVC-AFM order parameter (blue), and layer polarization in the ground (solid green) and normal states (dotted green) as a function of $E_z$ at $\nu=1$ for $\theta=4.25^\mathrm{o}$, $\epsilon=48$. The gray shadowed area indicates the $E_z$ where a full gap opens. 
    (c) IVC-AFM order parameter (blue), layer polarization in the ground state (solid green), and VP-FM order parameter (red) as a function of $\nu$ at $E_z=0$.}
    \label{Sfig:HF_theta=3}
\end{figure*}

Up to a global $U_{\mathrm{V}}(1)$ phase, the operators transforming according to the TRS-odd $\Gamma_3$ IVC-FM are (see also Fig.~\ref{Sfig:FM_real_space_gauges}(a)):
\begin{align}
    \hat{\Delta}_{\Gamma_3^{\mathrm{IVC}-}}^{\mathrm{TT}} &= \sum_{\boldsymbol{r}_\mathrm{T}} 
    \phi\dag_{\mathrm{T}\uparrow\boldsymbol{r}_\mathrm{T}} \phi_{\mathrm{T}\downarrow\boldsymbol{r}_\mathrm{T}}+\mathrm{h.c.},\\  
    \hat{\Delta}_{\Gamma_3^{\mathrm{IVC}-}}^{\mathrm{AA}} &= \hat{\Delta}_{\Gamma_3^{\mathrm{IVC}-}}^{\mathrm{BB}} = 0, \\
    \hat{\Delta}_{\Gamma_3^{\mathrm{IVC}-}}^{\mathrm{TA}} &= \sum_{\boldsymbol{r}_\mathrm{T},j} e^{-i\tfrac{2\pi}{3}(j+1)} \phi\dag_{\mathrm{T}\uparrow\boldsymbol{r}_\mathrm{T}} \phi_{\mathrm{A}\downarrow\boldsymbol{r}_\mathrm{T}-\boldsymbol{u}_j}+\mathrm{h.c.},\\  
    \hat{\Delta}_{\Gamma_3^{\mathrm{IVC}-}}^{\mathrm{TB}} &= - \sum_{\boldsymbol{r}_\mathrm{T},j} e^{-i\tfrac{2\pi}{3}(j+1)} \phi\dag_{\mathrm{T}\uparrow\boldsymbol{r}_\mathrm{T}} \phi_{\mathrm{B}\downarrow\boldsymbol{r}_\mathrm{T}+\boldsymbol{u}_j}+\mathrm{h.c.},\\ 
    \hat{\Delta}_{\Gamma_3^{\mathrm{IVC}-}}^{\mathrm{BA}} &= - \sum_{\boldsymbol{r}_\mathrm{T},j} e^{i\tfrac{2\pi}{3}(j+1)} \phi\dag_{\mathrm{B}\uparrow\boldsymbol{r}_\mathrm{T}+\boldsymbol{d}_\mathrm{B}} \phi_{\mathrm{A}\downarrow\boldsymbol{r}_\mathrm{T}+\boldsymbol{d}_\mathrm{B}+\boldsymbol{u}_j}+\mathrm{h.c.}.
\end{align}

Using the operators transforming according to an irrep $I$, we define the corresponding order parameter as:
\begin{equation}
\begin{split}
    \Delta_I =& \left[ \sum_{\alpha\beta} \frac{1}{2} \left( \left| \frac{1}{2} \langle \hat{\Delta}_I^{\mathrm{\alpha\beta}} + \hat{\Delta}_I^{\mathrm{\beta\alpha}} \rangle \right|^2 + \right. \right. \\
    &+ \left. \left. \left| \frac{1}{2} \langle \left( \hat{\Delta}_I^{\mathrm{\alpha\beta}} + \hat{\Delta}_I^{\mathrm{\beta\alpha}} \right)^* \rangle \right|^2 \right) \right]^{1/2}.
\end{split}
\end{equation}

We note that this symmetry classification in irreps permits a straightforward separation of the contribution of $\boldsymbol{q}=0$ orders ($\Gamma_n$ irreps) from that of $\boldsymbol{q}=\boldsymbol{\kappa}$ orders ($K_n$ irreps).

\section{Details of the Hartree-Fock calculations}

In our self-consistent iterative Hartree-Fock calculations in the $\sqrt{3}\times\sqrt{3}$ supercell, we use a Monkhorst-Pack $k$-grid with $100\times100$ $k$-points. We use one \textit{ansatz} for each irrep of the extended point group $C_{3v}^{(K)}$ (see Table~\ref{table:character_table_extended_group}), and three additional random \textit{ansatze}. We use a convergence criterion where the difference between the expectation values in two consecutive iterations is smaller than $10^{-5}$. Specifically, 
\begin{equation}
\begin{split}
    & \sum_{\alpha}\sum_{\sigma,\sigma'}\sum_{\boldsymbol{r}_\alpha}|\langle \phi\dag_{\alpha\sigma\boldsymbol{r}_{\alpha}} \phi_{\alpha\sigma'\boldsymbol{r}_{\alpha}} \rangle_{\mathrm{iter+1}} - \langle \phi\dag_{\alpha\sigma\boldsymbol{r}_{\alpha}} \phi_{\alpha\sigma'\boldsymbol{r}_{\alpha}} \rangle_{\mathrm{iter}}|^2 + \\
    & + \sum_{\alpha,\beta}\sum_{\sigma,\sigma'}\sum_{\langle\boldsymbol{r}_\alpha,\boldsymbol{r}_\beta\rangle}|\langle \phi\dag_{\alpha\sigma\boldsymbol{r}_{\alpha}} \phi_{\beta\sigma'\boldsymbol{r}_{\beta}} \rangle_{\mathrm{iter+1}} - \langle \phi\dag_{\alpha\sigma\boldsymbol{r}_{\alpha}} \phi_{\beta\sigma'\boldsymbol{r}_{\beta}} \rangle_{\mathrm{iter}}|^2 \\
    & < 10^{-10}.
\end{split}
\end{equation}

\section{Hartree-Fock phase diagram for $\theta=3^\circ$}
\label{app:HF_angle_3}

Fig.~\ref{Sfig:HF_theta=3}(a) shows the Hartree-Fock phase diagram for twist angle $\theta=3^\circ$ and dielectric constant $\epsilon=48$. The phase diagram is qualitatively similar to that at $\theta=3.5^\circ$ (see Fig.~\ref{fig:HF_phase_diagram_angle}(c)), but the IVC-AFM extends to $E_z=0$ at half-filling due to the smaller bandwidth in the $\theta=3^\circ$ case. This is further shown in the line cut at $\nu=1$ in Fig.~\ref{Sfig:HF_theta=3}(b), which shows that the full IVC-AFM is gapped at $\nu=1$, inducing spontaneous ferroelectricity at $E_z=0$~\cite{qiu_interaction-driven_2023,li_electrically_2024}. Notably, as shown in Fig.~\ref{Sfig:HF_theta=3}(c), the spontaneous ferroelectricity extends for a range of densities where the IVC-AFM is the ground state at $E_z=0$. The spontaneous ferroelectricity dies below the transition to the VP-FM phase which appears around the VH at $E_z=0$. This VP-FM phase is compatible with the magnetic circular dichroism signal of Ref.~\cite{knuppel_mak_correlated_2024}. Contrary to the case for larger angles, the maximum stability and maximum magnetization of the VP-FM are higher than those of the IVC-AFM for $\theta=3^\circ$. This suggests that, with decreasing twist angle or increasing interaction strength, the VP-FM will extend to half-filling, leading to the Quantum Anomalous Hall insulator found in tMoTe$_2$ at strong coupling~\cite{qiu_interaction-driven_2023,li_continuous_2021}, which is layer-hybridized, consistent with our VP-FM metal. Finally, we note that there is a a weak VP-AFM phase (gray) emerging from the approximate nesting point at $\nu=4/3$ and small $E_z$ (see Figs.~\ref{fig:chi_nonint}(d) and~\ref{Sfig:Intervalley_susceptibility_at_kappa_line_cuts_Ez_n_FS}(c)). This VP-AFM transforms as the intravalley TRS-odd $K_1$ irrep (see App.~\ref{app:symmetry_analysis}), and is characterized by modulations $\cos(\boldsymbol{\kappa}\cdot\boldsymbol{r})$ and $\sin(\boldsymbol{\kappa}\cdot\boldsymbol{r})$ of the out-of-plane spin in the T sites.

\begin{figure*}[!t]
    \centering
    \includegraphics[width=1\linewidth]{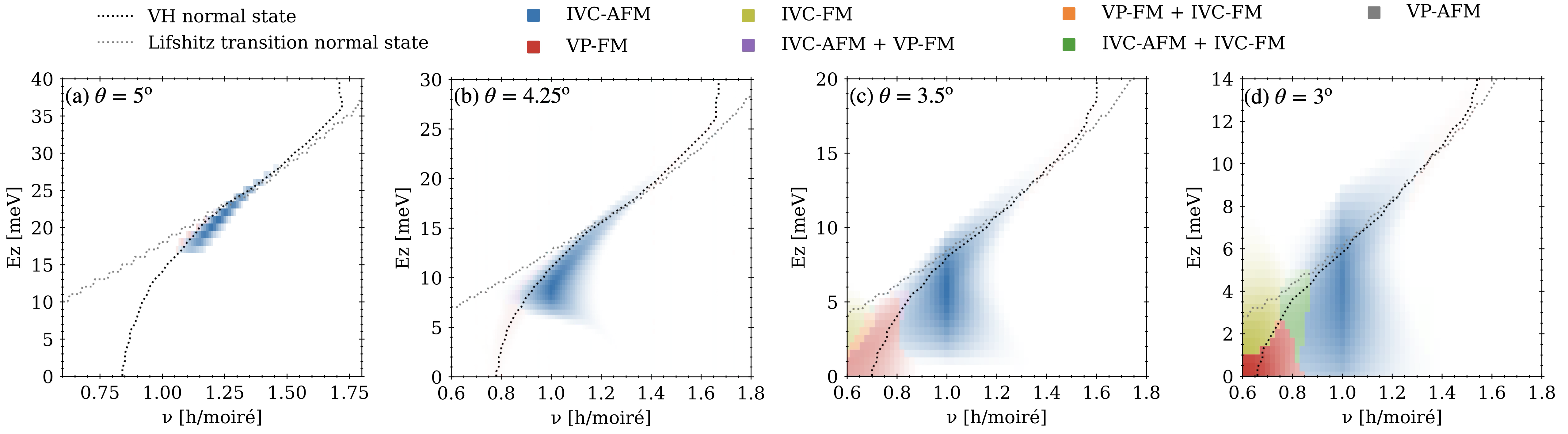}
    \caption{Hartree-Fock phase diagram of tWSe2 in a $\sqrt{3}\times\sqrt{3}$ supercell for dielectric constant $\epsilon=48$ and decreasing twist angle $\theta$: $\theta=5^\circ$ (a), $\theta=4.25^\circ$ (b), $\theta=3.5^\circ$ (c), and $\theta=3.0^\circ$ (d). Different phases are indicated by different colors: IVC-AFM (blue), VP-FM (red), IVC-FM (yellow), coexisting IVC-AFM and VP-FM (purple), coexisting IVC-AFM and IVC-FM (green), coexisting VP-FM and IVC-FM (orange), and VP-AFM (gray). 
    The intensity of each color is proportional to the energy gain of the ground state with respect to the normal state, normalized for each $\theta$. The normal state is defined as the self-consistent Hartree-Fock state which does not spontaneously break any symmetry. Black dotted lines indicate the VH line in the normal state, determined as the maximum of the DOS for $\eta=0.1\mathrm{meV}$. Gray dotted lines signal the Lifshitz transition or ``layer-polarization line'' in the normal state.}
    \label{Sfig:HF_phase_diagram_angle_energy}
\end{figure*}

\begin{figure*}[!t]
    \centering
    \includegraphics[width=1\linewidth]{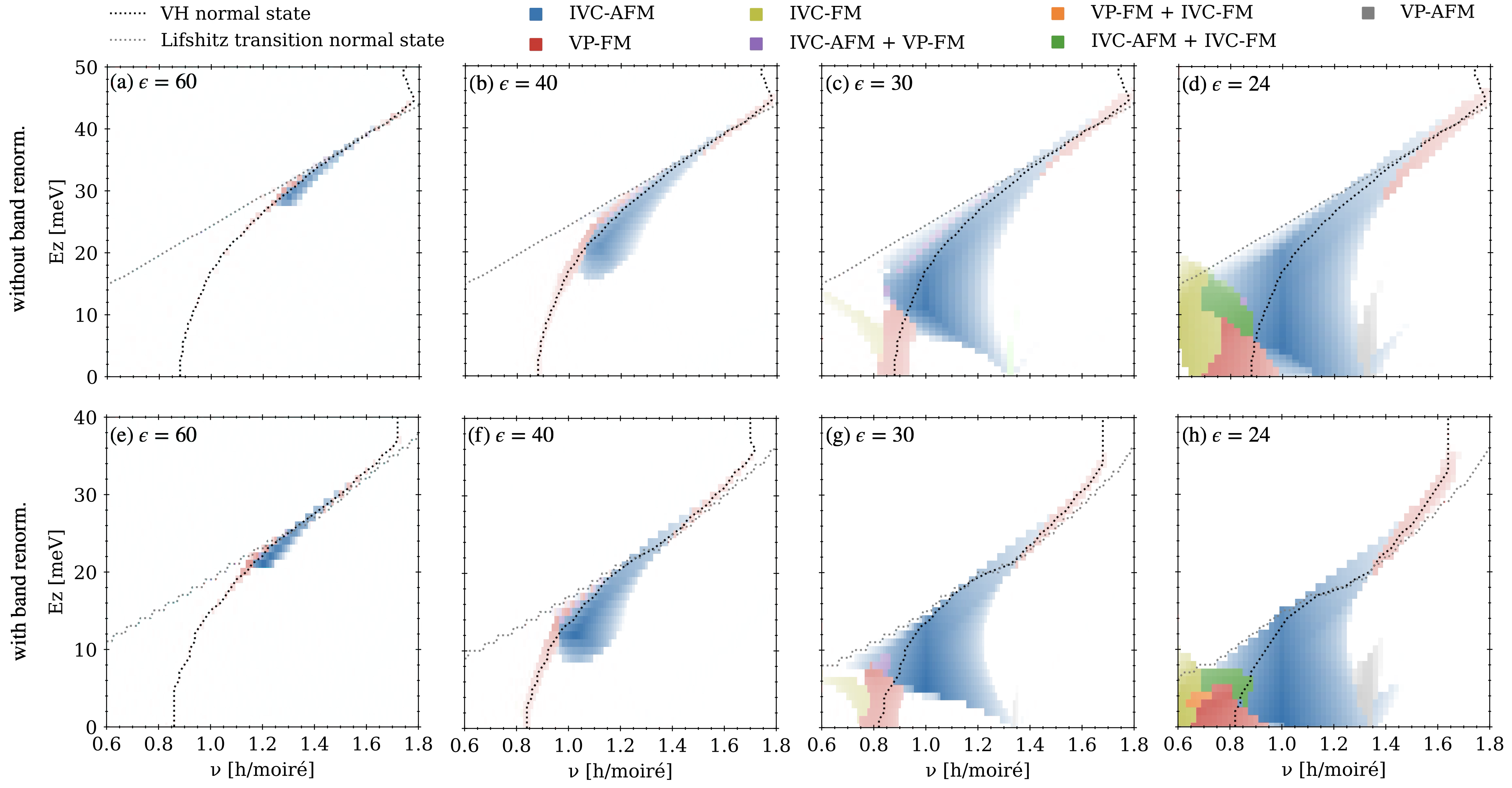}
    \caption{Hartree-Fock phase diagram of tWSe2 in a $\sqrt{3}\times\sqrt{3}$ supercell for $\theta=5^\circ$ and decreasing dielectric constant $\epsilon$. The first row neglects the symmetry-preserving band renormalizations induced by the interactions, which are included in the second row. Different phases are indicated by different colors: IVC-AFM (blue), VP-FM (red), IVC-FM (yellow), coexisting IVC-AFM and VP-FM (purple), coexisting IVC-AFM and IVC-FM (green), coexisting VP-FM and IVC-FM (orange), and VP-AFM (gray). The intensity of each color is proportional to the sum of the order parameters, normalized for each $\epsilon$. Black dotted lines indicate the VH line in the normal state, determined as the maximum of the DOS for $\eta=0.1\mathrm{meV}$. Gray dotted lines signal the Lifshitz transition or ``layer-polarization line'' in the normal state.}
    \label{Sfig:HF_phase_diagram}
\end{figure*}

\section{Twist-angle evolution of the Hartree-Fock phase diagram: stability and energy gain}
\label{app:HF_phase_diagram_angle_energy}

Fig.~\ref{Sfig:HF_phase_diagram_angle_energy} shows the Hartree-Fock phase diagram as a function of density and displacement field for decreasing twist angle. Unlike the phase diagrams of Figs.~\ref{fig:HF_phase_diagram_angle} and~\ref{Sfig:HF_theta=3}(a), where the intensity of the color indicates the magnitude of the order parameters (see App.~\ref{app:symmetry_analysis} for their definitions), the intensity of the colors in Fig.~\ref{Sfig:HF_phase_diagram_angle_energy} is proportional to the energy saved with respect to the normal state. This quantity provides a measure of the stability against thermal and quantum fluctuations. Comparing Figs.~\ref{fig:HF_phase_diagram_angle} and~\ref{Sfig:HF_theta=3}(a) with Fig.~\ref{Sfig:HF_phase_diagram_angle_energy}, we conclude that the IVC-AFM phase at $\nu=1$ is much more stable than the other orders at larger angles, and the VP-FM and IVC-FM phases only become robust at $\theta=3^\circ$. Moreover, the IVC-AFM order parameter shown in Figs.~\ref{fig:HF_phase_diagram_angle} and~\ref{Sfig:HF_theta=3}(a) displays a two-tail structure with one tail extending along the VH and another towards the approximate nesting point at $\nu=4/3$ and $E_z=0$. The energy gain in the IVC-AFM, however, is much stronger only along the VH tail, in agreement with the experimental evidence~\cite{ghiotto_stoner_2024,xia_superconductivity_2024,guo_superconductivity_2025,knuppel_mak_correlated_2024}.

\section{Evolution of the IVC-AFM with interaction strength}
\label{app:HF_interactions}

In Sec.~\ref{subsec:HF_phase_diag} and App.~\ref{app:HF_angle_3} and~\ref{app:HF_phase_diagram_angle_energy}, we presented the evolution of the Hartree-Fock phase diagram for decreasing twist angle and constant dielectric constant, with the IVC-AFM extending along the VH and detaching from it at strong coupling at $\nu=1$. Fig.~\ref{Sfig:HF_phase_diagram} shows that the same features describe its evolution with increasing dielectric constant at fixed twist angle, both neglecting and including the symmetry-preserving band renormalizations induced by the interactions, as displayed in the first and second rows of Fig.~\ref{Sfig:HF_phase_diagram}, respectively. Moreover, we can also conclude that including the symmetry-preserving band renormalizations (second row of Fig.~\ref{Sfig:HF_phase_diagram}), and therefore the direct magneto-electric coupling of the IVC-AFM to the layer polarization, favors the IVC-AFM phase. First, the IVC-AFM order parameter slightly increases. Second, at weak coupling the IVC-AFM extends to lower density for the same interaction strength, and at strong coupling it onsets at a smaller $E_z$ at half filling.

\section{Optimal layer polarization for the IVC-AFM}
\label{app:IVC-AFM-LP}

\begin{figure}
    \centering
    \includegraphics[width=1\linewidth]{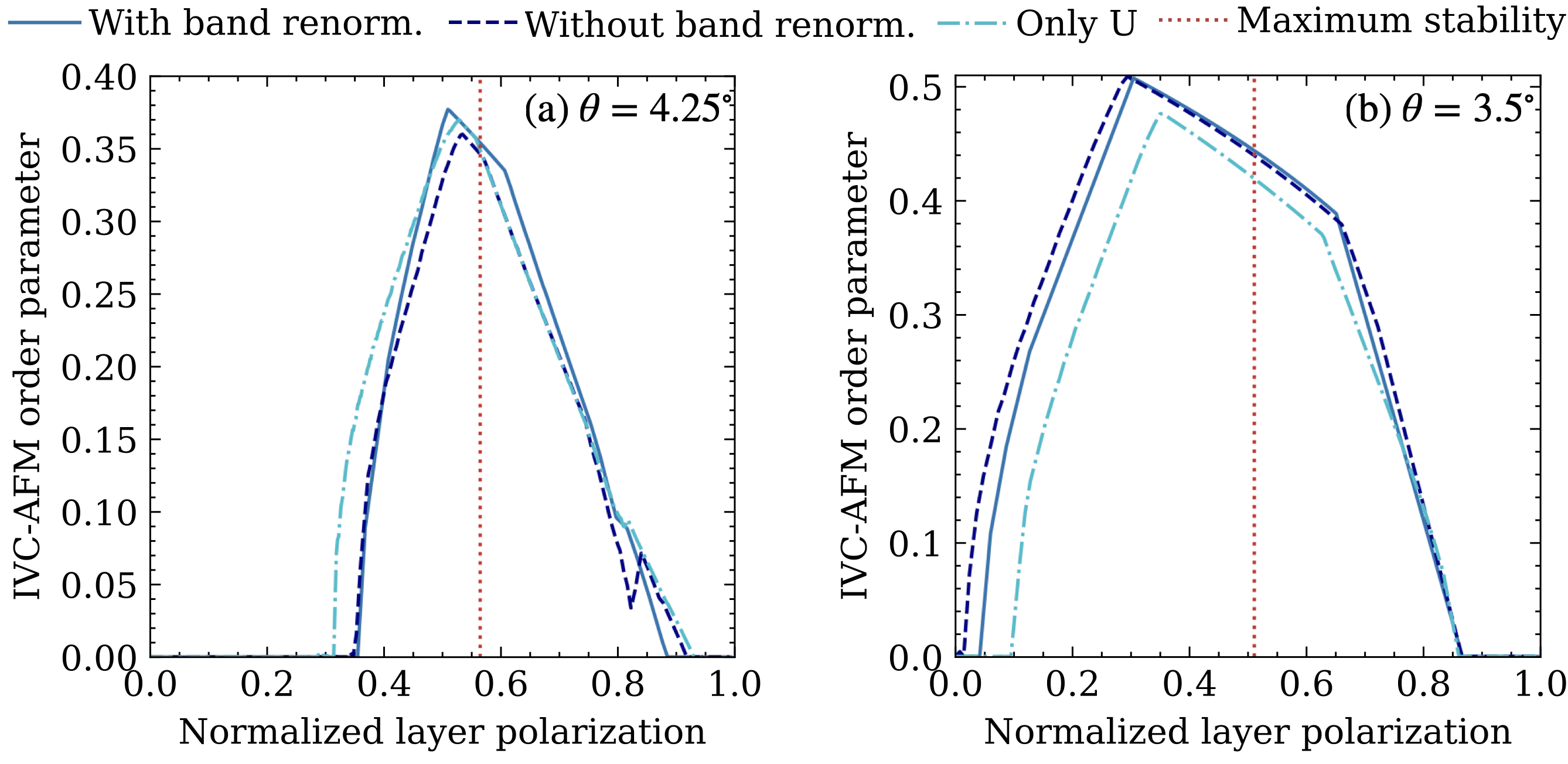}
    \caption{IVC-AFM order parameter at half-filling as a function of the normalized layer polarization $\langle n_B-n_A\rangle/\langle n_B+n_A\rangle$ for $\theta=4.25^\circ$ (a) and $\theta=3.5^\circ$ (b). 
    The blue solid line indicates the order parameter with nearest-neighbor density-density interactions $V$, the dark blue dashed line represents the order parameter with $V$ but neglecting the symmetry-preserving band renormalizations, and the cyan dash-dot line shows the order parameter with only onsite Hubbard $U$. When including the $V$, we have used $\epsilon=48$, while in the case with only $U$ we have used a smaller $\epsilon=29$ so that the magnitude of the IVC-AFM order parameter is similar in both cases. 
    The red vertical dotted line indicates the point where the energy gain of the IVC-AFM with respect to the normal state is maximum, which coincides for the three cases within $3\%$.}
    \label{Sfig:OP_vs_LP}
\end{figure}

In Sec.~\ref{sec:coupling_AFM-LP}, we have established that there is an optimal layer polarization for the stability of the IVC-AFM, which is both non-zero and non-maximal, which arises due to the layer structure of the part of the band mainly determining the energetics of the IVC-AFM. In Fig.~\ref{fig:1D_cut}, we showed the IVC-AFM order parameter as a function of $E_z$ for three cases: with nearest-neighbor density-density interactions $V$, with $V$ but neglecting the symmetry-preserving band renormalizations, and with only onsite Hubbard $U$. Due to the Hartree shift of $V$, which favors the layer polarization and renormalizes the normal state as detailed in App.~\ref{app:DOS_angles}, the IVC-AFM in the two latter cases emerges at a larger $E_z$ than in the former case. Nevertheless, when plotted against the layer polarization instead of $E_z$ as shown in Fig.~\ref{Sfig:OP_vs_LP}, the IVC order parameters for the three cases approximately collapse into an universal curve. Moreover, the insulating regions and the maximum stability of the IVC-AFM also occur for the same values of the layer polarization. Remarkably, these quantities remain at approximately the same value of the layer polarization for the twist angles $\theta=4.25^\circ$ (Fig.~\ref{Sfig:OP_vs_LP}(a)) and $\theta=3.5^\circ$ (Fig.~\ref{Sfig:OP_vs_LP}(b)). This further shows that the phase diagram is basically controlled by the attraction in the IVC-AFM channel and the self-consistently determined layer polarization, regardless of the other details of the interactions.   

\begin{figure}
    \centering
    \includegraphics[width=1\linewidth]{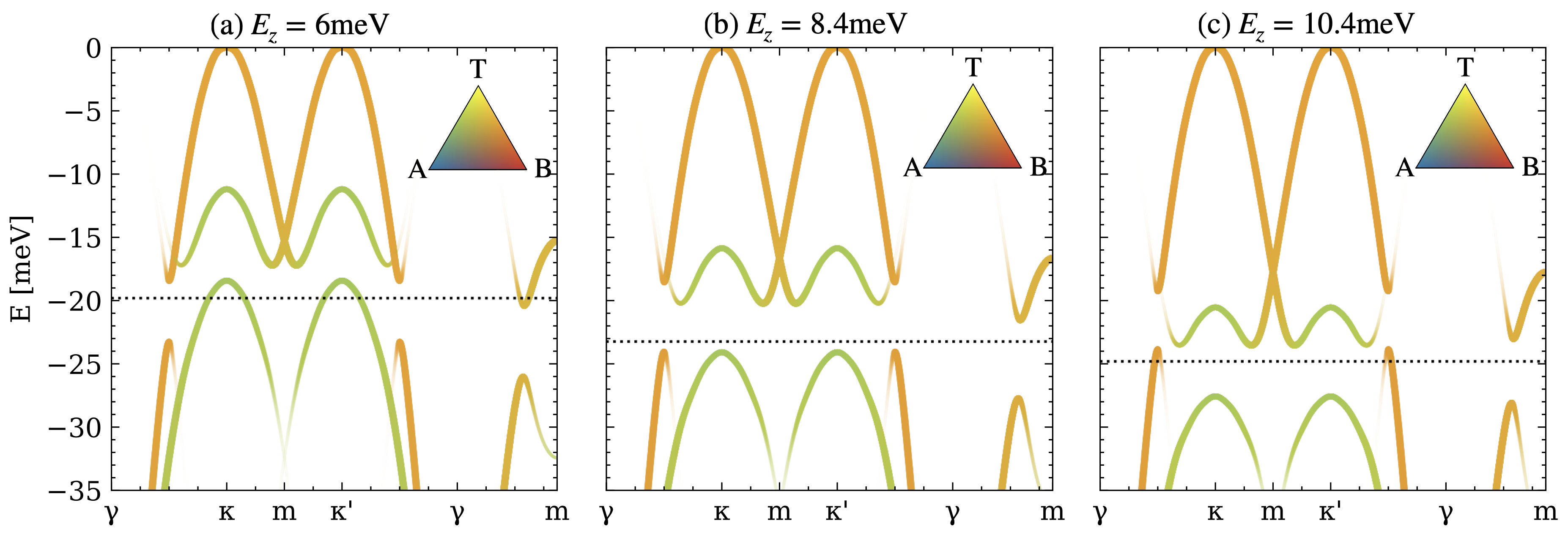}
    \caption{Reconstructed band structure within the IVC-AFM phase backfolded to the original Brilloin zone for $\theta=4.25^\circ$, $\nu=1$ and three different $E_z$.}
    \label{Sfig:Bands_half-filling_gap}
\end{figure}

\begin{figure*}[!t]
    \centering
    \includegraphics[width=\linewidth]{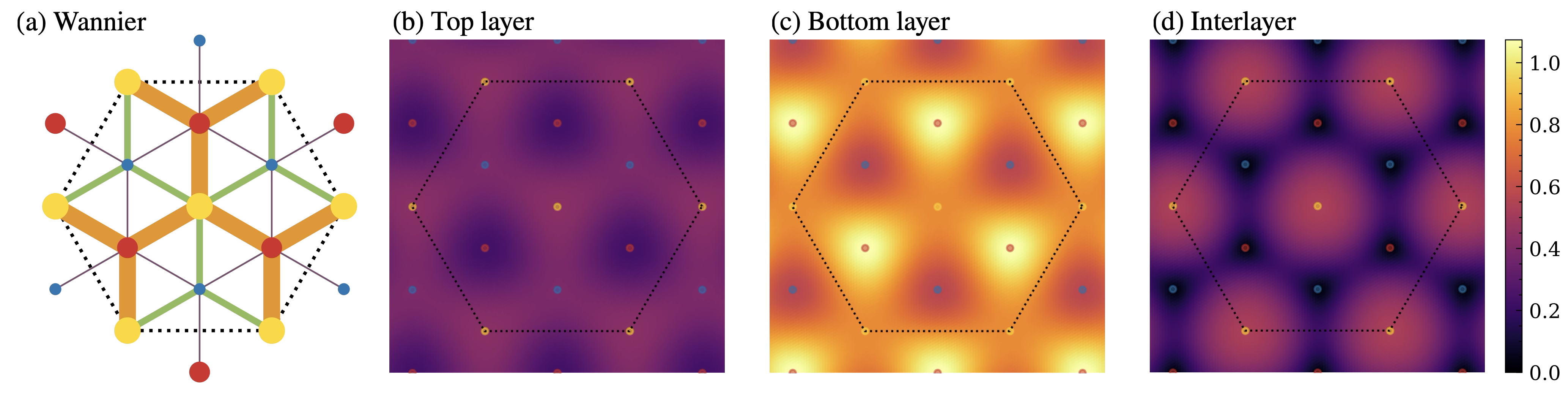}
    \caption{Real-space charge structure of the IVC-AFM ($\theta=4.25^\mathrm{o}$, $\epsilon=48$, $E_z=10\mathrm{meV}$, $\nu=1$). 
    (a) Onsite and bond charge expectation values in the Wannier orbital basis, $\langle \phi\dag_{i\alpha} \sigma_0 \phi\dag_{j\beta}\rangle$, with their magnitude indicated by the size of the dots and lines, respectively. Yellow, blue and red represent the position and onsite expectation values in the T, A and B sublattices, respectively, while green, orange and purple indicate the positions and spin expectation values in the TA, TB and BA bonds. The black dotted line corresponds to the $\sqrt{3}\times\sqrt{3}$ supercell. 
    (b,c) Total charge density in real space in the top (b), and bottom (c) layers, defined as the expectation value $\langle c\dag_l(\boldsymbol{r}) \sigma_0 c_{l}(\boldsymbol{r})\rangle$, where $c_{l\sigma}(\boldsymbol{r})$ is the quantum field operator annihilating a particle at layer $l$, valley $\sigma$ and position $\boldsymbol{r}$. 
    (d) Interlayer charge hybridization , defined as the expectation value $\sqrt{|\langle c\dag_l(\boldsymbol{r}) \sigma_0 c_{l}(\boldsymbol{r})\rangle|^2 + |\langle c\dag_l(\boldsymbol{r}) i\sigma_z c_{l}(\boldsymbol{r})\rangle|^2}$ (the $i\sigma_z$ component ensures the gauge invariance of this quantity, see App.~\ref{app:IVC_real_space}).}
    \label{Sfig:charge_density}
\end{figure*}

To further understand the connection between the layer polarization and the opening of a gap within the IVC-AFM phase, we show in Fig.~\ref{Sfig:Bands_half-filling_gap} the band structure within the IVC-AFM phase for $\theta=4.25^\circ$ at half filling and three different $E_z$. As discussed in Sec.~\ref{sec:FSR}, the main effect via which the IVC-AFM saves energy is the splitting of the smaller hole pocket centered around the $\kappa$ points coming from the $T$ and $A$ orbitals (green in Fig.~\ref{Sfig:Bands_half-filling_gap}). Due to the band folding, a partical gap also opens at the larger hole pockets around the $\kappa$ points coming from the $T$ and $B$ orbitals (orange in Fig.~\ref{Sfig:Bands_half-filling_gap}). However, for an arbitrary $E_z$ this partial gap occurs at a different energy from that of the smaller pockets due to the different chemical potential where the approximate nesting of these pockets occurs. This generically leads to a metallic band structure (see Figs.~\ref{Sfig:Bands_half-filling_gap}(a) and (c)), except for the $E_z$ where both partial gaps are aligned (see Fig.~\ref{Sfig:Bands_half-filling_gap}(b)), and the IVC-AFM is further stabilized due to the full gap opening. Due to the layer structure of these pockets, this corresponds to a certain optimal layer polarization, which is both non-zero and non-maximal.

\section{Real space structure of the IVC-AFM}
\label{app:charge_density}

\subsection{Charge distribution within the IVC-AFM}

To further understand the interplay between the spin and charge distributions within the IVC-AFM phase, Fig.~\ref{Sfig:charge_density} shows the real space distribution of the charge density for a given $E_z>0$. In the Wannier basis picture of Fig.~\ref{Sfig:charge_density}(a), the charge density is mainly located in the MM (T) and XM (B) sites. In real space, this corresponds to strong bottom layer polarization, in agreement with Figs.~\ref{Sfig:charge_density}(b,c). Notably, the spin density is mainly localized in the top layer (see Fig.~\ref{fig:AFM_real_space}), highlighting the importance of properly taking into account the layer structure for capturing the intricate details of the IVC-AFM. Fig.~\ref{Sfig:charge_density}(d) shows the interlayer charge hybridization, which, as expected, vanishes in the MX (A) and XM (B) positions.

\subsection{Real space spin structure of the IVC-AFM and continuum model gauges}
\label{app:IVC_real_space}

In this Appendix, we discuss a subtlety related to the real space structure of the magnetization in IVC phases. Its magnitude, showed in Fig.~\ref{fig:AFM_real_space} for the IVC-AFM, is a well-defined quantity. However, the direction of the magnetization of the envelope function described by the moir\'e continuum model depends on the origin of momentum chosen for each layer, which constitutes a gauge freedom within the continuum model. In IVC ordered states, besides the modulation of the magnetization of the envelope function described by the moir\'e continuum model, the physical magnetization has an additional modulation on the TMD lattice scale due to the coupling between the valleys. The gauge freedom of the continuum model described before changes the modulation at the moir\'e scale, but also that at the lattice scale, so that the direction of the physical magnetization is a well-defined observable. 

To be more specific, we can start from the moir\'e TMD continuum model in the gauge described in Refs.~\cite{wu2019topological,devakul2021magic}, where both layers are described with a common origin of momenta:
\begin{equation}
    H_\uparrow(\boldsymbol{r}) = \begin{pmatrix}
        -\frac{(\boldsymbol{k}-\boldsymbol{\kappa}_t)^2}{2m}+V_t(\boldsymbol{r}) & T(\boldsymbol{r}) \\
        T\dag(\boldsymbol{r}) & -\frac{(\boldsymbol{k}-\boldsymbol{\kappa}_b)^2}{2m}+V_b(\boldsymbol{r})
    \end{pmatrix},
\end{equation}
where $V_l(\boldsymbol{r})=2v\sum_{j=1,3,5} \cos(\boldsymbol{g}_j\cdot\boldsymbol{r}+l\psi)$ is the intralayer moir\'e potential, with $l=\pm$ the top/bottom layer, and $T(\boldsymbol{r})=w(1+e^{-i\boldsymbol{g}_2\cdot\boldsymbol{r}}+e^{-i\boldsymbol{g}_3\cdot\boldsymbol{r}})$ is the interlayer moir\'e tunneling, with $\boldsymbol{g}_j = C_6^j \boldsymbol{g}_1$ the reciprocal lattice vectors. In this gauge, due to the common momentum origin between the top and bottom layers, the Hamiltonian is translation invariant, but $C_3$ acts nontrivially due to choosing one of the three equivalent $\boldsymbol{\kappa}_t = \tfrac{1}{3}(\boldsymbol{g}_1-\boldsymbol{g}_3)$ ($\boldsymbol{\kappa}_b = \tfrac{1}{3}(\boldsymbol{g}_1+\boldsymbol{g}_2)$) to backfold to the $K$ point in the top (bottom) layer:
\begin{align}
        & C_3 H_{\uparrow}(\boldsymbol{r}) C_3^{-1} = \begin{pmatrix}
        e^{i\boldsymbol{g}_3\cdot\boldsymbol{r}} & 0 \\
        0 & e^{-i\boldsymbol{g}_1\cdot\boldsymbol{r}}
    \end{pmatrix} H_{\uparrow}(\boldsymbol{r}) \begin{pmatrix}
        e^{-i\boldsymbol{g}_3\cdot\boldsymbol{r}} & 0 \\
        0 & e^{i\boldsymbol{g}_1\cdot\boldsymbol{r}}
    \end{pmatrix} \nonumber \\
    & = \begin{pmatrix}
        -\frac{(\boldsymbol{k}-C_3 \boldsymbol{\kappa}_t)^2}{2m}+V_t(\boldsymbol{r}) & T(C_3^{-1}\boldsymbol{r}) \\
        T\dag(C_3^{-1}\boldsymbol{r}) & -\frac{(\boldsymbol{k}-C_3 \boldsymbol{\kappa}_b)^2}{2m}+V_b(\boldsymbol{r})
    \end{pmatrix}, \\
    & t(\boldsymbol{a}) H_{\uparrow}(\boldsymbol{r}) t(\boldsymbol{a})^{-1} = H_{\uparrow}(\boldsymbol{r}+\boldsymbol{a}) = H_{\uparrow}(\boldsymbol{r}), 
\end{align}
with $T(C_3^{-1}\boldsymbol{r}) = w (1+e^{i\boldsymbol{g}_1\cdot\boldsymbol{r}}+e^{i\boldsymbol{g}_2\cdot\boldsymbol{r}})$. 

The gauge degree of freedom is represented by unitary transformations of the form $\mathrm{diag}(e^{-i\boldsymbol{p}_t\cdot\boldsymbol{r}}, e^{-i\boldsymbol{p}_b\cdot\boldsymbol{r}})$, where $\boldsymbol{p}_l$ is the momentum shift of layer $l$. Another widely used gauge (see e.g. Ref.~\cite{po_origin_TBG_2018}) is the one where the origin of the momentum for each layer is in the corresponding $K$ point of each layer, which is related to $H_\uparrow(\boldsymbol{r})$ by a unitary transformation $U_\sigma(\boldsymbol{r}) = \mathrm{diag}(e^{-i\sigma\boldsymbol{\kappa}_t\cdot\boldsymbol{r}}, e^{-i\sigma\boldsymbol{\kappa}_b\cdot\boldsymbol{r}})$:
\begin{equation}
    \begin{split}
        \tilde{H}_\uparrow(\boldsymbol{r}) &= U_\uparrow(\boldsymbol{r}) H_\uparrow(\boldsymbol{r}) U^{-1}_\uparrow(\boldsymbol{r}) = \\ 
        &= \begin{pmatrix}
        -\frac{\boldsymbol{k}^2}{2m}+V_t(\boldsymbol{r}) & \tilde{T}(\boldsymbol{r}) \\
        \tilde{T}\dag(\boldsymbol{r}) & -\frac{\boldsymbol{k}^2}{2m}+V_b(\boldsymbol{r})
    \end{pmatrix},
    \end{split}
\end{equation}
where the interlayer moir\'e tunneling reads $\tilde{T}(\boldsymbol{r}) = w \sum_{n=0,1,2} e^{-iC_3^n(\boldsymbol{\kappa}_t-\boldsymbol{\kappa}_b)\cdot\boldsymbol{r}}$. This gauge is explicitly $C_3$ symmetric, but translations are nontrivial (it is translationally invariant in a $\sqrt{3}\times\sqrt{3}$ moir\'e supercell):
\begin{align}
    & t(\boldsymbol{a}) \tilde{H}_{\uparrow}(\boldsymbol{r}) t(\boldsymbol{a})^{-1} = U_\uparrow(\boldsymbol{a}) \tilde{H}_{\uparrow}(\boldsymbol{r}) U^{-1}_\uparrow(\boldsymbol{a}) = \tilde{H}_{\uparrow}(\boldsymbol{r}+\boldsymbol{a})= \nonumber \\
    & = \begin{pmatrix}
        -\frac{\boldsymbol{k}^2}{2m}+V_t(\boldsymbol{r}) & e^{-i(\boldsymbol{\kappa}_t-\boldsymbol{\kappa}_b)\cdot\boldsymbol{r}} \tilde{T}(\boldsymbol{r}) \\
        e^{i(\boldsymbol{\kappa}_t-\boldsymbol{\kappa}_b)\cdot\boldsymbol{r}} \tilde{T}\dag(\boldsymbol{r}) & -\frac{\boldsymbol{k}^2}{2m}+V_b(\boldsymbol{r})
    \end{pmatrix}, \\
        & C_3 \tilde{H}_{\uparrow}(\boldsymbol{r}) C_3^{-1} = \tilde{H}_{\uparrow}(C_3^{-1 }\boldsymbol{r}) = \tilde{H}_{\uparrow}(\boldsymbol{r}).
\end{align}

Under such gauge transformations, the eigenfunctions transform as $\tilde{\psi}_{l\sigma}(\boldsymbol{r}) = e^{-i\sigma\boldsymbol{p}_l\cdot\boldsymbol{r}} \psi_{l\sigma}(\boldsymbol{r})$. The intralayer intravalley operators, such as charge density, are invariant under such transformations, $\tilde{\psi}\dag_{l\sigma}(\boldsymbol{r}) \tilde{\psi}_{l\sigma}(\boldsymbol{r}) = \psi\dag_{l\sigma}(\boldsymbol{r}) \psi_{l\sigma}(\boldsymbol{r})$. However, interlayer and intervalley operators are not invariant, and in general $\tilde{\psi}\dag_{l\sigma}(\boldsymbol{r}) \tilde{\psi}_{l'\sigma'}(\boldsymbol{r}) = e^{i(\sigma\boldsymbol{p}_l-\sigma'\boldsymbol{p}_{l'})\cdot\boldsymbol{r}} \psi\dag_{l\sigma}(\boldsymbol{r}) \psi_{l'\sigma'}(\boldsymbol{r})$. The direction of the intralayer intervalley magnetization is therefore gauge dependent, but its magnitude is gauge invariant. 

\begin{figure}[!t]
    \centering
    \includegraphics[width=\linewidth]{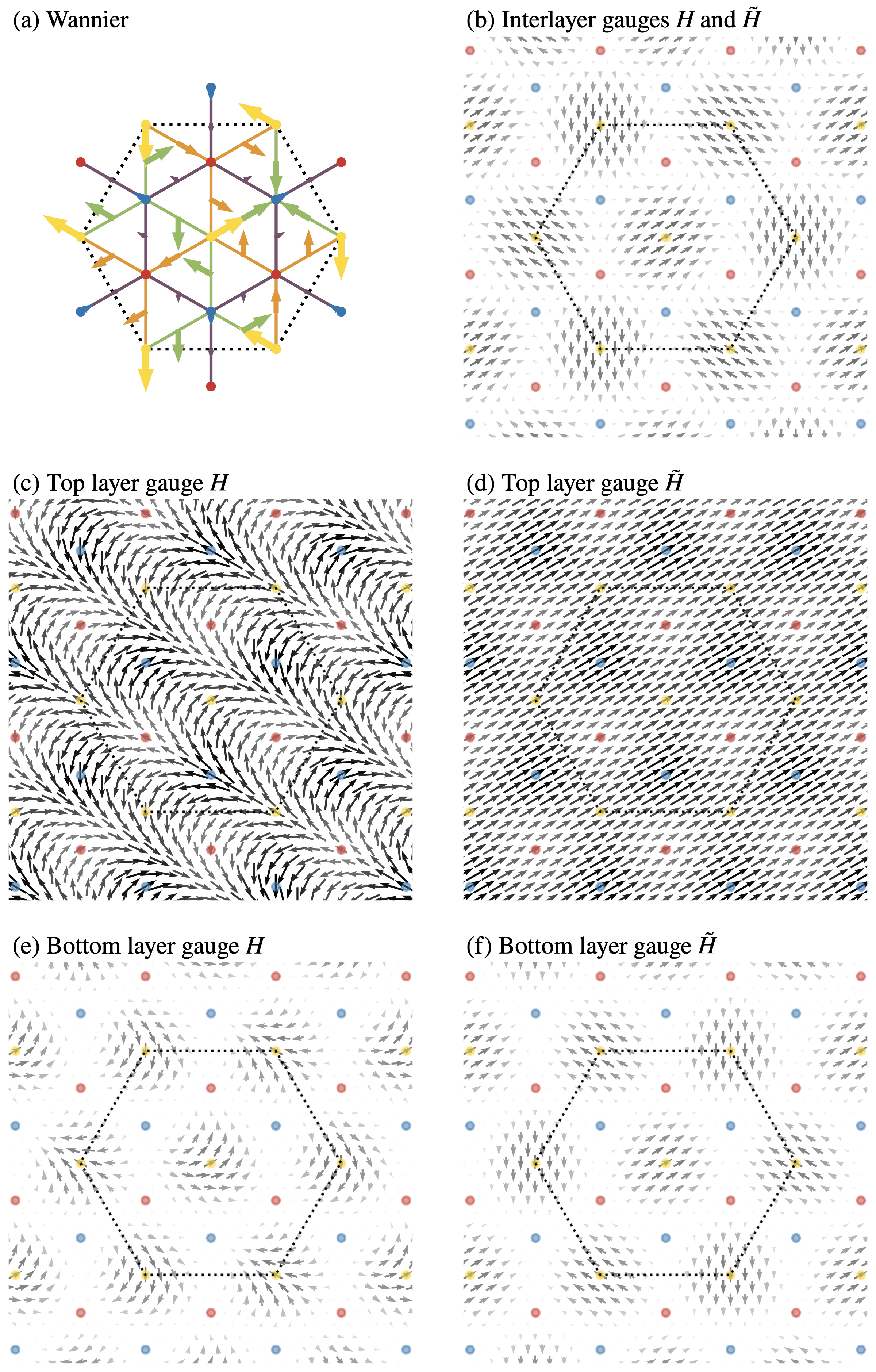}
    \caption{Real-space spin structure of the IVC-AFM in two different gauges ($\theta=4.25^\mathrm{o}$, $\epsilon=48$, $E_z=10\mathrm{meV}$, $\nu=1$). 
    (a) Onsite and bond spin expectation values in the Wannier orbital basis, $\langle \phi\dag_{i\alpha} \sigma_\mu \phi\dag_{j\beta}\rangle$, for $\mu=x,y$, with their direction and magnitude indicated by the direction and size of the arrows. Yellow, blue and red represent the position and onsite expectation values in the T, A and B sublattices, respectively, while green, orange and purple indicate the positions and spin expectation values in the TA, TB and BA bonds. The black dotted line corresponds to the $\sqrt{3}\times\sqrt{3}$ moir\'e supercell. 
    (b) Interlayer spin expectation value $\langle c\dag_t(\boldsymbol{r}) \sigma_\mu c_{b}(\boldsymbol{r})\rangle$, with $\mu=x,y$, which coincides in the $H$ and $\tilde{H}$ gauges. $c_{l\sigma}(\boldsymbol{r})$ is the quantum field operator annihilating a particle at layer $l$, valley $\sigma$ and position $\boldsymbol{r}$. 
    (c,d) Spin density in the top layer in the $H$ (c) and $\tilde{H}$ (d) gauges, defined as the expectation value $\langle c\dag_t(\boldsymbol{r}) \sigma_\mu c_{t}(\boldsymbol{r})\rangle$. 
    (e,f) Spin density in the bottom layer in the $H$ (e) and $\tilde{H}$ (f) gauges, defined as the expectation value $\langle c\dag_b(\boldsymbol{r}) \sigma_\mu c_{b}(\boldsymbol{r})\rangle$. The arrow length and color in (b-f) is normalized with respect to the maximum $\sqrt{\sum_{\mu=x,y} |\langle c\dag_l(\boldsymbol{r}) \sigma_\mu c_{l'}(\boldsymbol{r})\rangle|^2}$.}
    \label{Sfig:AFM_real_space_gauges}
\end{figure}

\begin{figure}[!t]
    \centering
    \includegraphics[width=\linewidth]{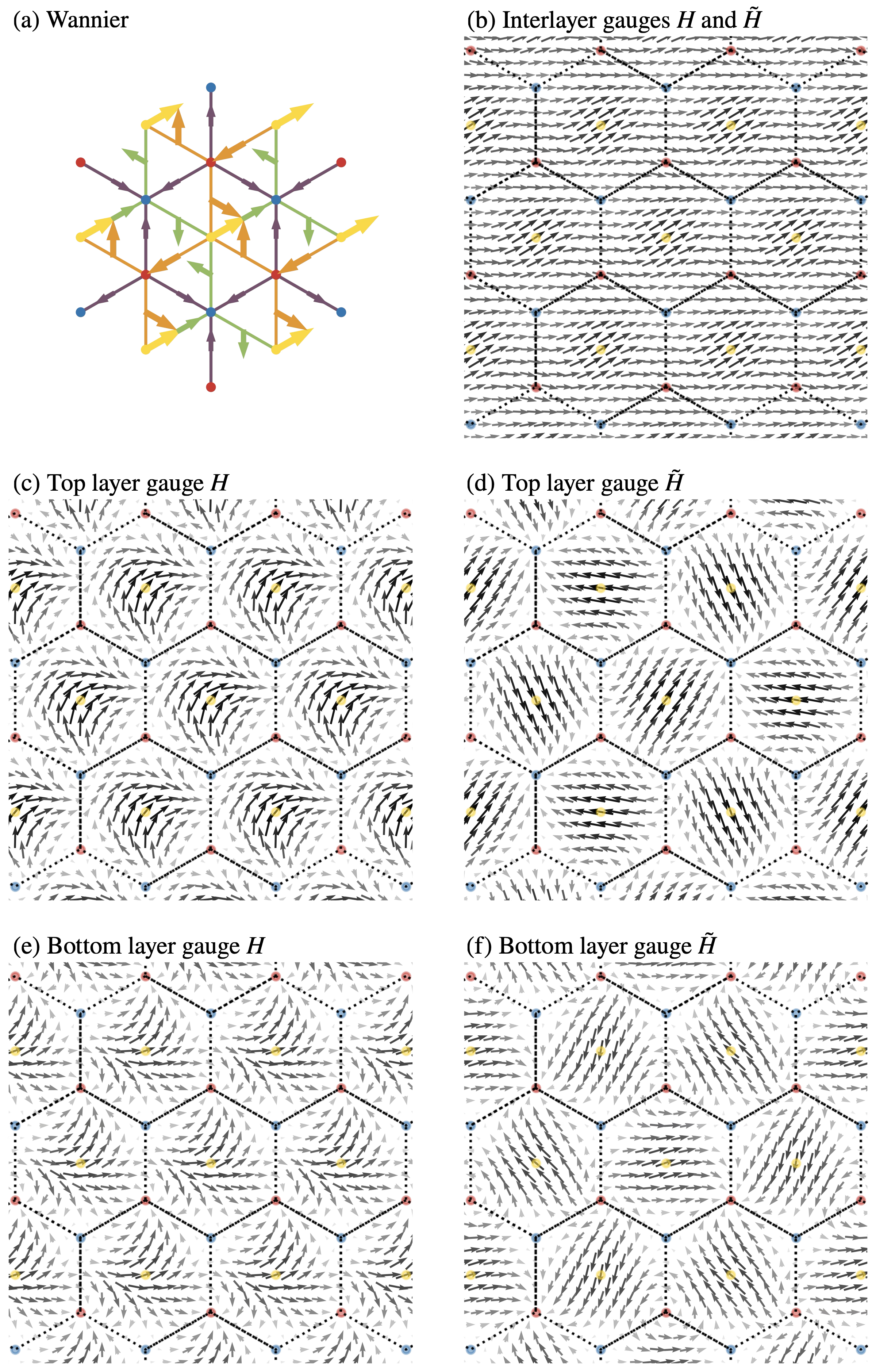}
    \caption{Real-space spin structure of the IVC-FM in two different gauges ($\theta=3.5^\mathrm{o}$, $\epsilon=48$, $E_z=3.5\mathrm{meV}$, $\nu=1$). 
    (a) Onsite and bond spin expectation values in the Wannier orbital basis, $\langle \phi\dag_{i\alpha} \sigma_\mu \phi\dag_{j\beta}\rangle$, for $\mu=x,y$, with their direction and magnitude indicated by the direction and size of the arrows. Yellow, blue and red represent the position and onsite expectation values in the T, A and B sublattices, respectively, while green, orange and purple indicate the positions and spin expectation values in the TA, TB and BA bonds. The purple BA bonds delimit the moir\'e unit cell. 
    (b) Interlayer spin expectation value $\langle c\dag_t(\boldsymbol{r}) \sigma_\mu c_{b}(\boldsymbol{r})\rangle$, with $\mu=x,y$, which coincides in the $H$ and $\tilde{H}$ gauges. $c_{l\sigma}(\boldsymbol{r})$ is the quantum field operator annihilating a particle at layer $l$, valley $\sigma$ and position $\boldsymbol{r}$. The black dotted line corresponds to the moir\'e unit cell.
    (c,d) Spin density in the top layer in the $H$ (c) and $\tilde{H}$ (d) gauges, defined as the expectation value $\langle c\dag_t(\boldsymbol{r}) \sigma_\mu c_{t}(\boldsymbol{r})\rangle$. 
    (e,f) Spin density in the bottom layer in the $H$ (e) and $\tilde{H}$ (f) gauges, defined as the expectation value $\langle c\dag_b(\boldsymbol{r}) \sigma_\mu c_{b}(\boldsymbol{r})\rangle$. The arrow length and color in (b-f) is normalized with respect to the maximum $\sqrt{\sum_{\mu=x,y} |\langle c\dag_l(\boldsymbol{r}) \sigma_\mu c_{l'}(\boldsymbol{r})\rangle|^2}$.}
    \label{Sfig:FM_real_space_gauges}
\end{figure}

Figs.~\ref{Sfig:AFM_real_space_gauges}(c-f) displays the magnetization of the IVC-AFM in the top and bottom layers in the gauges $H$ and $\tilde{H}$. Due to the unusual transformation under $C_3$ and translations in the $H$ and $\tilde{H}$ gauges, respectively, the magnetization pattern of the continuum model envelope function displays an intricate pattern. For instance, the bottom layer magnetization of Fig.~\ref{Sfig:AFM_real_space_gauges}(f) in the $\tilde{H}$ gauge is $C_3$ symmetric around the MX (A, blue) site by considering that the arrows rotate clockwise under a counterclockwise rotation, and including the phase $e^{i2\boldsymbol{\kappa}_b\cdot\boldsymbol{a}}$ obtained from the translation by a lattice vector $a$. The same applies to the top layer magnetization of Fig.~\ref{Sfig:AFM_real_space_gauges}(c) in the $\tilde{H}$ gauge, where one should also take into account that, due to the $|d_{x^2-y^2}\rangle+i|d_{xy}\rangle$ of the monolayer wavefunction at the $K$ point, the magnetization vanishes at the MX point, around which there is a vortex in the microscopic scale. The explicit $\sqrt{3}\times\sqrt{3}$ translation and $C_{3A}$ symmetries of the IVC-AFM would be recovered when including the lattice modulation of the magnetization. Notably, these gauge issues do not appear in the Wannier model, where the symmetries are represented as usual, which constitutes another advantage for using this approach in the Hartree-Fock calculations. We also note that the interlayer magnetization is the same in the $H$ and $\tilde{H}$ gauges (see ~\ref{Sfig:AFM_real_space_gauges}(b)), and explicitly reflects the symmetries of the IVC-AFM state as in the Wannier basis.

To give further intuition into the magnetization pattern in the different gauges, Fig.~\ref{Sfig:FM_real_space_gauges} shows the magnetization in the IVC-FM state. In the Wannier basis, the IVC-FM has a strong onsite contribution from the T orbital, and a slightly stronger contribution from the TB bonds than from the TA bonds for $E_z>0$. However, since the IVC-FM appears at small $E_z$, the value of the magnetization is similar in both layers. The moir\'e translational symmetry of the IVC-FM state is explicitly reflected in the translational symmetric $H$ gauge (Figs.~\ref{Sfig:FM_real_space_gauges}(c) and (e)). In the $\tilde{H}$ gauge, the translational symmetry with lattice vector $\boldsymbol{a}$ is recovered when including the phase shift $e^{i2\boldsymbol{\kappa}_l\cdot\boldsymbol{a}}$ in each layer $l$ (Figs.~\ref{Sfig:FM_real_space_gauges}(d) and (f)). 

Finally, we outline how the microscopic modulation on top of the envelope function could be incorporated. In momentum space, the state $|\boldsymbol{k}+\boldsymbol{g},l\rangle$ in the continuum model should be replaced by the monolayer Bloch state at layer $l$ and crystal momentum $\boldsymbol{Q}_l+\boldsymbol{k}+\boldsymbol{g}$ modulo the monolayer Brillouin zone, where $\boldsymbol{Q}_l$ is the origin of momentum chosen for layer $l$. In the $H$ gauge, $\boldsymbol{Q}_l = \boldsymbol{K}_l - \boldsymbol{\kappa}_l$ is the same for both layers, while in the $\tilde{H}$ gauge, $\boldsymbol{Q}_l = \boldsymbol{K}_l$.

\section{Comparison to previous theory works}
\label{app:comparison_theory}

There are other recent theory works studying the IVC-AFM in tWSe$_2$~\cite{tuo2024theorytopo,fischer2024theory} and tMoTe$_2$~\cite{qiu_interaction-driven_2023,li_electrically_2024}. Functional renormalization group (fRG) \cite{fischer2024theory} predicts IVC-AFM order whose $120^\mathrm{o}$ N\'eel structure and symmetry, predominant weight in the MM sublattice and bonds, and evolution in the phase diagram for twist angles $4^\mathrm{o}-5^\mathrm{o}$ are in agreement with our zero-temperature Hartree-Fock phase diagrams. The main quantitative difference is the dielectric constant used, which is $\epsilon=48$ in our calculations and $\epsilon=16$ in Ref. \cite{fischer2024theory}. While Ref. \cite{fischer2024theory} uses a finite screening length of $100\text{\AA}$ and we consider the onsite and nearest-neighbor interactions projected from the long-range Coulomb interaction, this might only account for a part of the difference since their screening length is larger than the moir\'e lattice constant. A sizable part of the difference might therefore arise from the overestimation of the tendency to order in Hartree-Fock theory. While Hartree-Fock theory is less reliable than fRG, it can be applied to stronger coupling and therefore smaller twist angles, where we have analyzed the evolution of the gap and the interplay with the layer polarization, and it grants access to the properties inside the ordered phase, such as the Fermi surface reconstruction and the DOS, which we have shown to explain the observed transport signatures. 

Ref.~\cite{tuo2024theorytopo} has performed Hartree-Fock calculations in a similar three-orbital model for twist angle $3.5^\mathrm{o}$ and filling $\nu=1$ using onsite Hubbard $U_H=35\mathrm{meV}$ and $U_\mathrm{T}=25\mathrm{meV}$ interactions, also obtaining a $120^{\mathrm{o}}$ IVC-AFM phase above a certain minimum displacement field and below the layer polarization, which becomes insulating in the central part of this region. However, its weight is larger on the A sublattice and nonvanishing on the B sublattice, in contrast to the predominant T and vanishing B contributions in our case, which points to a different symmetry of their IVC-AFM. Part of the discrepancies might arise from the smaller ratio $U_\mathrm{T}/U_H$ and the approximate modeling of the displacement field in Ref.~\cite{tuo2024theorytopo}, as opposed to our Wannierization of the continuum model at each $E_z$. Moreover, due to the nearest-neighbor interactions, we have also studied how these favor sharper transitions from low density, which might be testable experimentally ~\cite{zheng2020unconventional,yasuda2021stacking,niu2022giant,vizner2021interfacial,zhang2023visualizing,dolde_electric-field_2011,block_optically_2021,bian_nanoscale_2021}. Finally, we mention that our fully-gapped ferroelectric IVC-AFM at $\nu=1$ for twist angle $3^\mathrm{o}$ is compatible with the ``$O$-$120^\circ$ AFM'' state that Refs.~\cite{qiu_interaction-driven_2023,li_electrically_2024} found for tMoTe$_2$ at half-filling.

\bibliography{bib.bib}

\end{document}